\documentclass[journal]{IEEEtran}

\usepackage{amsmath,amssymb,amsfonts,bm,mathtools}
\usepackage{graphicx}
\usepackage{booktabs}
\usepackage{multirow}
\usepackage{array}
\usepackage{url}
\usepackage{cite}
\usepackage{xcolor}
\usepackage{hyperref}
\usepackage{tabularx}

\newcommand{\x}{\bm{x}}

\newcommand{\Sph}{\mathbb{S}}
\newcommand{\Tor}{\mathbb{T}}
\newcommand{\RP}{\mathbb{RP}}
\newcommand{\CP}{\mathbb{CP}}
\newcommand{\R}{\mathbb{R}}
\newcommand{\C}{\mathbb{C}}
\newcommand{\SPD}{\mathcal{S}_{++}}
\newcommand{\PSD}{\mathcal{S}_{+}}
\newcommand{\St}{\mathrm{St}}
\newcommand{\Gr}{\mathrm{Gr}}
\newcommand{\SO}{\mathrm{SO}}

\newcommand{\tr}{\mathrm{tr}}

\newcommand{\argmin}{\mathop{\mathrm{arg\,min}}}
\newcommand{\argmax}{\mathop{\mathrm{arg\,max}}}

\newcommand{\Orth}{\mathrm{O}}

\usepackage[most]{tcolorbox}
\usepackage{multicol}

\title{Directional Statistics in Signal Processing:\\ A Geometry Driven Overview of Models, Learning, and Applications }

\author{Anders~S.~Olsen, Kanti~V.~Mardia, Dimitri~Van~De~Ville, and Morten~M{\o}rup
\thanks{A. S. Olsen is with the Neurobiology Research Unit at Copenhagen University Hospital. K. V. Mardia is with the University of Leeds. D. Van De Ville is with Ecole Polytechnique Fédérale de Lausanne (EPFL) and University of Geneva. M. M{\o}rup is with the Department of Applied Mathematics and Computer Science at Technical University of Denmark. }
}

\begin{document}
\bstctlcite{IEEEexample:BSTcontrol}
\maketitle

\begin{abstract}
Many signal-processing objects are defined by periodicity, normalization, sign or global-phase ambiguity, basis invariance, or positive-semidefinite constraints and therefore cannot be modeled faithfully in ordinary Euclidean space. This paper develops a geometry driven framework for directional statistics in signal processing, connecting signal representations to their natural manifolds or quotient spaces and to probability models compatible with their invariances. After reviewing instances of directional statistics in the signal processing literature, we review circular, toroidal, spherical, projective, Stiefel, Grassmann, and covariance-valued distributions, emphasizing their statistical assumptions, computational requirements, and dimensional limitations relevant to signal processing. We describe various modeling and estimation strategies, including analytical and gradient-based maximum likelihood, structured low-rank parameterizations, mixtures and hidden Markov models (HMMs), score-based and contrastive estimation, and geometry-aware K-means. To exemplify the consequences of geometry-aware modeling, we study multivariate phase coherence in synthetic data and functional neuroimaging, comparing toroidal phase vectors to projective, Grassmann, and positive-semidefinite representations. We also study speaker separation in a car environment, showing that directional representations outperform a standard Euclidean model. These examples demonstrate that when the geometric representation, statistical distribution, inference method, and computational regime are considered together, they have the potential to lead to improved models across a broad range of signal processing applications. 
\end{abstract}


\section{Introduction}
\IEEEPARstart{S}{ignal} processing has historically been dominated by Euclidean models. Additive Gaussian noise, least-squares estimation, linear filtering, covariance estimation, and spectral analysis all rely on Euclidean vector-space structure with extraordinary success. However, many signal representations are not Euclidean: phases are periodic; orientations may be axial; normalized vectors live on spheres; eigenvectors are identifiable only up to sign or complex phase; subspaces are invariant to basis rotations; and covariance matrices are positive semidefinite~\cite{mardia_statistics_1975,mardiaDirectionalStatistics1999}. Applying conventional statistics to these objects can lead to phase discontinuities, invalid averages, incorrect uncertainty estimates, and the general use of models that ignore the invariances induced by the signal representation. A simple example is the arithmetic average of the two angles $0$ and $2\pi$ which incorrectly gives $\pi$.

\begin{figure*}[t]
\centering
    \includegraphics[width=1\textwidth]{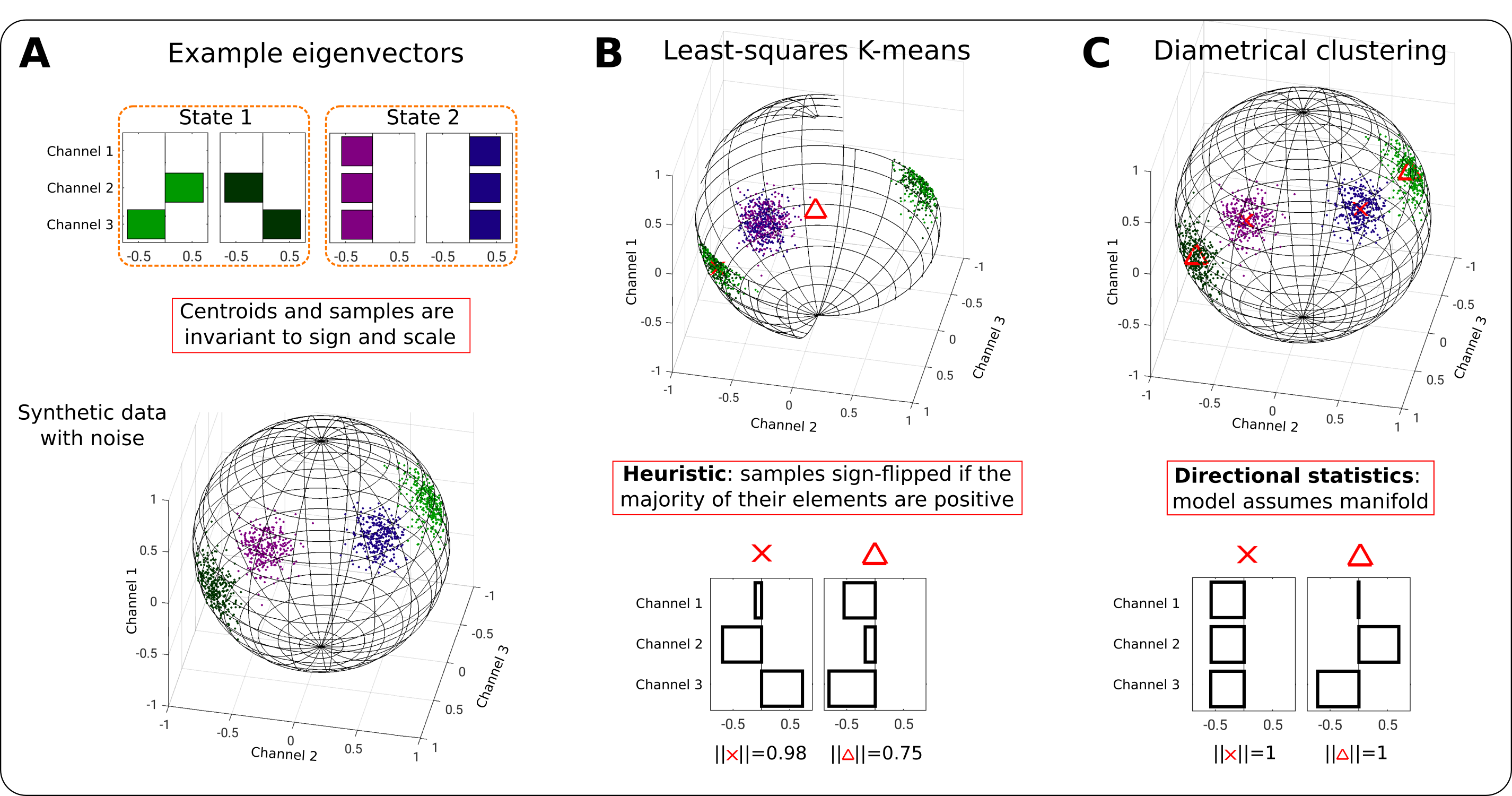}
    \caption{Comparison of clustering approaches for eigenvectors, which are defined only up to sign and scale. (A) Synthetic eigenvectors were generated around two ground-truth states on the projective space, i.e., with unit norm and random sign. (B) Least-squares $K$-means ignores the geometry in the data and requires a sign-fixing heuristic, which artificially partitions the projective space and yields distorted, scale-dependent centroids. (C) Diametrical clustering correctly uses a sign- and scale-invariant projective loss that respects the data manifold and recovers normalized prototypes. Thus, considering the geometry of the signal object rather than assumed Euclidean geometry may lead to entirely different results. }
    \label{fig:3dspheres}
\end{figure*}

As a more concrete example from neuroimaging, dynamic functional connectivity as derived from functional MRI (fMRI) concerns representing brain activity as metastable  ``brain states" reflective of intermittent brain network configurations~\cite{preti1701}. One prominent method clusters leading eigenvectors of the matrix of instantaneous pairwise cosine differences between regional time-series, which can be applied to both fMRI ~\cite{cabralCognitivePerformanceHealthy2017} and EEG~\cite{Heggli_2021} recordings. Eigenvectors are both unit norm and sign-indeterminant (in practice, the sign is completely random), see Fig.~\ref{fig:3dspheres}. The sign-indeterminacy is typically approached by heuristically flipping the sign of each observation based on the number of same-sign elements, and brain states obtained through least-squares $K$-means clustering (Fig.~\ref{fig:3dspheres}B). The sign-flipping leads to non-affine discontinuities of the sample space, potentially splitting natural modes in the data, and the centroids are not on the same sample space as the observation. If instead using \emph{diametrical clustering}, which correctly assumes that observations are sign-invariant and unit norm, computed centroids respect both the sign- and scale-invariance in the data (Fig.~\ref{fig:3dspheres}C)~\cite{psilocybin_olsen_2022}. A directional model for brain phase coherence is now state-of-the-art when considering unsupervised approaches to characterize tasks performed by subjects inside the MRI scanner~\cite{olsen_uncovering_2026}.

Directional statistics is the branch of statistics concerned with data on circles, spheres, tori, projective spaces, rotation groups, and related manifolds~\cite{mardiaDirectionalStatistics1999}. Canonical statistical distributions include the von Mises distribution on the circle and torus~\cite{vonMises1918Ganzzahligkeit}, the von Mises--Fisher distribution on the sphere~\cite{dispersion_fisher_1953}, and the Watson, Bingham, and angular central Gaussian distributions for axial data such as eigenvectors~\cite{equatorial_watson_1965,antipodally_bingham_1974,statistical_tyler_1987}. Matrix-variate models on Stiefel and Grassmann manifolds are available for orthonormal frames and subspaces~\cite{matrix_chikuse_1990}. We here also consider directional-adjacent distributions on covariance matrices on the symmetric positive definite (SPD) manifold. In this paper, we argue that matching signal representations with an appropriate \emph{sample space} guides the choice of a suitable \emph{model}, such as a statistical distribution or loss function for an optimizer, thereby improving models and uncertainty estimates. 

Some signal processing communities have adopted pieces of directional statistical theory; see Sec.~\ref{sec:applications} for an overview. However, despite the obvious benefits, directional statistics have not yet become a standard organizing language for signal processing. At the same time, parameter estimation in directional probability distributions is often more involved than Euclidean counterparts. Particularly, normalization constants can be prohibitively expensive and, as a consequence, likelihood functions may become intractable. High-dimensional regimes now common in neuroscience, communications, computer vision, and machine learning require computationally well-behaved models in non-trivial dimensionalities. In the case where repeated evaluation of normalization constants is necessary, such as in mixture model fitting using the EM algorithm, the computation time can be unacceptable. We therefore describe the computational difficulty of various models and alternative estimation schemes. 

The purpose of this overview is to provide a unified signal-processing framework connecting signal representations and their invariances with statistical models and computationally feasible learning strategies. We organize distribution families by their statistical construction and compare their assumptions, parameterizations, and computational requirements, with particular attention to high-dimensional applications. Through phase-coherence and speaker-diarization case studies, we illustrate how the information retained by a representation, the flexibility of the fitted model, and the treatment of nuisance variation jointly affect unsupervised state discovery. Specifically, we introduce sample spaces and list signal processing fields where directional statistics has been adopted in Sec.~\ref{sec:applications}. We review core distribution families and provide indications of dimensionality regimes in Sec.~\ref{sec:distributions} and discuss unsupervised methods including directional $K$-means and mixture models in Sec.~\ref{sec:state_discovery}. We describe learning strategies including EM, numerical optimization in modern libraries, and approaches for unnormalized distributions in Sec.~\ref{sec:learning}. As case studies we describe the clustering of phases in functional brain imaging data to obtain brain networks, highlighting the consequences of the choice of sample space in Sec.~\ref{sec:phase_case}, and speaker diarization in audio measurements in a car in Sec.~\ref{sec:audio}. Sec.~\ref{sec:discussion} discusses our most important points and additional relevant topics beyond this overview.

\section{Sample spaces and applications of directional statistics in signal processing}
\label{sec:applications}

\begin{figure}
    \centering
    \includegraphics[width=1\linewidth]{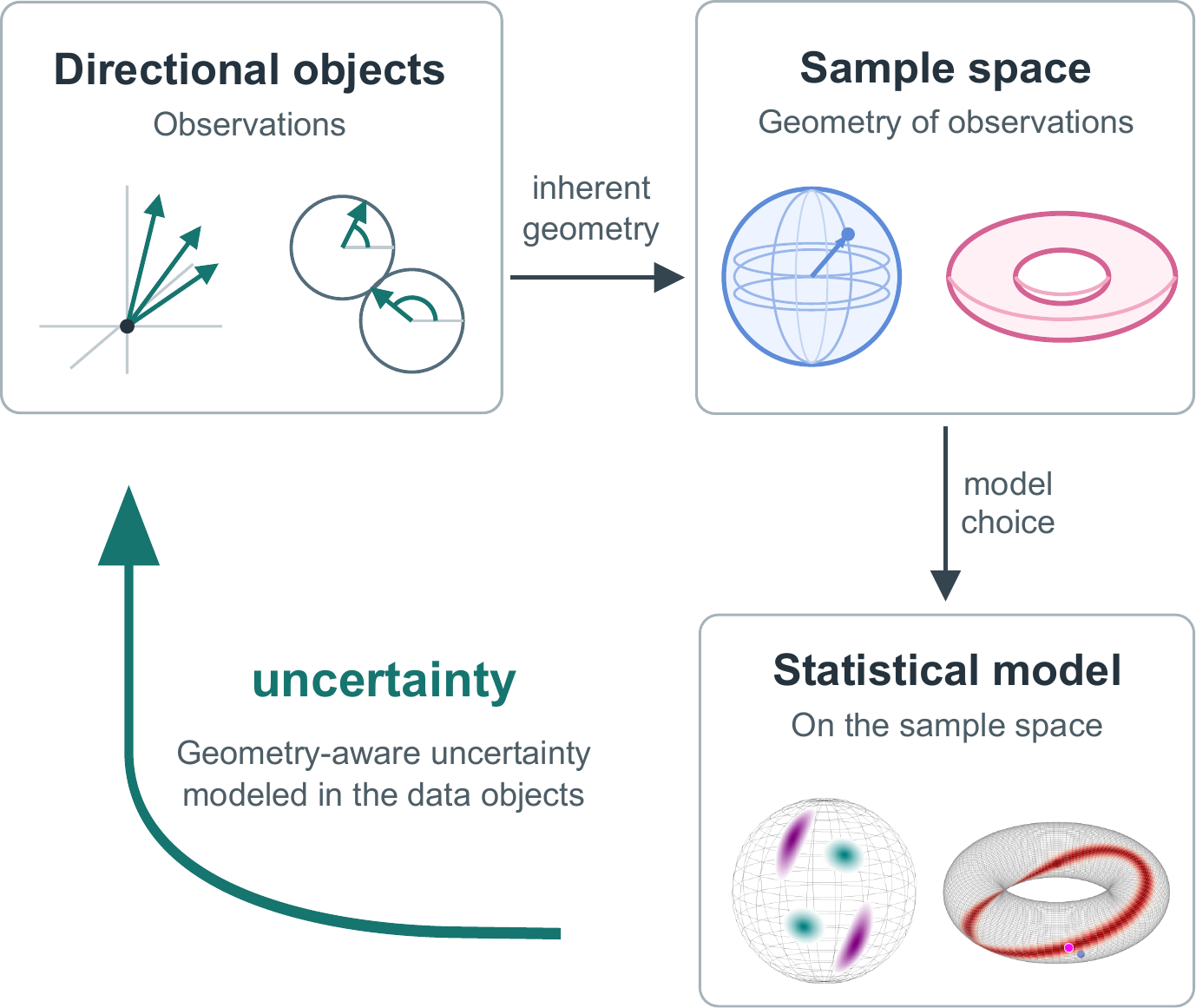}
    \caption{Geometry-aware statistical modeling of directional observations. The directional structure of the data objects identifies their natural sample space — for example, the sphere \(\Sph^2\) for three-dimensional bearings or the torus \(\Tor^p=(\Sph^1)^p\) for vectors of \(p\) phases. This geometry guides the selection of a compatible probability model. The model, in turn, characterizes uncertainty in the original data objects while respecting their geometric constraints.}
    \label{fig:framework}
\end{figure}

In many signal-processing problems, several numerical representations correspond to the same physical or scientifically meaningful state. A phase is unchanged by adding an integer multiple of $2\pi$; an eigenvector is unchanged by reversing its sign; a normalized complex beamformer is unchanged by multiplication with a global phase; and a subspace is unchanged by rotating the orthonormal basis used to represent it. These transformations define the \emph{invariances} of the signal object, and the invariances in turn define the \emph{sample space} that the signal object lives on. Knowing the sample space allows selecting a particular model or probability distribution that respects the invariances defined by the sample space, thereby providing geometry-aware modes and uncertainty in the data. This framework is illustrated in Fig.~\ref{fig:framework}.

Many common sample spaces are well known. A univariate phase lives on the circle $\Sph^1$, while a multivariate phase lives on the torus, defined as the product of independent circles $\Tor^p=(\Sph^1)^p$. A normalized vector of $p$ (ambient) dimensions lives on the $(p-1)$-dimensional sphere $\Sph^{p-1}$, where one degree of freedom is removed by the normalization. If the vector is also sign-invariant, it lives on the real projective space $\RP^{p-1}$, i.e., an antipodally defined unit sphere. A matrix with orthonormal columns generally lives on the Stiefel manifold but if the general orthogonal rotation of the basis is irrelevant, including column sign and ordering, it lives on the Grassmann manifold. SPD matrices such as covariances occupy a cone. Accounting for invariance in data observations is consequential because statistical operations should be invariant to the same operations that the signal objects are invariant to. Technical Box~1 collects formal definitions of the geometric sample spaces considered in this paper, while Fig.~\ref{fig:manifolds} illustrates some low-dimensional manifolds.

\begin{figure}
    \centering
    \includegraphics[width=1\linewidth]{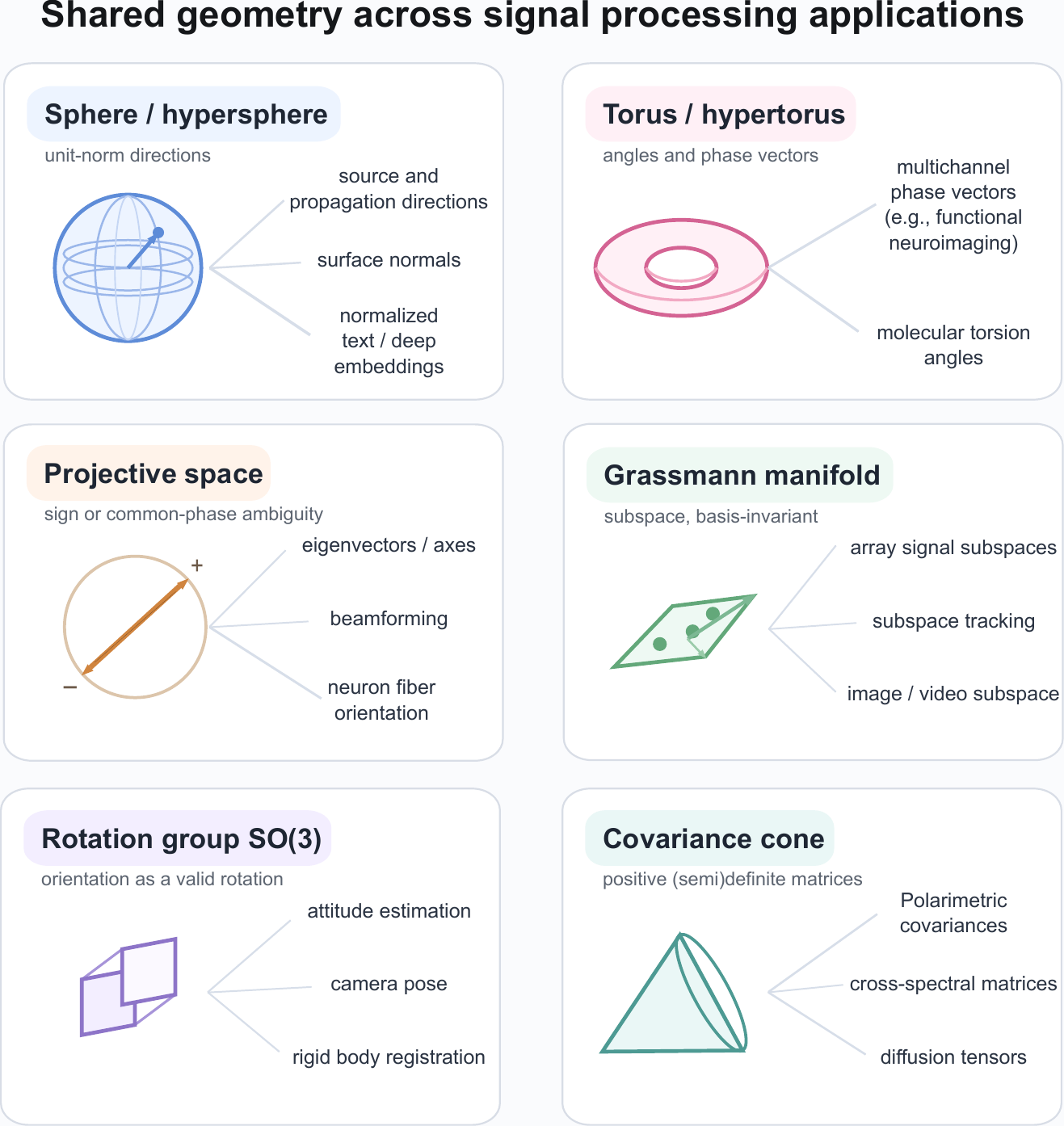}
    \caption{Geometric sample spaces arising in signal processing. Spheres, tori, projective spaces, Grassmann manifolds, rotation groups, and covariance manifolds encode distinct invariances and constraints that the listed applications naturally follow.}
    \label{fig:application_atlas}
\end{figure}

Geometric structures recur across signal-processing fields that otherwise share little terminology. A source bearing in navigation, a surface normal in 3D computer vision, and a normalized embedding in representation learning are all spherical and benefit from modeling this geometry. Similarly, an eigenvector of instantaneous fMRI connectivity, a quaternion representing a 3D rotation, and a normalized complex beamformer in array processing may all be considered projective. In this section, we review applications of directional statistics across these fields; Fig.~\ref{fig:application_atlas} summarizes how the cited literature uses these geometries. The overview refers to several statistical distributions, which are introduced in detail in Sec.~\ref{sec:distributions}.

\begin{figure*}[t]
\begin{tcolorbox}[
    colback=blue!5,
    colframe=blue!40,
    title=\textbf{Technical Box 1}: Common geometric sample spaces
]
\begin{multicols}{2}

\paragraph{Circle.}
The unit circle is
\begin{equation*}
    \Sph^1 = \{\bm x\in\R^2:\|\bm x\|_2=1\} \cong \R/(2\pi\mathbb Z).
\end{equation*}
It represents scalar phases and planar directions under the equivalence
$\theta\sim\theta+2\pi m,\quad m\in\mathbb{Z}$. 

\paragraph{Torus.}
The $p$-torus is the product
\begin{equation*}
    \Tor^p = (\Sph^1)^p\cong\R^p/(2\pi\mathbb Z)^p.
\end{equation*}
It represents $p$ component-wise periodic variables.
Removing a common phase gives
$\Tor^p/\Sph^1\cong\Tor^{p-1}$.

\paragraph{Sphere.}
The real unit sphere is
\begin{equation*}
    \Sph^{p-1} = \{\bm x\in\R^p:\|\bm x\|_2=1\}.
\end{equation*}
It represents normalized, oriented real vectors. While its ambient dimension is $p$, its intrinsic dimension is $p-1$.

\paragraph{Real projective space.}
The real projective space is
\begin{equation*}
    \RP^{p-1} = \Sph^{p-1}/\{\pm1\},\qquad \bm x\sim-\bm x.
\end{equation*}
Equivalently, it is the set of one-dimensional real subspaces of $\R^p$.
It represents axial data such as sign-indeterminate eigenvectors.

\paragraph{Complex projective space.}
The complex projective space is
\begin{equation*}
    \CP^{p-1} = \{\bm z\in\C^p:\bm z^{\mathrm H}\bm z=1\}/\Sph^1,\qquad\bm z\sim e^{i\alpha}\bm z.
\end{equation*}
It is the set of one-dimensional complex subspaces of $\C^p$ and represents
normalized complex vectors with arbitrary global phase. Its real dimension is
$2(p-1)$.

\paragraph{Special orthogonal group.}
The rotation group is
\begin{equation*}
    \SO(p) = \{\bm R\in\R^{p\times p}: \bm R^\top\bm R=\bm I_p,\ \det(\bm R)=1\}.
\end{equation*}
It represents orientation-preserving rotations and has dimension
$p(p-1)/2$. In three dimensions,
$\SO(3)\cong\Sph^3/\{\pm1\}\cong\RP^3$.

\paragraph{Stiefel manifold.}
The Stiefel manifold is
\begin{equation*}
    \St(q,p) = \{\bm U\in\R^{p\times q}: \bm U^\top\bm U=\bm I_q\}.
\end{equation*}
It represents ordered orthonormal $q$-frames, for which the individual basis
vectors remain meaningful, including their sign. Its dimension is
$pq-q(q+1)/2$.

\paragraph{Grassmann manifold.}
The Grassmann manifold is
\begin{equation*}
    \Gr(q,p) = \St(q,p)/\Orth(q), \qquad \bm U\sim\bm U\bm Q, \quad \bm Q\in\Orth(q).
\end{equation*}
It represents $q$-dimensional subspaces independently of the selected
orthonormal basis and has dimension $q(p-q)$.

\paragraph{Positive-semidefinite matrices.}
The symmetric positive-semidefinite cone is
\begin{equation*}
    \PSD(p) = \{\bm C=\bm C^\top: \bm x^\top\bm C\bm x\geq0 \ \text{for all }\bm x\in\R^p\}.
\end{equation*}
Its positive-definite interior is denoted $\SPD(p)$. The rank-$r$ stratum is
\begin{equation*}
    \PSD^{(r)}(p) = \{\bm C\in\PSD(p):\operatorname{rank}(\bm C)=r\},
\end{equation*}
and every such matrix admits
$\bm C=\bm U\bm\Lambda\bm U^\top$ with
$\bm U\in\St(r,p)$ and $\bm\Lambda$ positive diagonal. The eigenspace encodes
directional information, while the eigenvalues encode scale or variance.

\end{multicols}
\end{tcolorbox}
\end{figure*}

\begin{figure*}[t]
\centering
    \includegraphics[width=1\textwidth]{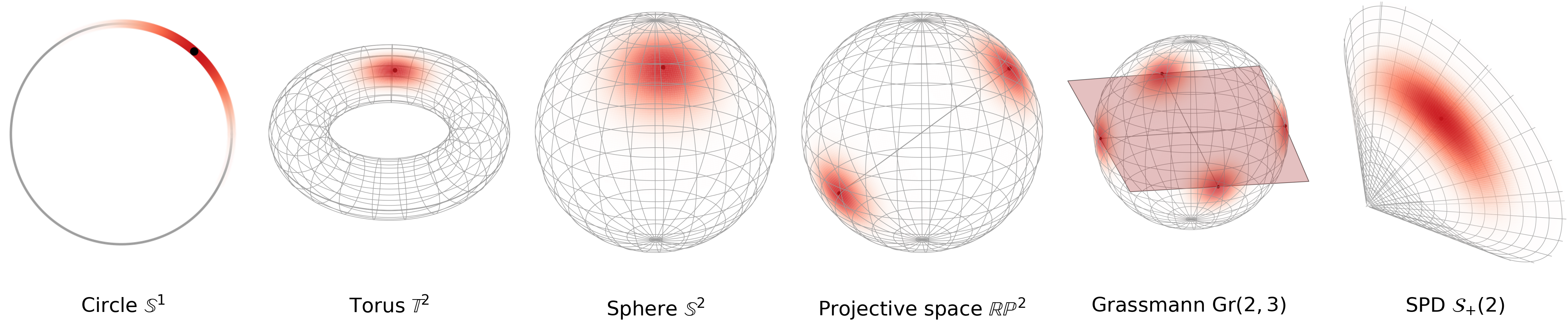}
    \caption{
        Selected geometric sample spaces relevant for directional statistics: the circle $\Sph^1$, torus $\Tor^2$, sphere $\Sph^2$, real projective space $\RP^2$, Grassmann manifold $\Gr(2,3)$, and cone of $2\times2$ symmetric positive-definite matrices $\SPD(2)$. Each panel includes a schematic probability density concentrated around a mode. Some panels are representations rather than literal Euclidean embeddings: $\RP^2$ is shown through antipodal identification on $\Sph^2$, $\Gr(2,3)$ through representative planes in $\R^3$, and $\SPD(2)$ as an open cone in the vector space of symmetric $2\times2$ matrices.
    }
    \label{fig:manifolds}
\end{figure*}

\subsection{Acoustics, microphone arrays, and localization.}
Directionality appears when the observation is a source bearing, which is circular in two dimensions and spherical in three dimensions. Directional measurement models such as the von Mises--Fisher distribution have been used for angle-of-arrival positioning and target tracking respecting this geometry, giving a more reasonable uncertainty quantification than Euclidean models~\cite{nurminen_aoa_2018,gaussian_garcafernndez_2019}. 

Microphone-array observations are commonly represented in the short-time Fourier domain as complex multichannel vectors. Upon bin-wise normalization across sensors, overall energy is removed and only the relative amplitudes and phases remain. If a shared complex phase is also irrelevant, the observation is a complex-projective object rather than an unconstrained vector. Mixture models using statistical distributions that respect this invariance, such as the Bingham and angular central Gaussian distribution, have been used for blind source separation, source counting, and time--frequency mask estimation~\cite{blind_vu_2010,source_drude_2014,modeling_ito_2016,complex_ito_2016}. Multiple-speaker tracking and diarization likewise model speaker directions on the sphere or complex projective space~\cite{multiple_traa_2014,multichannel_traa_2014,dubey_speaker_2018}. This example is treated as a case study in Sec.~\ref{sec:audio}.

\subsection{Array subspaces, beamforming, and noncoherent communications.}
In applications such as radar, sonar, and wireless communications, sensor arrays use measurements from multiple sensors or antennas to resolve incoming signals. Array-processing algorithms such as MUSIC and ESPRIT use signal and noise subspaces to estimate source directions~\cite{schmidt_music_1986,roy_esprit_1989}. These subspaces can be represented by orthonormal bases, where an orthogonal change of basis leaves the underlying span unchanged. The geometric object is therefore a point on a Grassmann manifold. Grassmannian clustering and online subspace tracking operate directly on this span, making them invariant to the choice of basis~\cite{grassmann_gruber_2006,balzano_grouse_2010}.

In wireless beamforming, a beamforming vector specifies the complex weights applied across transmitting antennas. A normalized beamforming vector has a related rank-one ambiguity: multiplying all weights by the same complex phase leaves the beam pattern unchanged. Limited-feedback MIMO systems, in which a receiver reports a preferred beamforming configuration using a small number of bits, consequently use codebooks that pack complex lines rather than ordinary vectors~\cite{love_grassmannian_2003}. In noncoherent MIMO communication, the receiver does not know the channel realization. Under the subspace signaling formulation, the information-bearing object over a coherence block is the transmitted column space, leading naturally to communication on a Grassmann manifold~\cite{zheng_grassmann_2002}.

\subsection{Attitude and pose.}
A rigid-body orientation is a rotation and is therefore part of the special orthogonal group. When 3-dimensional, they may also be considered as quaternions on the real projective space. Global orientation filters and uncertainty models therefore operate either on the rotation group or on antipodally identified quaternions. Bingham-based orientation filtering and matrix-Fisher filtering on $\SO(3)$ are geometry-respecting examples~\cite{gilitschenski_unscented_2016,lee_bayesian_2018}. 

The same rigid-body structure appears in robotics and computer vision, where a neural network may estimate the pose of a camera or object based on the Bingham or von Mises distribution~\cite{prokudin_deep_2018,deng_deepbingham_2022}. Rotation-valued uncertainty has also been used in semi-supervised pose learning~\cite{yin_fishermatch_2022}. 

\subsection{Oscillatory signals and phase coupling.}
The phase of an oscillatory signal is circularly periodic, and a vector of multichannel phases is periodic in every channel and therefore lives on the torus; specifically, the quotient torus (see Technical Box 1). In neuroscience and other multivariate oscillatory systems, joint phase models have been used to estimate coupling parameters and to distinguish direct from indirect interactions~\cite{phase_cadieu_2010,torus_klein_2019}. For high-dimensional mixture modeling of phases, the torus becomes computationally difficult due to an intractable normalizing constant. Instead, it may be modeled on the complex projective space in which the quotient torus is embedded~\cite{olsen_uncovering_2026}.

In digital communication, oscillator phase noise is a latent circular state; mixtures of circular components have been used in iterative phase tracking and decoding~\cite{shayovitz_tikhonov_2016}. Circular filtering of interferometric synthetic-aperture radar avoids artifacts at the wrap boundary compared to a Euclidean representation~\cite{huber_mork_vonmises_2001}. Exploiting multiple acquisitions from different radars, phase linking estimates a mutually consistent vector of wrapped phase differences and can be formulated as optimization over phase-only complex vectors~\cite{vu_cofi_phase_2025}. Although not directly related to signal processing, toroidal models are also commonly used for the modeling of torsion angles in proteins~\cite{carriere2026circulabasedmultivariatedistributionsflat,protein_mardia_2007}. 

\subsection{Computer vision and three-dimensional sensing.}
Several standard visual quantities are directional. Image gradients have circular orientation; surface normals from depth images or point clouds are unit vectors and may be axial when their sign is unreliable. Axial and spherical mixture models have consequently been used to cluster normals and segment depth data~\cite{unsupervised_hasnat_2014,straub_hypersphere_2015,bae_surface_2021}. Image sets and videos also produce subspace-valued representations; Grassmann-based statistical methods and deep architectures exploit this geometry in recognition tasks~\cite{statistical_turaga_2011,building_huang_2018}. Normalized visual embeddings used in recognition, retrieval, detection, tracking, and re-identification are typically spherical. Notably, directional models have here been used to represent within-class variation and uncertainty in such embeddings~\cite{shaban_directional_2022,directional_bouzid_2022,nonisotropic_kirchhof_2022}.

\subsection{Diffusion and biomedical imaging.}
Diffusion-weighted magnetic resonance imaging measures how water diffuses along different spatial directions. Fiber orientation is axial because diffusion along $\bm x$ and $-\bm x$ corresponds to the same underlying tract direction. Orientation distribution functions therefore describe one or more fiber populations on an antipodally identified sphere, i.e., the real projective space. Such models have been applied to orientation-distribution estimation and disambiguating crossing fibers~\cite{rathi_directional_2009,modeling_tabelow_2012}.

Diffusion tensor imaging provides a compact representation. Instead of retaining only one or more orientations, it summarizes diffusion within each voxel by a symmetric positive-definite matrix. The tensor eigenvectors indicate the principal diffusion directions, while the eigenvalues quantify the amount of diffusion along them~\cite{basser_diffusion_1994}. This has motivated statistical averaging, interpolation, and comparison methods that respect the geometry of positive-definite matrices rather than treating their entries as unconstrained Euclidean coordinates~\cite{fletcher_riemannian_2007}. 

\subsection{Radar, remote sensing, and covariance-valued sensing.}
Polarimetric synthetic-aperture radar represents each image location by a complex covariance or coherency matrix formed from several ``polarization" channels. This matrix is constrained to be Hermitian positive semidefinite, and its eigenvalues and eigenspaces all carry physical information. Complex Wishart models preserve this full matrix structure and have long been used for terrain classification~\cite{lee_wishart_1994}; region-level classification using distances between Wishart models has also been reported to outperform Gaussian models of vectorized amplitudes~\cite{silva_wishart_2013}. 

\subsection{Directional representations in deep learning.}
Deep networks frequently normalize hidden features or classifier weights so that cosine similarity becomes the operative score. This representation is spherical, and a directional likelihood can attach an interpretable uncertainty or concentration to it. This idea has been used for confidence-aware face recognition, probabilistic embedding objectives, and regularization under limited data, imbalance, or label noise~\cite{li_spherical_2021,scott_vmf_2021,kobayashi_tvmf_2021}. Hyperspherical variational autoencoders replace a Euclidean Gaussian latent space by a spherical one when the latent factor is naturally directional~\cite{davidson_hyperspherical_2018}. 

\subsection{Further areas.}
Directional and related manifold-valued models also occur in environmental and animal-movement data~\cite{applied_ley_2019}, statistical shape analysis~\cite{statistical_dryden_2016}, and computer graphics, where spatio-directional mixtures can guide light-transport sampling~\cite{path_dodik_2021}. 

To sum up, directionality arises when a signal object is periodic, normalized, sign-indeterminate, phase-indeterminate, basis-invariant, rotation-valued, or matrix-constrained, see Fig~\ref{fig:application_atlas}. Once the invariance and the sample space has been identified, the statistical model can be chosen to match the task and the required form of uncertainty, which is the topic of the next section.

\section{Directional and manifold-valued distribution families}
\label{sec:distributions}

We review the principal distribution families associated with the sample spaces introduced in Section~\ref{sec:applications}. The purpose is not to provide an exhaustive catalogue of directional statistics, for which we refer to the following books and reviews~\cite{PewseyGarciaPortugues2021,mardiaDirectionalStatistics1999,topics_jammalamadaka_2001}, but to expose the relations among distributions that are useful in choosing an appropriate model. We therefore organize the section by statistical construction rather than by sample space: canonical mean and scalar/trace concentration models, quadratic and anisotropic extensions, radially projected elliptical models as computationally cheap alternatives to quadratic models, wrapped and product-space toroidal models, and covariance-generated matrix models\footnote{Technical note: The circle and sphere densities below are written with respect to the ordinary angular or surface-area measure. Densities on compact matrix manifolds are written with respect to the normalized invariant probability measure unless stated otherwise. Full-rank positive-definite matrix densities are taken with respect to the Lebesgue measure on the independent symmetric entries, whereas singular positive-semidefinite densities require an explicit lower-dimensional volume or Hausdorff measure.}.

A recurring difficulty with directional distributions is the evaluation of the normalizing constant and its parameter derivatives. The difficulty varies substantially with the form of the normalizer. Some distributions have normalizers expressed through scalar special functions, such as the modified Bessel or scalar hypergeometric functions, which are usually inexpensive to evaluate, except in extreme dimensions or concentrations~\cite{koev_efficient_2005}. Hypergeometric functions often have simple derivative definitions~\cite{abramowitz_handbook_1965}. Stiefel- and Grassmann distributions are sometimes normalized by hypergeometric functions of matrix argument, which provide a useful analytic representation and are often more tractable than direct integration over the sample space, but their cost and numerical sensitivity generally increase drastically with the matrix dimension and concentration. In high dimensions or highly concentrated regimes, they may therefore be evaluated more efficiently using approximations, much like normalizers without a convenient special-function representation, which must instead be treated through numerical integration or approximations. Table~\ref{tab:manifold_distribution_comparison} summarizes the main distributions treated here, including a heuristic on practical computational difficulty. In the end of this section, we provide an overview of approximate computation times for most of the distributions introduced.  

\begin{table*}[t]
\centering
\caption{Signal objects, sample spaces, invariances, and representative
distribution families including computational considerations. We distinguish four levels of normalization complexity: \textsf{E}, an explicit expression; \textsf{S}, a scalar special function that is reasonably fast to evaluate; \textsf{M}, a hypergeometric function of matrix argument requiring approximations in high dimensions; and \textsf{N}, numerical integration or a truncated infinite sum whose cost is strongly parameter- and dimension-dependent. }
\label{tab:manifold_distribution_comparison}

\scriptsize
\renewcommand{\arraystretch}{1.25}
\setlength{\tabcolsep}{4pt}

\begin{tabularx}{\textwidth}{
    >{\raggedright\arraybackslash}m{0.12\textwidth}
    >{\raggedright\arraybackslash}m{0.09\textwidth}
    >{\raggedright\arraybackslash}m{0.15\textwidth}
    >{\raggedright\arraybackslash}m{0.16\textwidth}
    >{\raggedright\arraybackslash}X
    >{\raggedright\arraybackslash}X
}
\toprule
\textbf{Signal object}
&
\textbf{Sample space}
&
\textbf{Invariance or constraint}
&
\textbf{Distribution families}
&
\textbf{Structure represented}
&
\textbf{Normalization comment}
\\
\midrule

Scalar phase $\theta$
&
$\Sph^1$
&
$\theta \equiv \theta+2\pi$
&
von Mises (S); wrapped normal (N)
&
Symmetric unimodal concentration for von Mises; wrapped families arise by
reducing Euclidean variables modulo $2\pi$.
&
Univariate von Mises involves a scalar hypergeometric function. The wrapped normal is normalized by construction, but density evaluation
requires truncating an infinite series. 
\\\hline

Phase vector $\bm{\theta}$
&
$\Tor^p=(\Sph^1)^p$
&
Componentwise periodicity
&
Multivariate von Mises (N); multivariate wrapped normal (N); torus graph (N); one-factor multivariate circula (S)
&
Marginal phase concentration and circular dependence; globally shift-invariant models depend only on phase differences, whereas more general toroidal models may also contain absolute-phase or phase-sum terms.
&
Exponential-family von Mises and torus-graph models
generally have an intractable log-partition function. The wrapped normal density requires a sum over
$\mathbb{Z}^p$. The one-factor von Mises circula has a closed form involving
scalar $I_0$. 
\\\hline

Oriented unit vector $\bm{x}/\|\bm{x}\|$
&
$\Sph^{p-1}$
&
Unit norm; $\bm{x}$ and $-\bm{x}$ remain distinct
&
von Mises--Fisher (S); Fisher--Bingham (N)
&
The vMF family gives rotationally symmetric concentration about a mean
direction; Fisher--Bingham models add general linear and quadratic anisotropy.
&
The vMF normalizer involves a scalar hypergeometric function.
Fisher--Bingham normalizers require numerical integration methods.
\\\hline

Unit real axis $[\bm{x}]=\{\bm{x},-\bm{x}\}$ or unit complex ray $[\bm{z}]$
&
$\RP^{p-1}$ or $\CP^{p-1}$
&
Unit norm and sign invariance, $\bm{x}\sim-\bm{x}$; or unit norm and global
phase invariance, $\bm{z}\sim e^{\mathrm{i}\phi}\bm{z}$
&
Watson (S); Bingham (M); angular central Gaussian (ACG) (E)
&
Watson gives rotationally symmetric axial concentration; Bingham and ACG permit
general anisotropy.
&
Watson involves a scalar hypergeometric function; The Bingham
normalizer is a confluent hypergeometric function of matrix argument. ACG has an explicit determinant-based normalizer.
\\\hline

Rotation $\bm{R}$
&
$\SO(p)$
&
$\bm{R}^{\mathsf T}\bm{R}=\bm{I}$ and
$\det(\bm{R})=1$
&
Matrix Fisher (M/N)
&
Represents anisotropic uncertainty around preferred rotations.
&
On $\SO(3)$ the matrix-Fisher normalizer can be evaluated using
specialized one-dimensional
methods. General $\SO(p)$ evaluation becomes substantially more costly as
$p$ grows. 
\\\hline

Ordered orthonormal frame $\bm{U}$
&
$\St(q,p)$
&
$\bm{U}^{\mathsf T}\bm{U}=\bm{I}_q$; column ordering and orientation retained
&
matrix Langevin (matrix vMF) (M); matrix Fisher--Bingham (N)
&
The matrix Langevin family gives linear concentration
about a preferred ordered frame; matrix Fisher--Bingham models additionally
include quadratic interactions.
&
The matrix Langevin normalizer is a hypergeometric function of matrix
argument. General matrix Fisher--Bingham normalizers are harder and commonly
require approximations. 
\\\hline

Subspace $\operatorname{span}(\bm{U})$
&
$\Gr(q,p)$
&
Basis invariance,
$\bm{U}\sim\bm{U}\bm{Q}$ for $\bm{Q}\in\mathrm{O}(q)$
&
Matrix Bingham (M); matrix ACG (MACG) (E)
&
Models anisotropic concentration around preferred subspaces.
&
Matrix Bingham normalization is a hypergeometric function of matrix argument; MACG has an explicit determinant-based density.
\\\hline

Covariance or low-rank PSD matrix $\bm{C}$
&
$\SPD(p)$ or $\PSD^{(q)}(p)$
&
Symmetry and positive definiteness on $\SPD(p)$; or positive semidefiniteness
with fixed rank $q$ on $\PSD^{(q)}(p)$
&
Wishart (E); singular Wishart (E)
&
Represents joint eigenvalue--eigenspace variation induced by covariance
generative constructions.
&
Closed-form densities are available under the corresponding full-rank or
singular generative assumptions.
\\

\bottomrule
\end{tabularx}
\end{table*}

\subsection{Linear and rotationally symmetric concentration models}

Much like the isotropic multivariate Gaussian distribution, the simplest directional models use first-order directional or trace statistics. The von Mises on the circle, von Mises--Fisher on the sphere,  Watson on the projective space, and matrix--Fisher on the Stiefel manifold are all rotationally symmetric about a preferred direction, axis, or frame. 

For a scalar phase or planar direction $\theta\in[0,2\pi)\in\Sph^1$, the canonical unimodal model is the von Mises distribution~\cite{vonMises1918Ganzzahligkeit},
\begin{equation}
f_{\operatorname{vM}}(\theta\mid\mu,\kappa)=\frac{1}{C_{\operatorname{vM}}(\kappa)}\exp\left\{\kappa\cos(\theta-\mu)\right\}.
\end{equation}
Its normalizing constant is
\begin{equation}
C_{\operatorname{vM}}(\kappa)=2\pi I_0(\kappa),
\end{equation}
where $\mu\in[0,2\pi)$ is the mean direction, $\kappa\geq0$ is the concentration, and $I_0$ is the modified Bessel function of the first kind of order zero~\cite{abramowitz_handbook_1965}. The model becomes circularly uniform as $\kappa\to0$ and increasingly concentrated around $\mu$ as $\kappa$ increases. 

For oriented normalized real vectors on the sphere $\bm x\in\Sph^{p-1}$, the corresponding model is the von Mises--Fisher (vMF) distribution~\cite{dispersion_fisher_1953},
\begin{equation}
f_{\operatorname{vMF}}(\bm x\mid\bm\mu,\kappa)=\frac{1}{C_{\operatorname{vMF}}(\kappa)}\exp\left\{\kappa\bm\mu^\top\bm x\right\},\qquad \bm\mu\in\Sph^{p-1}.
\end{equation}
The normalizing constant is
\begin{equation}
C_{\operatorname{vMF}}(\kappa)=\frac{(2\pi)^{p/2}I_{p/2-1}(\kappa)}{\kappa^{p/2-1}}.
\end{equation}
The vMF distribution is rotationally symmetric about $\bm\mu$. It can be obtained by restricting an isotropic Gaussian density with nonzero mean to the unit sphere: if the ambient Gaussian has mean $\bm m$ and covariance $\sigma^2\bm I$, the conditional directional density has $\bm\mu=\bm m/\lVert\bm m\rVert_2$ and $\kappa=\lVert\bm m\rVert_2/\sigma^2$~\cite{mardiaDirectionalStatistics1999}. 

When the observation is an axis defined on the real or complex projective space, $\RP^{p-1}$ or $\CP^{p-1}$, the likelihood must be invariant to sign in the real case and to global phase in the complex case. Let $\mathbb F$ denote either $\mathbb R$ or $\mathbb C$, and define
\begin{equation}
a=\begin{cases}1/2,&\mathbb F=\mathbb R,\\1,&\mathbb F=\mathbb C.\end{cases}
\end{equation}
For $\bm x\in\mathbb F^p$ with $\bm x^\mathrm H\bm x=1$, the Watson distribution~\cite{equatorial_watson_1965,complex_mardia_1999} is
\begin{equation}
f_{\operatorname{W}}(\bm x\mid\bm\mu,\kappa)=\frac{1}{\alpha(ap)M(a,ap,\kappa)}\exp\left\{\kappa\left|\bm\mu^\mathrm H\bm x\right|^2\right\},
\end{equation}
where
\begin{equation}
\alpha(ap)=\frac{2\pi^{ap}}{\Gamma(ap)}
\end{equation}
and $M(a,ap,\kappa)$ is Kummer's confluent hypergeometric function~\cite{ber_kummer_1836}. For $\kappa>0$, the density concentrates around $\bm\mu$. For $\kappa<0$, it concentrates near the equatorial sphere (girdle) orthogonal to $\bm\mu$ in the real case or the projective hyperplane orthogonal to $\bm\mu$ in the complex case. The scalar hypergeometric normalizer requires evaluating an infinite sum but is typically inexpensive, except in extreme parameter regimes~\cite{SablicaEtAl2025Watson}. 

An analogous linear trace model applies to ordered orthonormal frames and rotations. Let $\mathcal M$ denote either the Stiefel manifold $\St(q,p)$ or the rotation group $\SO(3)$. For $\bm X\in\mathcal M$, the matrix linear exponential family is
\begin{equation}
f_{\operatorname{MF}}(\bm X\mid\bm F)=\frac{1}{C_{\mathcal M}(\bm F)}\exp\left\{\operatorname{tr}(\bm F^\top\bm X)\right\}.
\end{equation}
On $\St(q,p)$ this model is commonly called the matrix Langevin or matrix Fisher distribution~\cite{downs_orientation_1972}, whereas on $\SO(3)$ it is usually called the matrix Fisher distribution. The density has the same kernel on both spaces, but the normalizing constants differ because the supports and reference measures differ. On the Stiefel manifold,
\begin{equation}
C_{\St(q,p)}(\bm F)={}_0F_1\left(\frac{p}{2};\frac{1}{4}\bm F^\top\bm F\right),
\end{equation}
is the hypergeometric function of matrix argument. The Stiefel model is appropriate when the ordering and orientation of the frame columns are meaningful. Through $\SO(3)\cong\RP^3$, the matrix Fisher is equivalent to a Bingham distribution in projective unit-quaternion coordinates after matching parameterizations and reference measures. We note that the matrix Fisher can represent anisotropic interactions between frame columns but not interactions between channels. The normalization constant has been approached using saddlepoint approximations~\cite{saddlepoint_kume_2013}. If the data are rotations, the matrix Langevin normalizer does not apply. For $\SO(3)$, the matrix-Fisher normalizer can instead be evaluated using specialized one-dimensional integrals, series expansions, or holonomic-gradient methods~\cite{sei_properties_2013}. 

\subsection{Quadratic and anisotropic extensions}

Canonical concentration models are deliberately restrictive in that they describe rotationally symmetric variation around one preferred object and are therefore insufficient when the dispersion is coupled across variables. Quadratic and anisotropic distributions are much more flexible but also come at the cost of quadratically increasing number of estimated parameters. The Euclidean analogy is the vast difference between the isotropic (scalar variance) versus full-covariance multivariate Gaussian.

On the sphere $\Sph^{p-1}$, the Fisher--Bingham distribution~\cite{fisherbingham_kent_1982} combines linear directional preference with quadratic anisotropy,
\begin{equation}
f_{\operatorname{FB}}(\bm x\mid\bm b,\bm A)=\frac{1}{C_{\operatorname{FB}}(\bm b,\bm A)}\exp\left\{\bm b^\top\bm x+\bm x^\top\bm A\bm x\right\}.
\end{equation}
Here, $\bm b\in\mathbb R^p$ and $\bm A$ is symmetric. Because $\bm x^\top(\bm A+c\bm I)\bm x=\bm x^\top\bm A\bm x+c$ on the unit sphere, $\bm A$ is identifiable only up to addition of a scalar multiple of the identity; a constraint such as $\operatorname{tr}(\bm A)=0$ is therefore required. The linear term controls orientation, while the quadratic term allows anisotropic structure. The family contains the vMF distribution for $\bm A=\bm 0$, the Bingham distribution for $\bm b=\bm 0$, and the Kent distribution as an important special case on $\Sph^2$. Equivalently, it is the family obtained by restricting a general multivariate Gaussian density to the unit sphere. The flexibility is accompanied by a more demanding normalizing constant, which does not have a general hypergeometric series. Instead, saddlepoint approximations, holonomic-gradient methods, and numerical integration have been developed~\cite{saddlepoint_kume_2005,exact_kume_2018,parametric_ali_2019,maximum_chen_2021}. Chen and Tanaka (2021) report normalization constant computation time linear in $p$, although the number of integration nodes still depends on the dimensionality~\cite{maximum_chen_2021,sato_probabilistic_2023}. 

For real or complex projective data, $\x\in\RP^{p-1}$ or $\x\in\CP^{p-1}$, the corresponding quadratic family is the Bingham distribution~\cite{antipodally_bingham_1974,complex_kent_1994},
\begin{equation}
f_{\operatorname{B}}(\bm x\mid\bm A)=\frac{1}{\alpha(ap){}_1F_1(a;ap;\bm A)}\exp\left\{\bm x^\mathrm H\bm A\bm x\right\}.
\end{equation}
Here ${}_1F_1(a;ap;\bm A)$ denotes the real or complex hypergeometric function of matrix argument. The parameter $\bm A$ is symmetric in the real case and Hermitian in the complex case. Similarly to the Fisher--Bingham distribution, $\bm A$ is identifiable only up to addition of a scalar multiple of the identity. The Watson distribution is the rank-one special case $\bm A=\kappa\bm\mu\bm\mu^\mathrm H$. The Bingham model allows anisotropic axial concentration, but its normalizer is a confluent hypergeometric function of matrix argument. Therefore, evaluation has also been approached using the same methods as for the Fisher--Bingham normalizer. 

For ordered frames $\bm U\in\St(q,p)$, linear and quadratic terms can be combined in the matrix Fisher--Bingham family~\cite{von_khatri_1977} (treated more extensively in~\cite{statistics_chikuse_2003,simulation_hoff_2009})
\begin{equation}
\begin{aligned}f_{\operatorname{MFB}}(\bm U\mid\bm F,\bm A,\bm B)&=\frac{1}{C_{\operatorname{MFB}}(\bm F,\bm A,\bm B)}\\&\quad\times\exp\left\{\operatorname{tr}(\bm F^\top\bm U)+\operatorname{tr}(\bm B\bm U^\top\bm A\bm U)\right\}.\end{aligned}
\end{equation}
The matrices $\bm A$ and $\bm B$ are symmetric and of compatible dimensions; if one of them is zero, the matrix Fisher or matrix Langevin distribution is obtained. The first term in the exponential specifies a preferred ordered frame, while the quadratic term permits direction-specific dispersion and, when $\bm B$ is non-diagonal, interactions among frame directions. The general normalizing constant has no universally convenient hypergeometric form and is commonly treated using saddlepoint, holonomic-gradient, series, or asymptotic methods~\cite{saddlepoint_kume_2013,complete_bagyan_2024}.

When only the subspace is meaningful, the model must be invariant under the basis transformation $\bm U\mapsto\bm U\bm Q$ for every $\bm Q\in\Orth(q)$. The matrix Bingham distribution on $\Gr(q,p)$ satisfies this requirement~\cite{jupp_maximum_1979},
\begin{equation}
f_{\operatorname{MB}}(\bm U\mid\bm A)=\frac{1}{C_{\operatorname{MB}}(\bm A)}\exp\left\{\operatorname{tr}(\bm U^\top\bm A\bm U)\right\},
\end{equation}
where $\bm A$ is symmetric and is identifiable only up to addition of $c\bm I$. The normalizing constant is
\begin{equation}
C_{\operatorname{MB}}(\bm A)={}_1F_1\left(\frac{q}{2};\frac{p}{2};\bm A\right).
\end{equation}
The trace is unchanged under $\bm U\mapsto\bm U\bm Q$, and hence the density depends only on the represented subspace on the Grassmannian. Even though the normalizing constant is expressed through a special function, it is still complicated and evaluation has therefore been approached using the same methods as for the matrix Fisher--Bingham distribution. 

\subsection{Radially projected models}

The families above are obtained by restricting an ambient Gaussian density to the respective manifold. A different construction is to discard the radius or basis of an elliptical random variable. This projection gives the projected normal distribution~\cite{statistics_mardia_1972} and its matrix variant as direct alternatives to the Fisher--Bingham and matrix Fisher--Bingham distributions. Let $\bm y\sim\mathcal{N}(\bm y|\bm\mu,\bm\Sigma)$ be the $p$-dimensional multivariate normal distribution with density $\phi_p(\bm y|\bm\mu,\bm\Sigma)$. 
The projected normal distribution for the projection $\bm x=\frac{\bm y}{\|\bm y\|}$ then has density~\cite{statistics_mardia_1972,wang_directional_2013}
\begin{align}
    &f_{\operatorname{PN}}(\bm x|\bm \mu, \bm \Sigma)=\int_0^\infty r^{p-1}\phi_p(r\bm x|\bm\mu,\bm\Sigma)\ dr\nonumber\\
    &=\frac{1}{(2\pi)^{(p-1)/2}|\Sigma|^{1/2}(\bm x^\top\bm\Sigma^{-1}\bm x)^{p/2}}\exp\left[-\frac{\bm\mu^\top\bm\Sigma^{-1}\bm\mu-\lambda^2}{2}\right]M_{p-1}(\lambda),
\end{align}
where $\lambda=\frac{\bm x^\top\bm\Sigma^{-1}\bm\mu}{\sqrt{\x^\top\bm\Sigma^{-1}\x}}$ and 
\begin{equation*}
    M_k(\lambda)=\int_0^\infty t^k\phi(t-\lambda)\ dt
\end{equation*}
is the truncated normal, which is relatively easily computed without having to resort to numerical quadratures. The projected normal distribution is identifiable up to $f_{\operatorname{PN}}(\x|\bm\mu,\bm\Sigma)=f_{\operatorname{PN}}(\x|s\bm\mu,s^2\bm\Sigma),\ s>0$, which is typically resolved by imposing a trace constraint such as $\tr(\bm\Sigma)=p$.

If the ambient-Euclidean Gaussian is \emph{zero-mean}, the angular central Gaussian (ACG) family and its matrix analogue are obtained~\cite{statistical_tyler_1987,matrix_chikuse_1990} as popular direct alternatives to the Bingham and matrix Bingham distributions. 

For $\bm x\in\mathbb F^p$ with $\bm x^\mathrm H\bm x=1$, the real or complex angular central Gaussian distribution is~\cite{statistical_tyler_1987,data_kent_1997}
\begin{equation}
f_{\operatorname{ACG}}(\bm x\mid\bm\Psi)=\frac{1}{\alpha(ap)|\bm\Psi|^a}\left(\bm x^\mathrm H\bm\Psi^{-1}\bm x\right)^{-ap},
\end{equation}
where $\bm\Psi$ is positive definite in the real case and Hermitian positive definite in the complex case. The scale is similarly identifiable up to multiplication of $\bm\Psi$ by a positive scalar which leaves the density unchanged; a convention such as $\operatorname{tr}(\bm\Psi)=p$ is therefore imposed. The ACG density requires only a determinant and a quadratic form and is therefore particularly convenient for (repeated) high-dimensional likelihood evaluation. Additionally, it is unimodal. 

Imposing instead $\bm\Sigma\bm\mu=\bm\mu$ and $|\bm\Sigma|=1$ on the projected normal yields the elliptically symmetric angular Gaussian (ESAG) distribution~\cite{elliptically_paine_2018}, a direct alternative to the Fisher--Bingham distribution that has elliptical contours about the mean direction $\bm\mu/\|\bm\mu\|$ while retaining the closed-form density above, and that reduces to the ACG when $\bm\mu=\bm 0$.

The matrix angular central Gaussian (MACG) distribution~\cite{matrix_chikuse_1990} provides the corresponding model for subspaces $\bm U\in\Gr(q,p)$,
\begin{equation}
f_{\operatorname{MACG}}(\bm U\mid\bm\Psi)=\frac{1}{|\bm\Psi|^{q/2}}\left|\bm U^\top\bm\Psi^{-1}\bm U\right|^{-p/2}.
\end{equation}
If $\bm X$ follows a centered matrix-variate normal distribution with row scatter $\bm\Psi$ and uncorrelated columns, orthonormalization gives
\begin{equation}
\bm U=\bm X(\bm X^\top\bm X)^{-1/2}, \quad \operatorname{span}(\bm U)\sim\operatorname{MACG}(\bm\Psi).
\end{equation}
As for the vector ACG, $\bm\Psi$ is identifiable only up to a positive scale. The determinant form makes MACG models substantially more practical than matrix Bingham models in high-dimensional subspace applications, although the two families represent different stochastic assumptions rather than merely different numerical implementations of the same model.

The projected alternative to the vMF distribution is the projected isotropic normal distribution~\cite{mardia_projected_2026}. When measuring univariate EEG signals repeatedly in a time-locked experiment and extracting only the phase, the phase is appropriately modeled using the projected isotropic normal distribution~\cite{mardia_projected_2026}.

\begin{figure}[t]
    \centering
    \includegraphics[width=1\linewidth]{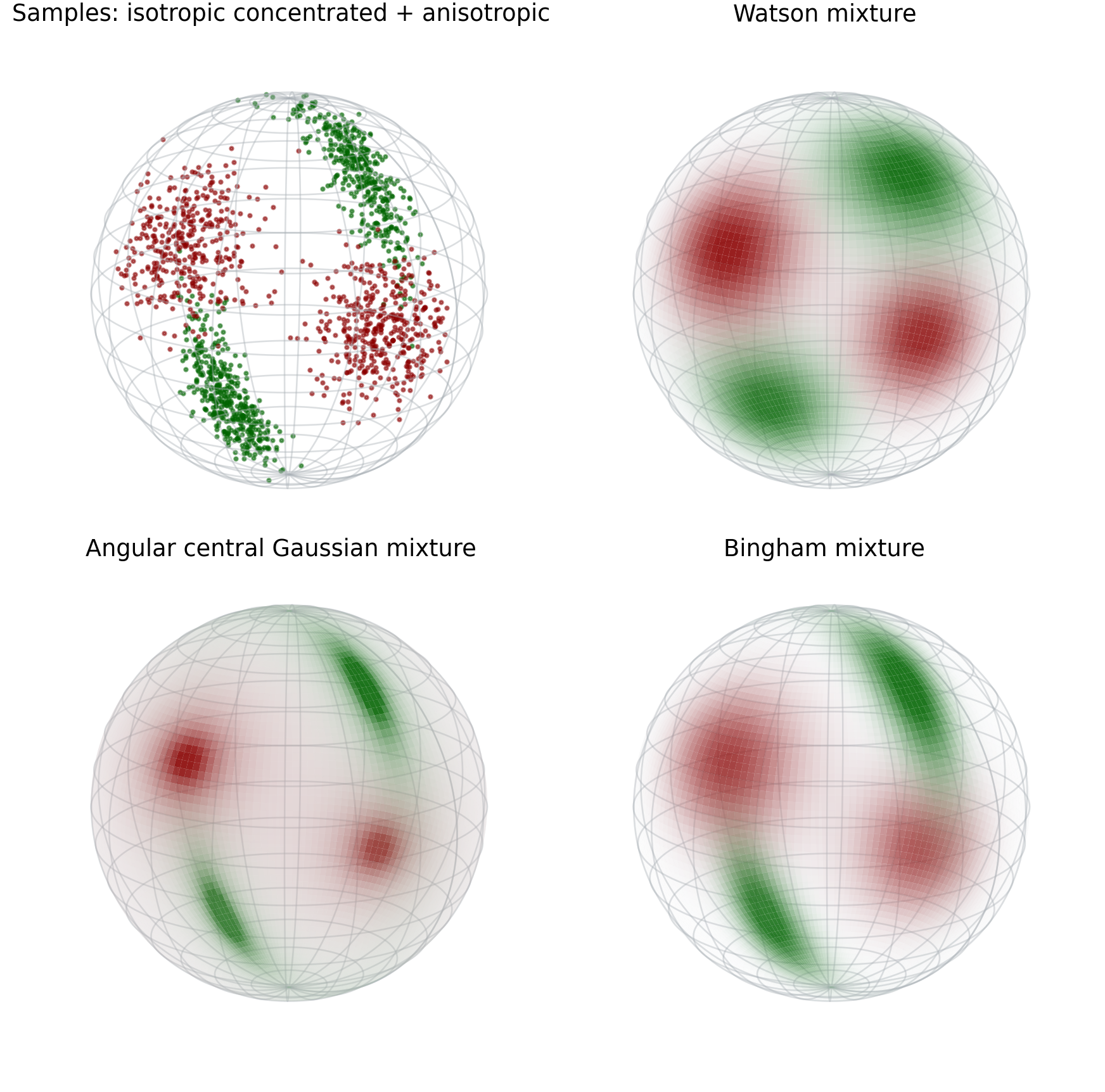}
    \caption{Samples from isotropic and anisotropic components are shown alongside fitted Watson, angular central Gaussian, and Bingham mixtures. Watson and Bingham are exponential-family distributions with somewhat involved normalization constants, whereas the angular central Gaussian is non-exponential with an explicit normalizer, leading to a different density shape while being computationally viable in arbitrary dimensions.}
    \label{fig:ACG_vs_bingham}
\end{figure}

In summary, radially projected distributions are useful alternative families to their manifold-conditioned counterparts while respecting the same invariances. Parameters of projected distributions are simpler to estimate via likelihood optimization due to their much simpler normalization constants. However, they are constructed differently and therefore have different density shapes; see Fig.~\ref{fig:ACG_vs_bingham} for a comparison between the Watson, Bingham, and ACG distributions. 

\subsection{Toroidal models}

Circular and toroidal observations may also arise by reducing an ambient Euclidean variable modulo $2\pi$. For example, wrapping a univariate Gaussian variable gives the wrapped normal density
\begin{equation}
f_{\operatorname{WN}}(\theta\mid\mu,\sigma^2)=\frac{1}{\sqrt{2\pi\sigma^2}}\sum_{m=-\infty}^\infty\exp\left\{-\frac{(\theta-\mu+2\pi m)^2}{2\sigma^2}\right\}.
\end{equation}
The infinite sum can be very hard to evaluate in non-trivial dimensions as winding scales with $(2M+1)^p$ where $M$ is the number of wrappings~\cite{estimation_nodehi_2020}. Wrapped normal and wrapped Cauchy models can capture tail behavior that differs from the von Mises family. The general wrapped Cauchy distribution of Mardia and Sriram (2020) extends to any dimension, but its normalising constant is unknown, reflecting the difficulty of high‑dimensional toroidal likelihoods~\cite{families_mardia_2020}. 

For a multichannel phase vector $\bm\theta\in\Tor^p$, pairwise torus models can represent both marginal phase preference and conditional dependence among channels. A general torus density can be written as~\cite{directions_mardia_2005,torus_klein_2019}

\begin{equation}
f_{\operatorname{TG}} \left(\bm\theta\mid\{\bm h_j\},\{\bm\phi_{jk}\}\right) = \frac{1}{C_{\operatorname{TG}} \left(\kappa,\eta\right)} \exp\left\{g_{\operatorname{TG}}(\bm\theta)\right\},
\end{equation}
where
\begin{equation}\label{eq:TG}
\begin{aligned}
g_{\operatorname{TG}}(\bm\theta) &= \sum_{j=1}^p \bm h_j^\top
\begin{bmatrix}
\cos\theta_j\\
\sin\theta_j
\end{bmatrix}
+ \sum_{j<k} \bm f_{jk}(\theta_j,\theta_k)^\top\bm\phi_{jk}.
\end{aligned}
\end{equation}
The pairwise sufficient statistics may include phase differences and phase reflections,
\begin{equation}
\bm f_{jk}(\theta_j,\theta_k)=\begin{bmatrix}\cos(\theta_j-\theta_k)\\\sin(\theta_j-\theta_k)\\\cos(\theta_j+\theta_k)\\\sin(\theta_j+\theta_k)\end{bmatrix}.
\end{equation}
We note that in the case of phase time-series with sufficient number of phase periods, the channel-wise marginals may readily be assumed uniform. Furthermore, considering signals sharing a common positive frequency, phase reflections are not meaningful. These two restrictions allows assuming uniform marginals and zero phase reflections, leaving only the phase difference terms. Such a globally phase-shift-invariant phase-difference submodel has been considered as directional Boltzmann machines~\cite{zemel_directional-unit_1992} and coupled oscillators~\cite{phase_cadieu_2010,canolty_detecting_2012}. Another submodel is the sine model, a particularly widely used model in protein bioinformatics~\cite{protein_mardia_2007, multivariate_mardia_2008, fundamental_mardia_2014}.

The main limitation of toroidal models is computational: $C_{\operatorname{TG}}(\bm\kappa,\bm\eta)$ generally lacks a finite elementary closed form; specialized series and quadrature are possible in some bivariate cases, while direct evaluation becomes increasingly difficult in higher dimensions. High-dimensional estimation therefore relies on restricted subfamilies or alternative estimation such as score matching or noise-contrastive estimation. Furthermore, in its full parametrization the torus family cannot be considered unimodal, as noted by Mardia \& Patrangenaru (2005)~\cite{directions_mardia_2005}. This is also the case for the sine submodel, where Mardia, Taylor, \& Subramanian noted that the deciding factor for unimodality is the ratio of marginal concentration to coupling strength: higher marginal concentration leads to more unimodality~\cite{protein_mardia_2007, fundamental_mardia_2014}. 

A recent preprint constructs multivariate copula distributions using a single latent uniform phase together with marginal and binding densities~\cite{carriere2026circulabasedmultivariatedistributionsflat}. For von Mises or wrapped-Cauchy binding densities, this construction yields explicitly normalized densities in arbitrary finite dimension. Its dependence structure is nevertheless restricted to the off-diagonal circular correlations having the one-factor form
\[
\bm R_c = \bm w\bm w^\top - \operatorname{diag}(\bm w\bm w^\top) + \bm I.
\]
Thus, the off-diagonal component is one-factor, and not every multivariate dependence structure can be represented. Extensions into higher-rank models, called \emph{vine copula} is attractive and copula distributions for the torus is in general an active research direction, e.g., the trivariate wrapped Cauchy copula provides a closed‑form, fully normalized model on the 3‑torus~\cite{kato_trivariate_2026}. 

A further projection-based construction is the toroidal projected normal (TPN) distribution~\cite{interpretable_kato_2025}. It is obtained from independent latent Gaussian vectors $\bm X_c\sim\mathcal N_p(\bm\kappa,\bm\Sigma_c)$ and $\bm X_s\sim\mathcal N_p(\bm 0,\bm\Sigma)$ by setting
$$
\theta_j=\mu_j+\operatorname{atan2}(X_{s,j},X_{c,j}),
$$
under the constraint $[\bm\Sigma_c]_{jk}=|[\bm\Sigma]_{jk}|$. Both $\bm\Sigma$ and $\bm\Sigma_c$ should be symmetric positive definite. The resulting family is closed under marginalization and has symmetric unimodal projected-normal marginals; setting $\bm\kappa=\bm0$ gives uniform marginals and a toroidal copula model. Although its multivariate angular density generally involves an intractable radial integral, the underlying Gaussian density is fully normalized, and latent-radius augmentation permits Bayesian estimation without evaluating a toroidal partition function. This makes TPN models, including mixture extensions, computationally attractive relative to exponential-family torus graphs.

\subsection{Covariance-generated matrix models}

Covariance matrices belong to positive definite or positive semidefinite sample spaces. Wishart distributions are justified by a specific Gaussian covariance construction; other matrix-valued signal representations may require different families even when they have the same positivity and symmetry constraints.

For a full-rank covariance matrix $\bm S\in\SPD(p)$ generated from Gaussian observations, the real Wishart density is~\cite{wishart_generalised_1928}
\begin{equation}
f_{\operatorname{Wi}}(\bm S\mid\bm P,m)=\frac{|\bm S|^{(m-p-1)/2}\exp\left\{-\frac{1}{2}\operatorname{tr}(\bm P^{-1}\bm S)\right\}}{C_{\operatorname{Wi}}(\bm P,m)}.
\end{equation}
The normalizing constant is
\begin{equation}
C_{\operatorname{Wi}}(\bm P,m)=2^{mp/2}|\bm P|^{m/2}\Gamma_p(m/2),
\end{equation}
where
\begin{equation}
\Gamma_p(a)=\pi^{p(p-1)/4}\prod_{j=1}^p\Gamma\left(a+\frac{1-j}{2}\right).
\end{equation}
Here $m$ is the number of degrees of freedom and $\bm P$ is the scale matrix. For the displayed Lebesgue density one requires $m>p-1$; under the usual Gaussian sampling construction, $m$ is an integer and $\bm S$ is positive definite almost surely when $m\geq p$. 

Low-rank covariance-like objects require singular Wishart-type models~\cite{singular_uhlig_1994,singular_srivastava_2003}. Let $\bm S=\bm U\bm\Lambda\bm U^\top$ have rank $q<p$, where $\bm\Lambda=\operatorname{diag}(\lambda_1,\ldots,\lambda_q)$ contains the positive eigenvalues. The real central singular Wishart density can be written as
\begin{equation}
\begin{aligned}f_{\operatorname{sWi}}(\bm S\mid\bm P,q)&=\frac{\pi^{q(q-p)/2}|\bm\Lambda|^{(q-p-1)/2}}{2^{pq/2}\Gamma_q(q/2)|\bm P|^{q/2}}\\&\quad\times\exp\left\{-\frac{1}{2}\operatorname{tr}(\bm P^{-1}\bm S)\right\}.\end{aligned}
\end{equation}
Singular Wishart models are relevant when rank deficiency follows from the sampling regime or from the construction of the signal object itself, such as instantaneous low-rank covariance or coherence matrices. Their spectral decomposition also makes explicit the distinction between eigenvalue information and eigenspace information: discarding the eigenvalues leads to a Grassmann-valued representation, while retaining only one eigenvector leads to a projective representation.

\subsection{Parameter computation times}

%The distribution families above can be summarized by their constructions. Von Mises and vMF models, positive-concentration Watson models, and linear matrix concentration models provide interpretable concentration about preferred directions, axes, frames, or rotations. Fisher--Bingham, Bingham, and their matrix variants introduce quadratic anisotropy but require increasingly specialized normalizer evaluation. Projected normal distributions including the ACG family arise from radial projection of gaussian variables and replace matrix-argument partition functions by simpler normalizers, such as determinants. Wrapped-normal models are normalized by construction but require infinite or lattice-sum density evaluation, whereas coupled multivariate von Mises and torus-graph models generally have difficult partition functions and are not guaranteed to be unimodal. 

To give an indication of the practical difficulty of some of the directional distributions mentioned above, we provide parameter estimation times for distributions on data sampled from the same distribution as they are estimated on. We generated $n=400$ synthetic datapoints for each distribution family in dimensionalities $p\in\{4,8,16,32,64\}$ using a shared random orthogonal basis to align preferred directions across families. We tested two scenarios: A ``high-concentration" mode with modal alignment 0.80 and a ``low-concentration" mode with modal alignment 0.20 to test if fitting time increased with concentration. 

A full Methods description is given in our Github repository and provided here only briefly. Samples from the Wrapped Normal, ACG, and MACG were generated by wrapping or projecting Gaussian samples. vMF samples were generated using the rejection method of~\cite{simulation_wood_1994} while Watson, Bingham, and matrix--Bingham sampling followed~\cite{unified_kent_2018}. Fisher--Bingham, matrix Fisher and matrix Fisher--Bingham samples were generated following~\cite{simulation_hoff_2009,habeck_geodesic_2025}. All matrix distributions were sampled and estimated with $q=2$ columns. Matrix Fisher, matrix Bingham, and matrix Fisher--Bingham normalizers were estimated using a second-order saddlepoint approach~\cite{saddlepoint_kume_2013}, Bingham and Fisher--Bingham were approached using quadrature integration~\cite{maximum_chen_2021}, vMF and Watson used their special function formulation, and the rest of the models have trivial normalizers. Parameters were estimated by optimizing the log-likelihood in PyTorch using ADAM~\cite{kingma_adam_2014} with learning rate $0.1$, using only one CPU thread, and stopping after 25 iterations without a log-likelihood improvement of more than $10^{-5}/n$. The estimation framework used our own python-toolbox described in further detail in Sec.~\ref{sec:software}. We note that this ``benchmark" should only be interpreted heuristically due to the heterogeneity of models---a detailed comparison would require assessing the goodness of fit, aligning normalizer methods, and stopping models depending on the number of parameters. 

\begin{figure}
    \centering
    \includegraphics[width=1\linewidth]{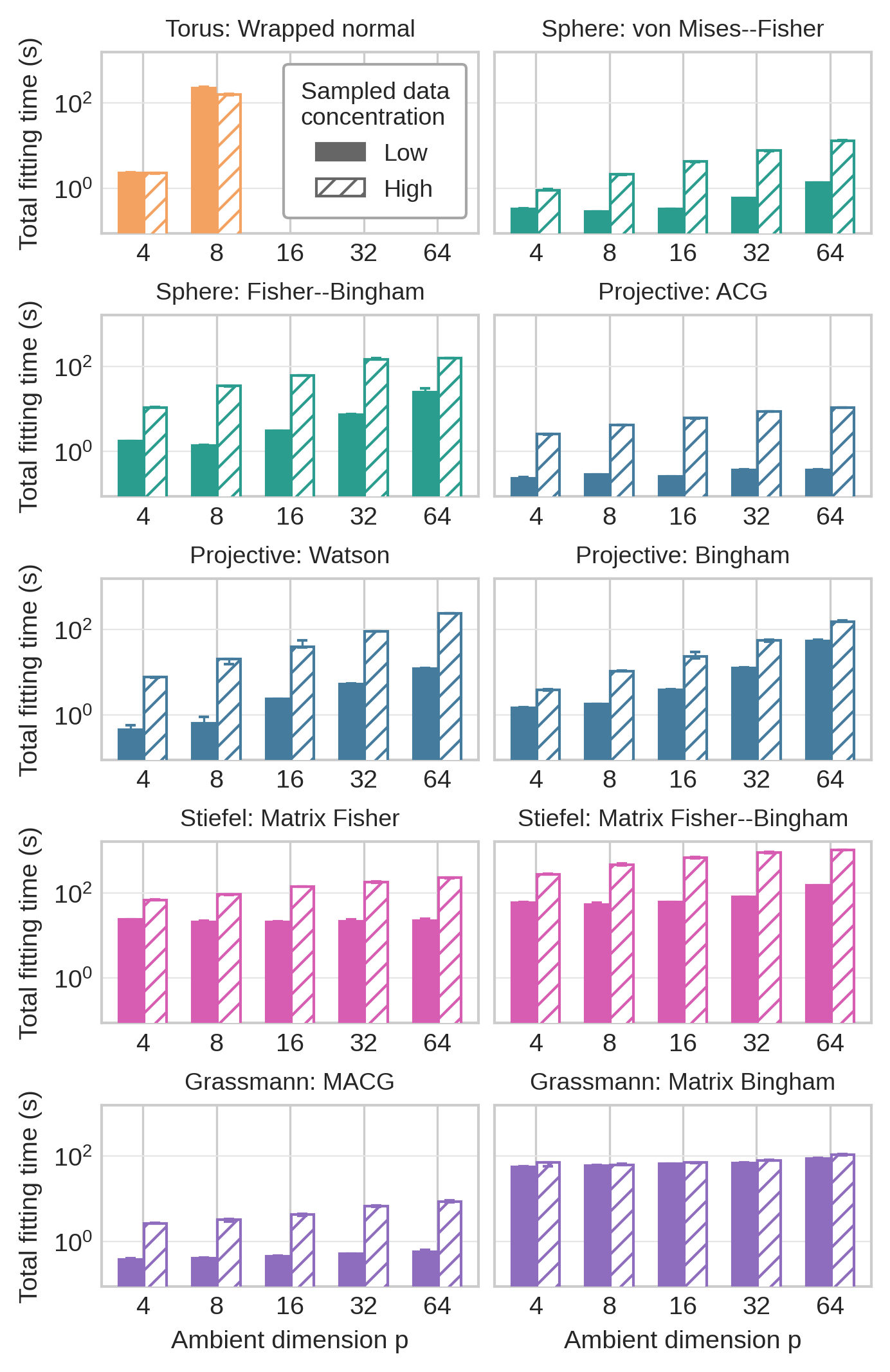}
    \caption{Computational fitting times for various directional statistics distributions. The distributions are implemented differently given their different constructions. }
    \label{fig:computationtimes}
\end{figure}

Fitting times are shown in Fig.~\ref{fig:computationtimes}. As expected, vMF, ACG and MACG had the shortest computation times given their simple normalizers. The Watson, Bingham, and Fisher--Bingham distributions were slightly more involved, while the matrix Fisher, matrix Bingham and matrix Fisher--Bingham were much more computationally expensive even in moderate dimensions. The Wrapped Normal distribution estimation used a winding of $M=1$, making it very difficult to estimate for even relatively low dimensions. All models became slower to estimate with higher dimensionality, though this effect was more clear for vector than matrix distributions. Model parameters took longer to estimate on high-concentration data for all models except the Wrapped Normal and this aspect seemed to matter more than dimensionality in our, admittedly simple, study. In conclusion, some directional models are prohibitively expensive in high-dimensional applications, rendering alternative models, such as the ACG distribution instead of the Bingham distribution and the MACG instead of the matrix Bingham distribution, computationally attractive. In the mixture model case, where individual distributions need to be estimated alongside a partition variable, fitting times for individual components above, say, $1-10$ seconds per mixture iteration, may be unacceptable. 

\section{Unsupervised state discovery}
\label{sec:state_discovery}

Unsupervised learning concerns the identification of structure in observations without predefined class labels or target variables. In signal processing, a common objective is to identify a finite collection of recurring signal states and to assign observations to these states. Examples include recurrent patterns of time-resolved brain connectivity~\cite{cabralCognitivePerformanceHealthy2017} and speaker states in diarization and source-separation systems~\cite{multiple_traa_2014,multichannel_traa_2014,dubey_speaker_2018}. Although the term \emph{state} may imply temporal persistence, most methods considered here treat observations as an unordered collection. Temporal dependence is introduced at the end of the section through hidden Markov models.

Two principal approaches to unsupervised state discovery are $K$-means clustering and finite mixture modeling. Both describe the observations using $K$ recurring components, but the interpretation of these components differs. $K$-means partitions observations according to their dissimilarity from representative centroids. It produces hard assignments and requires only a suitable dissimilarity metric and a corresponding mean function. Mixture models instead assume that observations arise from one of several probability distributions. They provide soft assignments through posterior component probabilities and additionally describe within-state dispersion and the overall probability density. Furthermore, unsupervised decomposition, e.g., MINT-AGE provides a principled alternative when both low‑dimensional structure and state partitioning are required~\cite{mardia_principal_2022}.

The choice between the approaches therefore depends on the scientific objective. If the main goal is to partition observations or construct representative signal objects, $K$-means can be computationally attractive. If likelihoods, probabilistic assignments, rank-structure/anisotropy, uncertainty quantification, outlier detection, or surrogate-data generation are required, a mixture model is likely more appropriate. The additional information provided by mixtures comes at a computational cost, particularly when the component distributions have expensive normalization constants, as iterative model fitting requires potentially many likelihood evaluations. The approaches are nevertheless closely related. Euclidean $K$-means is linked to Gaussian mixtures with equal isotropic covariance, spherical $K$-means to von Mises--Fisher mixtures with common concentration, and diametrical clustering to Watson mixtures under similar restrictions~\cite{clustering_banerjee_2005,multivariate_sra_2013}. In these settings, the $K$-means objective can be interpreted as a hard-assignment or high-concentration limit of the corresponding mixture model.

\subsection{Geometry-aware \textit{K}-means}
\label{sec:geometry_aware_kmeans}

When the scientific objective is to partition observations into recurring states rather than to estimate a normalized probability density, geometry-aware clustering is a suitable approach. For observations $\x_i$ on a sample space $\mathcal M$, $K$-means minimizes
\begin{equation}
\min_{z_1,\ldots,z_N,\bm c_1,\ldots,\bm c_K}\sum_{i=1}^N d^2(\x_i,\bm c_{z_i}),\qquad z_i\in{1,\ldots,K}.
\end{equation}
Here, $d$ is a dissimilarity that respects the constraints and invariances of the represented signal object. Fitting alternates between assigning each observation to its nearest centroid,
\begin{equation}
z_i\leftarrow\argmin_k d^2(\x_i,\bm c_k),
\end{equation}
and updating each centroid by minimizing the within-cluster dissimilarity,
\begin{equation}
\bm c_k\leftarrow\argmin_{\bm c\in\mathcal M}\sum_{i:z_i=k}d^2(\x_i,\bm c).
\end{equation}
Consequently, a $K$-means algorithm requires only a suitable dissimilarity and a corresponding centroid-update function. Neither step requires evaluation of an actual probability density or its normalization constant, which can be advantageous for toroidal, Fisher--Bingham, and matrix Fisher--Bingham models whose normalizers may be expensive or unavailable. 

The dissimilarity may be constructed intrinsically or extrinsically. Let $\x,\bm y\in\mathcal M$ be observations on a Riemannian manifold $\mathcal M$. The intrinsic distance between two points is the length of the shortest path that remains on the manifold, such as a geodesic,
\begin{equation}
d_g(\x,\bm y)=\inf_{\substack{\gamma(0)=\x\\\gamma(1)=\bm y}}\int_0^1|\dot{\gamma}(t)|_{\gamma(t)},\mathrm dt.
\end{equation}
The corresponding intrinsic sample mean is the Fréchet mean, however, a closed-form solution is generally unavailable. An extrinsic construction instead represents $\mathcal M$ through an embedding $\Phi:\mathcal M\rightarrow\mathbb R^D$ and measures Euclidean separation between the embedded points,
\begin{equation}
d_\Phi^2(\x,\bm y)=\|\Phi(\x)-\Phi(\bm y)\|_2^2.
\end{equation}
The extrinsic sample mean minimizes this distance and is obtained by projecting the ambient arithmetic mean onto the embedded manifold. Projection distances compare invariant projection representations, such as $\x\x^\mathrm{H}$ or $\bm U\bm U^\mathrm{H}$, whereas chordal distances measure the straight-line Euclidean distance between embedded points rather than the geodesic distance along the manifold. When the projection is unique and analytically available, this gives a closed-form mean function. Intrinsic distances may more directly reflect the geometry of the sample space, whereas embedding- or transform-based objectives are often attractive in large-scale clustering because their mean functions are computationally simpler. 

Table~\ref{tab:distance_mean_functions} summarizes computationally attractive representative choices for the sample spaces considered in this paper, including torus $K$-means~\cite{jung_clustering_2021}, cosine $K$-means for $\Sph^{p-1}$, diametrical clustering for $\RP^{p-1}$~\cite{diametrical_dhillon_2003} and $\CP^{p-1}$~\cite{olsen_uncovering_2026}, Stiefel-clustering using a Procustean mean function~\cite{cetingul_intrinsic_2009}, and Grassmann clustering~\cite{grassmann_gruber_2006}. In the below, we introduce two additional $K$-means functions for the globally phase-invariant torus and the PSD manifold. 

\begin{table*}[t]
\centering
\caption{Representative distance and mean functions for geometry-aware clustering. With the exception of the log-Euclidean positive-definite construction, the listed methods are extrinsic.}
\label{tab:distance_mean_functions}
\scriptsize
\renewcommand{\arraystretch}{1.5}
\begin{tabularx}{\textwidth}{@{}>{\raggedright\arraybackslash}m{0.15\textwidth}>{\raggedright\arraybackslash}m{0.13\textwidth}>{\raggedright\arraybackslash}m{0.25\textwidth}>{\raggedright\arraybackslash}X@{}}
\toprule
\textbf{Sample space} & \textbf{Construction} & \textbf{Typical dissimilarity} & \textbf{Mean function} \\
\midrule
Circle $\Sph^1$ and product torus $\Tor^p$ & Product-circle chordal & \(d_c^2(\bm\theta_i,\bm\mu)=2\sum_{j=1}^p\{1-\cos(\theta_{i,j}-\mu_j)\}\). & Componentwise circular mean: \(\bar{\mu}_j=\arg\left\{\sum_{i=1}^n\exp(i\theta_{ij})\right\},\ j=1,\ldots,p\). \\
Sphere \(\Sph^{p-1}\) & Spherical similarity & \(d^2(\x_i,\bm\mu)=1-\bm\mu^\top\x_i\). & Normalized arithmetic mean: \(\bm\mu=\frac{\sum_{i=1}^n\x_i}{\left\|\sum_{i=1}^n\x_i\right\|_2}\). \\
Real and complex projective spaces $\mathbb{P}^{p-1}$ & Projection & \(d^2(\x_i,\bm\mu)=1-\left|\bm\mu^\mathrm H\x_i\right|^2\). & If \(\bm A=\sum_{i=1}^n\x_i\x_i^\mathrm H\), then \(\bm\mu=\operatorname{eig}_{\max}(\bm A)\). \\
Stiefel manifold $\St(q,p)$ & Chordal & \(d^2(\bm U_i,\bm C)=\frac{1}{2}\|\bm U_i-\bm C\|_F^2\). & If \(\sum_{i=1}^n\bm U_i=\bm P\bm\Sigma\bm Q^\top\), then \(\bm C=\bm P\bm Q^\top\). \\
Grassmann manifold \(\Gr(q,p)\) & Projection & \(d^2(\bm U_i,\bm C)=\frac{1}{2}\left\|\bm U_i\bm U_i^\mathrm H-\bm C\bm C^\mathrm H\right\|_F^2\). & If \(\sum_{i=1}^n\bm U_i\bm U_i^\mathrm H=\bm V\boldsymbol\Lambda\bm V^\mathrm H\), then \(\bm C=\bm V^{(q)}\), the leading \(q\) eigenvectors. \\
Positive-definite matrices & Log-Euclidean & \(d_{\operatorname{LE}}^2(\bm S_i,\bm C)=\|\log(\bm S_i)-\log(\bm C)\|_F^2\). & \(\bm C=\exp\left\{\frac{1}{n}\sum_{i=1}^n\log(\bm S_i)\right\}\). \\
\bottomrule
\end{tabularx}
\end{table*}

\subsubsection{Quotient torus clustering}
\label{sec:quotient_torus_clustering}

Besides the manifolds listed in Table~\ref{tab:distance_mean_functions}, the quotient torus is of particular relevance when working with time series of phases where the marginal distribution may be assumed uniform. This case is studied in detail in Sec.~\ref{sec:phase_case}. When a common ``carrier" phase is irrelevant, observations belong to $\Tor^p/\Sph^1\cong\Tor^{p-1}$. The orbit distance is
\begin{equation}
\begin{aligned}d_{\operatorname{qT}}^2([\bm\theta],[\bm\mu])&=\min_{\alpha\in[0,2\pi)}2\sum_{j=1}^p{1-\cos(\theta_j-\mu_j-\alpha)},\\&=2p-2\left|\sum_{j=1}^p\exp{i(\theta_j-\mu_j)}\right|.\end{aligned}
\end{equation}
The pairwise optimal global-phase alignment is available in closed form, whereas the sample mean generally requires iterative optimization; no general closed-form mean function is available for this distance. Instead, this objective can be optimized by alternating global-phase alignment and componentwise circular averaging. Given a current representative $\bm\mu$, observation $i$ is aligned by
\begin{equation}
\rho_i=\exp\!\left\{-i\arg\left(\sum_j\exp\{i(\theta_{ij}-\mu_j)\}\right)\right\}.
\end{equation}
The updated representative is then
\begin{equation}
\mu_j=\arg\sum_i\rho_i\exp\{i\theta_{ij}\},\qquad j=1,\ldots,p.
\end{equation}

\subsubsection{\textit{K}-medoids: When an average is unavailable}

When an appropriate mean is computationally expensive or unavailable for the chosen dissimilarity, $K$-medoids replaces each centroid by an observed exemplar from the data pool. It requires only dissimilarities between pairs of observations. Constructing the complete dissimilarity matrix can be expensive, however; sampling and accelerated variants can reduce this cost. The medoid automatically belongs to the observed sample space and respects all representation constraints.

\subsection{Mixture modeling}
\label{sec:mixture_modeling}

Mixture models are directly relevant for unsupervised discovery of recurring signal states as they provide soft clustering by representing observations as arising from one of several component distributions. Furthermore, they provide geometry-aware density estimation by combining simple (unimodal) distributions to form mixture densities useful for characterizing complex data distributions. For observations $\x_i$ on a manifold $\mathcal M$, where, for notational simplicity, $\x_i$ may denote a vector- or matrix-variate observation, a $K$-component mixture density is
\begin{equation}
f_{\operatorname{MM}}(\x_i\mid\boldsymbol\theta,\bm w)=\sum_{k=1}^K w_k f(\x_i\mid\boldsymbol\theta_k),
\end{equation}
where $w_k\geq0$ and $\sum_{k=1}^K w_k=1$. The component parameters depend on the chosen family, for example $\boldsymbol\theta_k=(\bm\mu_k,\kappa_k)$ for vMF or Watson components or $\boldsymbol\theta_k=\bm A_k$ for Bingham components.

For observations $\mathcal X={\x_i}_{i=1}^N$, the log-likelihood is
\begin{equation}
\mathcal L_{\operatorname{MM}}\left(\boldsymbol\theta,\bm w\mid\mathcal X\right)=\sum_{i=1}^N\log\left(\sum_{k=1}^K w_k f(\x_i\mid\boldsymbol\theta_k)\right).
\end{equation}
Unlike $K$-means, a mixture model assigns each observation a probability of belonging to each component. It can therefore represent ambiguous observations near state boundaries and can accommodate components with different concentrations, anisotropies, or other distribution-specific dispersion parameters. The overall density can also represent more complicated multimodal structure than any individual component.

The computational feasibility of mixture modeling depends strongly on the selected component distributions. Component densities and their normalizing constants must generally be evaluated repeatedly. Mixture estimation, including the treatment of constrained parameters, initialization, low-rank parameterizations, and unavailable normalizing constants, is considered in Section~\ref{sec:learning}.

\subsection{Hidden Markov models}
\label{sec:hidden_markov_models}

The $K$-means and mixture formulations above treat the state assignments as independent across observations. However, when observations form a time series recurring signal states may persist over several samples and transitions may occur preferentially between particular pairs of states. A hidden Markov model extends a mixture model by placing a first-order Markov model on the latent state sequence, thereby directly modeling sequential smoothness while retaining a state-specific ``emission" distribution.

Directional distributions can be used directly as state-conditional emission models. Relative to an independent mixture model, the transition matrix provides information about state persistence and switching structure and may stabilize state inference when individual observations are noisy.
The forward recursion, posterior state probabilities, decoding, and parameter-estimation equations require substantial additional notation and are therefore provided in the Appendix (Sec.~\ref{sec:app_hidden_markov_models}). Their estimation follows the general strategies described in Section~\ref{sec:learning}.

\section{Estimation and learning strategies}
\label{sec:learning}

In this section, we discuss various strategies for estimating the parameters of the statistical distributions and mixtures described above. We first describe analytical, expectation-maximization, and gradient-based maximum-likelihood estimation procedures, and subsequently consider low-rank-plus-isotropic parameterizations for high-dimensional matrix parameters as well as initialization heuristics. 

\subsection{Maximum-likelihood estimation}

Maximum-likelihood estimation remains a central approach for learning distribution parameters. However, for many directional distributions the likelihood contains special functions or normalizing constants of matrix argument, and the maximum-likelihood estimator is therefore not always available in closed form. For example, the MLE for the Watson distribution alters between updating $\bm\mu$ in closed form and optimizing $\kappa$ using a non-linear optimizer, for which several bounds have been derived to make estimation easier~\cite{multivariate_sra_2013,mixture_bijral_2007,complex_mardia_1999}. For the angular central Gaussian distribution, the estimator is an iterative fixed-point update that needs to be evaluated until convergence~\cite{statistical_tyler_1987}. 

\subsection{Gradient-based optimization}

Analytical or fixed-point estimators can be efficient when they are available, but model modifications such as low-rank parameterizations, regularization, or additional constraints typically require new analytical derivations. An alternative is gradient-based maximum-likelihood estimation using automatic differentiation libraries such as PyTorch. This approach still performs maximum-likelihood estimation; the difference is that the likelihood is optimized numerically rather than through a distribution-specific analytical or fixed-point update. The same framework can be used for unnormalized distributions using noise-contrastive estimation (defined in Sec.~\ref{sec:unnormalized_estimation}), mixture likelihoods, and hidden Markov models through the forward recursion. Its main practical requirements are selection of optimizer and learning-rate (ADAM~\cite{kingma_adam_2014} with learning rate set to 0.1 typically works well) and convergence diagnostics, which may require cumbersome trial-and-error development.

Gradient-based optimization requires specifying a loss function, which corresponds to the negative log-likelihood for maximum-likelihood estimation. Optimization is most convenient when parameters are optimized unconstrained and mapped to the required parameter space before evaluating the loss. For a directional parameter in a vMF or Watson distribution, one may optimize an unconstrained vector $\tilde{\bm\mu}$ and normalize it according to
\begin{equation}
\bm\mu=\frac{\tilde{\bm\mu}}{\|\tilde{\bm\mu}\|_2}.
\end{equation}
A nonnegative concentration parameter for the vMF distribution may similarly be obtained from an unconstrained scalar $\tilde\kappa$ using the softplus function
\begin{equation}
\kappa=\log\left(1+\exp(\tilde\kappa)\right).
\end{equation}
Furthermore, a positive-definite real or complex covariance-type matrix parameter can also be constructed from an unconstrained matrices, see Sec.~\ref{sec:low_rank_diagonal}. These parameterizations generally need to be continuously differentiable for standard gradient-based optimizers.

\subsection{Mixture model estimation}

In the EM algorithm for mixture models, the estimation alternates between an expectation step and a maximization step. In the E-step, the posterior responsibility of component $k$ for observation $i$ is
\begin{equation}
\beta_{ki}=\frac{w_k f(\x_i\mid\boldsymbol\theta_k)}{\sum_{k'=1}^K w_{k'} f(\x_i\mid\boldsymbol\theta_{k'})}.
\end{equation}
The M-step updates the component weights according to
\begin{equation}
w_k=\frac{1}{N}\sum_{i=1}^N\beta_{ki},
\end{equation}
and updates each $\boldsymbol\theta_k$ by weighted maximum likelihood. When a component-specific M-step is available, EM is attractive because each iteration generally is guaranteed to not decrease the observed-data likelihood (subject to convergence of iterative updates for those distributions where such is required). 

Alternatively, the mixture log-likelihood can be optimized directly using automatic differentiation. The weights can be parameterized by a softmax transformation, while the distribution-specific constraints are handled as explained above. Likewise, HMMs are easily implemented using automatic differentiation by using the forward function as loss and using row-wise softmax transformation on the transition matrix.

Both EM and direct optimization are sensitive to initialization and may converge to local optima; multiple random restarts or selective initialization can therefore be useful (see Sec.~\ref{sec:initialization}). 

\subsection{Low-rank plus isotropic parameterizations of matrix parameters}
\label{sec:low_rank_diagonal}

Several distributions considered in this paper are parameterized by a full real symmetric or complex Hermitian positive-definite covariance-type matrix, e.g., $\bm A$. In high-dimensional data, the effective variation may be concentrated in a low-dimensional subspace because of collinearity, limited sample size, or genuine low-dimensional structure. In such cases, an unrestricted covariance-parameter estimate may be poorly conditioned or may fail to exist.

A common workaround is to apply dimensionality reduction, for example by principal component analysis, before fitting the directional model. However, this introduces an additional preprocessing step and a separate heuristic of target dimensionality. For directional data, this also requires considering directional decomposition methods such as spherical PCA or torus PCA~\cite{torus_eltzner_2018} to ensure the low-dimensional reconstruction retains the directional invariances of the input data. An alternative is to impose structure directly in the model through a low-rank plus isotropic parameterization~\cite{angular_olsen_2023},
\begin{equation}
\bm A\approx\bm Z=\eta\left(\bm M\bm M^\mathrm{H}+\bm I\right),
\end{equation}
where
\begin{equation}
\bm M\in\R^{p\times r}\quad\text{or}\quad\bm M\in\C^{p\times r}.
\end{equation}
The matrix $\bm M\bm M^\mathrm{H}$ represents a rank-$r$ deviation from isotropy, while the identity matrix guarantees that $\bm Z$ is positive definite. The rank $r$ is therefore a tunable model complexity hyperparameter rather than a preprocessing dimension. The factor $\bm M$ is not unique, because $\bm M$ and $\bm M\bm Q$ generate the same matrix for any orthogonal or unitary $\bm Q$, but the induced parameter $\bm Z$ is identifiable subject to scale properties of the reparameterized distribution.

For scale-invariant distributions such as the ACG and MACG distributions, multiplying $\bm Z$ by a positive scalar does not change the density. The scalar $\eta$ can therefore be fixed by a convention such as $\operatorname{tr}(\bm Z)=p$~\cite{statistical_tyler_1987}, which gives
\begin{equation}
\eta=\frac{p}{\operatorname{tr}\left(\bm M\bm M^\mathrm{H}+\bm I\right)}.
\end{equation}
For models in which scale is identifiable, such as a Gaussian or Wishart under the usual parameterization, $\eta$ should instead be estimated jointly with $\bm M$.

This parameterization changes the optimization problem. Instead of optimizing over all positive-definite matrices $\bm A$, one optimizes over $\bm M$ and, where required, $\eta$. Distribution-specific analytical or EM updates must then be re-derived for these lower-dimensional parameters. Gradient-based optimization provides a simpler alternative: $\bm M$ is treated as an unconstrained parameter and $\bm Z$ is constructed inside the likelihood.

The low-rank plus isotropic form is also computationally attractive. Matrix inverses and determinants can be evaluated using the Woodbury identity and the matrix determinant lemma,
\begin{equation}
\left(\bm I+\bm M\bm M^\mathrm{H}\right)^{-1}=\bm I-\bm M\left(\bm I+\bm M^\mathrm{H}\bm M\right)^{-1}\bm M^\mathrm{H}.
\end{equation}
\begin{equation}
\left|\bm I+\bm M\bm M^\mathrm{H}\right|=\left|\bm I+\bm M^\mathrm{H}\bm M\right|.
\end{equation}
Thus, the most expensive matrix operations can often be reduced from ambient dimension $p$ to latent dimension $r$. 

\subsection{Estimating models with unavailable normalizing constants}
\label{sec:unnormalized_estimation}

Many directional models have densities of the form
\begin{equation}
f(\x\mid\boldsymbol\theta)=\frac{1}{C(\boldsymbol\theta)}\exp\{-E_{\boldsymbol\theta}(\x)\},
\end{equation}
where
\begin{equation}
C(\boldsymbol\theta)=\int_{\mathcal M}\exp\left\{-E_{\boldsymbol\theta}(\x)\right\}d\x
\end{equation}
and $d\x$ is a suitable measure. When $C(\boldsymbol\theta)$ is unavailable or expensive to evaluate, direct maximum-likelihood estimation becomes difficult. Besides precomputing normalizing constants in lookup tables or using approximations, several objectives avoid evaluating $C(\boldsymbol\theta)$ explicitly.

Score matching estimates the derivatives of the log-density rather than the density itself~\cite{hyvarinen_estimation_2005,score_mardia_2016}. On a Euclidean domain, the score is $\nabla_{\x}\log f(\x\mid\boldsymbol\theta)$; on a manifold, the corresponding object is the Riemannian gradient $\nabla_{\mathcal M}\log f(\x\mid\boldsymbol\theta)$. Since
\begin{equation}
\nabla_{\mathcal M}\log f(\x\mid\boldsymbol\theta)=-\nabla_{\mathcal M}E_{\boldsymbol\theta}(\x),
\end{equation}
the normalizing constant disappears. Under appropriate regularity and boundary conditions, the score-matching objective can be written using the Laplace--Beltrami operator $\Delta_{\mathcal M}$ as
\begin{equation}
J(\boldsymbol\theta)=\frac{1}{N}\sum_{i=1}^N\left[\frac{1}{2}\left\|\nabla_{\mathcal M}E_{\boldsymbol\theta}(\x_i)\right\|^2-\Delta_{\mathcal M}E_{\boldsymbol\theta}(\x_i)\right],
\end{equation}
up to constants independent of $\boldsymbol\theta$. This is attractive on compact manifolds without boundary. The main practical limitation is that the required differential operators may be nontrivial to derive and implement. 

An alternative strategy is noise-contrastive estimation, which turns density estimation into a classification problem between observed data and samples from a known noise distribution $g(\x\mid\bm\phi)$~\cite{gutmann_noise-contrastive_2010}. Matsuda and Hyvärinen extended the approach to finite mixtures of unnormalized component models~\cite{matsuda_estimation_2020}. An unnormalized mixture can be written as
\begin{equation}
\tilde f_{\operatorname{MM}}(\x\mid\boldsymbol\theta,\bm a)=\sum_{k=1}^K\exp\left\{a_k-E_{\boldsymbol\theta_k}(\x)\right\}.
\end{equation}
For a normalized mixture, the combined log-amplitude satisfies~\cite{matsuda_estimation_2020}
\begin{equation}
a_k=\log w_k-\log C(\boldsymbol\theta_k).
\end{equation}
If $\nu$ noise samples are drawn per data sample, the model probability that an observation came from the data distribution is
\begin{equation}
p(D=1\mid\x)=\frac{\tilde f_{\operatorname{MM}}(\x)}{\tilde f_{\operatorname{MM}}(\x)+\nu g(\x\mid\bm\phi)}.
\end{equation}
The parameters are then estimated by maximizing the binary classification likelihood for data and noise samples. The noise distribution must have support wherever the data distribution has support. Uniform noise is readily available on most compact directional domains, whereas broader geometry-matched distributions, such as weakly concentrated vMF distributions, may improve finite-sample efficiency. Noise-contrastive estimation can estimate the combined log-amplitudes $a_k$, including their common additive offset. Score matching can estimate relative log-amplitudes, with the common additive offset fixed by convention. Recovering the mixture weights $w_k$ additionally requires the component normalizing constants $C(\boldsymbol\theta_k)$, or their ratios.

\subsection{Initialization Heuristics}\label{sec:initialization}

When model complexity is high, selective initialization strategies can help lead to faster convergence and greater stability in optimization than starting from random values. Such strategies usually involve initializing a complex model using estimates from a simpler model. 

As an example, a $K$-means model may be initialized using probabilistic farthest-first seeding, also known as $K$-means++~\cite{arthur_k-means_2007}. In this process, the first of $K$ centroids is selected randomly from the pool of data samples. The remaining centroids are sequentially sampled from the remaining samples with a probability depending on the (geometry-aware) distance to existing centroids, ensuring initial centroids are spread far apart. Similarly, a mixture-model can be initialized using a $K$-means solution. In some cases, such as the vMF or Watson distributions, mean parameters may be initialized directly from centroids. In general, however, the $K$-means partition of data into $K$ groups may be used to estimate $K$ \textit{separate} probabilistic distributions, that are then used as initial parameters for the combined mixture. The mixture weighting parameter $\bm w$ may trivially be initialized as the cluster proportion. Finally, hidden Markov models may be initialized from a mixture model estimate, where the initial-observation weighting parameter $\bm\delta$ and transition matrix $\bm T$ are initialized from the mixture posterior. We note that parameter reparameterization for gradient-based numerical optimization using, e.g., softplus, softmax, or normalization requires the inverse transformation to be used when initializing. 

The low-rank plus isotropic matrix $\bm M$ may be initialized as~\cite{tipping_probabilistic_1999} 
\begin{equation*}
    \bm M=\bm U_r\left[\frac{1}{\epsilon}\left(\bm\Lambda_r-\epsilon\bm I_r\right)\right]^{1/2},
\end{equation*}
where $\bm U$ and $\bm\lambda$ contain the leading $r$ eigenvectors and eigenvalues of the sample covariance matrix and $\epsilon=\frac{\sum_{l=r+1}^p\lambda_{l}}{p-r}$, assuming the number of assigned samples is $n>p$. 

The low-rank plus isotropic form introduced above may also be sequentially initialized (``warm start"). For instance, given the parameter $\mathbf M_h$ of a rank $h<r$ model, the first $h$ columns of $\mathbf M$ could be initialized as $\mathbf M_h$ and the remaining $r-h$ columns as the singular value decomposition on the residual matrix after projecting out the contributions from the existing columns. 

\subsection{Model selection and validation}

Model selection may concern the number of components $K$ and/or the low-rank dimension $r$. Candidate models can be compared using full-sample likelihood-based criteria such as AIC or BIC, or by cross-validation on held-out data. The simplest validation strategy is a 50 \% hold-out split, where each candidate is fitted on one half and evaluated on the other.
%The perhaps simplest validation strategy is the 50 \% hold-out procedure splitting the dataset into two equal sized  training and test halves, each model candidate is estimated using only the training half, and its performance is evaluated on the test half. 
The evaluation measure should reflect the objective and may include held-out log-likelihood when a normalized density is available, overlap with a known partition when reference labels exist, or performance on a downstream task. In model-based clustering, held-out likelihood provides a natural criterion, whereas other clustering methods require an appropriate task-specific measure. Repeated splits or $K$-fold cross-validation can reduce dependence on a particular partition of the data, while the size of the training set can be increased by increasing $K$.

\subsection{Software}\label{sec:software}

Software for directional statistics is available across programming languages. For classical circular analysis, \texttt{CircStat} provides a widely used MATLAB implementation of descriptive statistics, visualization, hypothesis tests, regression, and basic distributional methods~\cite{Berens2009CircStat}. Comparable functionality is available in R through \texttt{circular}, while \texttt{Directional} extends the R ecosystem to higher-dimensional spherical data and includes simulation, likelihood estimation, kernel methods, testing, regression, classification, and clustering, and \texttt{Riemann} provides similar for general-purpose Riemannian manifolds~\cite{Riemann}. Python alternatives for classical circular analysis include \texttt{PyCircStat2}. More specialized R packages include \texttt{movMF} for maximum-likelihood estimation of von Mises--Fisher mixtures~\cite{HornikGrun2014movMF}, \texttt{watson} for sampling and fitting mixtures of Watson distributions~\cite{SablicaEtAl2025Watson}, \texttt{BAMBI} for Bayesian mixtures of univariate and bivariate angular distributions~\cite{ChakrabortyWong2021BAMBI}, and \texttt{rstiefel} for simulation from matrix Bingham--von Mises--Fisher distributions on Stiefel manifolds~\cite{simulation_hoff_2009}. Spatially and temporally dependent circular observations are addressed by \texttt{CircSpaceTime}, which implements Bayesian wrapped- and projected-Gaussian process models~\cite{JonaLasinioEtAl2020CircSpaceTime}. Of particular relevance to signal processing, \texttt{libDirectional} combines distributions on circles, hypertori, and real and complex hyperspheres with recursive filtering algorithms~\cite{KurzEtAl2019libDirectional}. 

Our Python toolbox Phase Coherence Mixture Modeling (\texttt{PCMM}) implements the log-density for most distributions covered in this manuscript including mixtures and hidden Markov models, where the latter was implemented through the forward function (Sec.~\ref{sec:app_hidden_markov_models}).  Estimation is performed using either EM (only for Watson, ACG, MACG, Singular Wishart, and not HMMs) or autodifferentiation via PyTorch (all models). Fisher--Bingham and Bingham normalizers were implemented using the numerical integration formulation by Chen \& Tanaka~\cite{maximum_chen_2021}, while matrix Fisher--Bingham and its submodels were implemented using saddlepoint approximations~\cite{saddlepoint_kume_2013}. We also provide implementations of all $K$-means functions covered here. The PCMM toolbox was originally developed for multivariate phase modeling using the complex ACG distribution in~\cite{olsen_uncovering_2026} and extended with this paper to cover a wider range of directional distributions. The toolbox is publicly available from: \href{https://github.com/anders-s-olsen/PCMM}{https://github.com/anders-s-olsen/PCMM}.

\section{Multivariate phase-coherence states in fMRI}\label{sec:phase_case}

\begin{figure}
\centering
\includegraphics[width=1\linewidth]{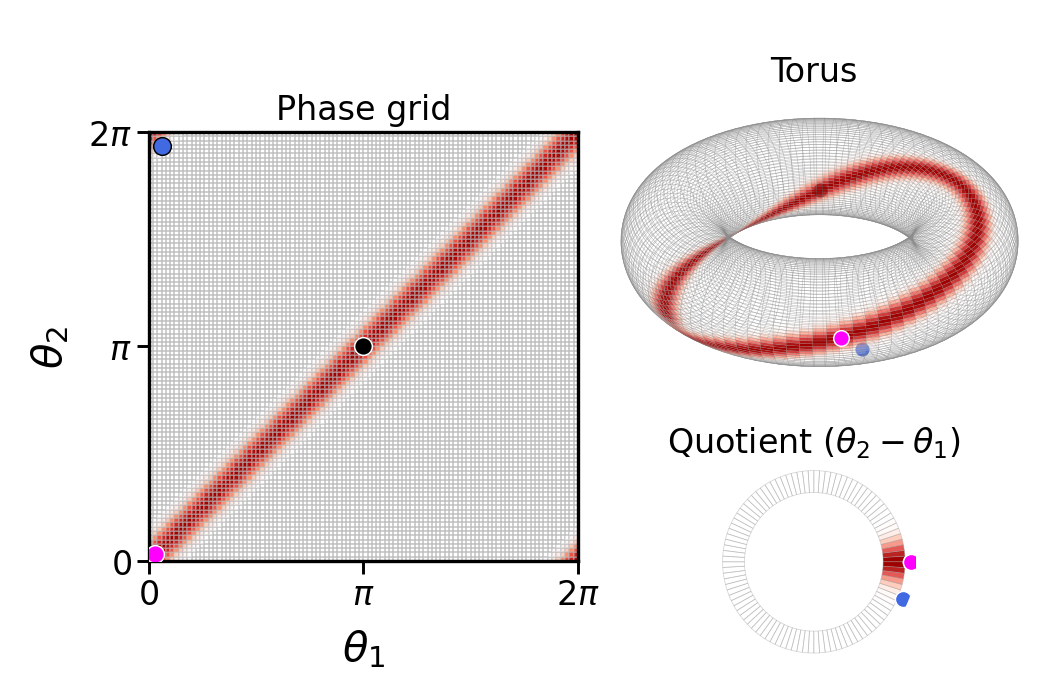}
\caption{Removing common phase while preserving relative phase. Left: two-region phase configurations differing only by a common rotation lie on a diagonal orbit in the phase grid, with periodic boundary identification. Top right: the same orbit represented on the torus $\Tor^2$. Bottom right: identifying all configurations along each orbit retains only the relative phase $\theta_2-\theta_1\in\Sph^1$, giving $\Tor^2/\Sph^1\cong\Sph^1$. Thus, a coherent or phase-offset relationship remains unchanged as both regional signals advance through their cycles together.}
\label{fig:torus}
\end{figure}

\subsection{Context}
As a particularly useful case study that incorporates many directional models, we here describe the estimation of dynamic phase coherence states in functional neuroimaging data. Functional brain connectivity describes correlations between regional blood-oxygen-level-dependent (BOLD) fMRI signals. In addition to estimating these relationships over an entire recording, time-resolved analyses seek to identify recurring patterns of interregional coordination and characterize how the brain moves between them using unsupervised state discovery methods. Traditionally, time-varying covariances were estimated using windowed correlations\footnote{It is usually modeled as a Wishart distribution~\cite{nielsen_modeling_2017}.}. However, phase-based approaches, where regional phases are computed; e.g., using the Hilbert transform, discard the (potentially noisy) amplitude of BOLD fMRI data, thereby providing a synchronization-based, rather than amplitude-based approach to estimating connectivity. Phase relationships can be evaluated instantaneously without estimating correlations over a sliding window~\cite{glerean_functional_2012}. 

\begin{figure*}
\centering
\includegraphics[width=1\linewidth]{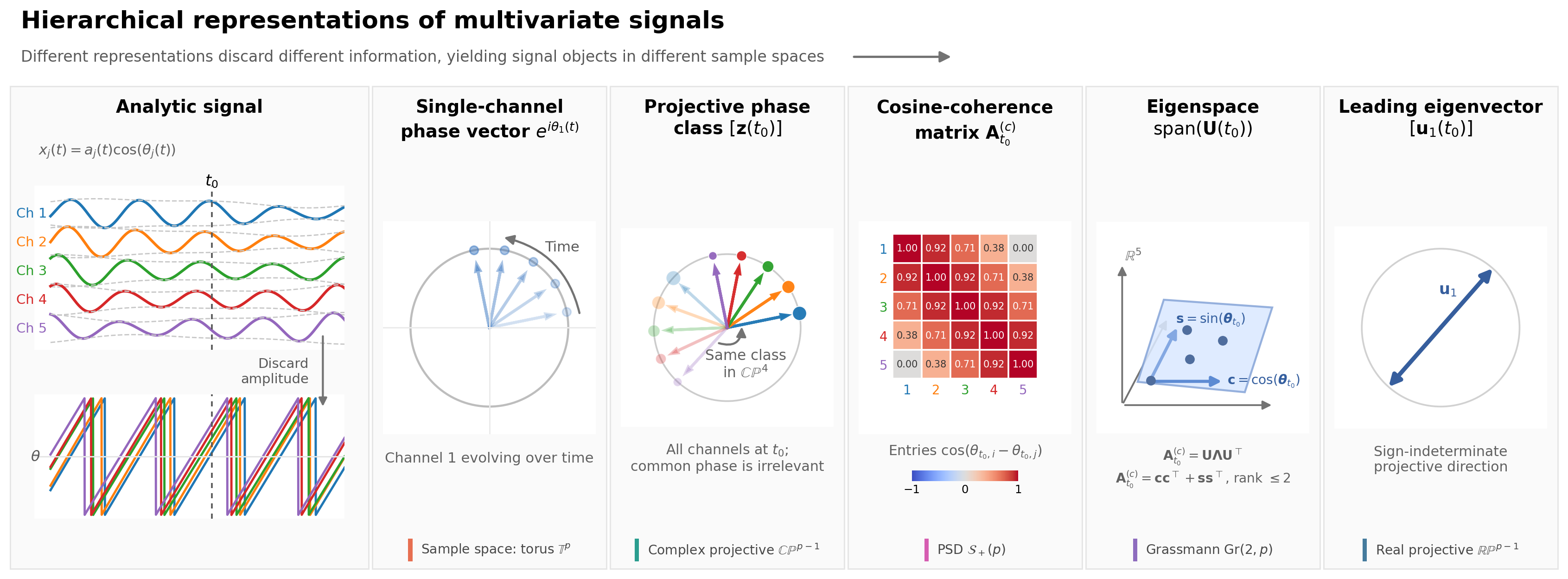}
\caption{Representations of multivariate phase relationships obtained by discarding different sources of variation. Removing regional amplitudes gives a phase-only configuration, and identifying common phase rotations retains its relative-phase structure. Taking the real part of the pairwise complex phase differences gives a cosine-coherence matrix of rank at most two. The matrix can be further summarized by its range or its sign-indeterminate leading eigenvector; these are alternative reductions that retain different aspects of its spectral structure. Colors identify corresponding channels across panels.}
\label{fig:representation_ladder}
\end{figure*}

To obtain regional phases, the BOLD signals are filtered within a narrow frequency band and combined with their Hilbert transforms to form analytic signals,
\begin{equation}
y_j(t)=a_j(t)\exp\{i\theta_j(t)\}, \quad j=1,\ldots,p,
\end{equation}
where $x_j(t)$ is the filtered BOLD signal, $a_j(t)$ is its instantaneous amplitude, and $\theta_j(t)$ describes its position within the oscillatory cycle. The resulting normalized phase vector is
\begin{equation}
\bm z(t)=\frac{1}{\sqrt p}\exp\{i\bm\theta(t)\}.
\end{equation}
Each regional phase is circular, and the joint phase configuration lies on the torus $\Tor^p=(\Sph^1)^p$. However, when considering phase time-series, the phase-coherence states concern relationships among regions rather than their shared position within an oscillatory cycle. For example, a group of regions fluctuating in phase expresses the same coherent configuration when those signals are near their peaks as when they are near their troughs. More generally, advancing every regional phase by the same angle leaves every pairwise phase difference unchanged,
\begin{equation}
(\theta_j(t)+\alpha)-(\theta_k(t)+\alpha)=\theta_j(t)-\theta_k(t).
\end{equation}
A common phase rotation therefore does not distinguish the connectivity states targeted by this analysis. Removing this variation preserves global coherence: a configuration in which all regions are aligned remains distinguishable from one in which their phases are dispersed. What is discarded is the common phase position, not the degree of alignment. If the data were time-locked, e.g., to a stimulus, a common phase would not be irrelevant.

Geometrically, identifying configurations related by a common rotation replaces $\Tor^p$ with the quotient $\Tor^p/\Sph^1\cong\Tor^{p-1}$. Fig.~\ref{fig:torus} illustrates this for two channels: configurations that move together around the torus share the same relative phase and become a single point after the common phase is removed.

The same relative-phase configuration can be represented by the projective class $[\bm z(t)]\in\CP^{p-1}$, which identifies $\bm z(t)$ and $e^{i\alpha}\bm z(t)$~\cite{olsen_uncovering_2026}. However, the complex projective space has larger support than the quotient torus, e.g., the vector $\bm z=[1,0,\ldots,0]$ is a valid point on $\CP^{p-1}$ but not the torus. As such, phase-only observations occupy an equal-magnitude subset of the complex projective space. However, using a model on $\CP^{p-1}$ provides access to computationally attractive distributions, including the complex ACG distribution~\cite{statistical_tyler_1987}.

Using toroidal or projective spaces, all pairwise phase differences are retained in the Hermitian outer product
\begin{equation}
\bm H(t)=p\bm z(t)\bm z(t)^{\mathrm H}, \quad H_{jk}(t)=\exp\{i(\theta_j(t)-\theta_k(t))\}.
\end{equation}
A common further simplification in fMRI is to retain only its real part, giving the instantaneous cosine-coherence matrix,
\begin{equation}
\bm A^{(c)}(t)=\operatorname{Re}\bm H(t), \quad A^{(c)}_{jk}(t)=\cos(\theta_j(t)-\theta_k(t)).
\end{equation}
Values near $1$ indicate aligned regional fluctuations, values near $-1$ indicate opposing fluctuations, and values near zero indicate approximately quarter-cycle offsets. This representation emphasizes the spatial pattern of coherence and anticoherence but discards other consistent offsets, and configurations with opposite directions of phase offset (e.g., $\pi/2$ and $3\pi/2$) become indistinguishable~\cite{olsen_uncovering_2026}.

Despite containing $p\times p$ entries, the cosine-coherence matrix has rank at most two, since~\cite{psilocybin_olsen_2022}
\begin{equation}
\bm A^{(c)}(t)=\bm c(t)\bm c(t)^\top+\bm s(t)\bm s(t)^\top,
\end{equation}
where $\bm c(t)=\cos\bm\theta(t)$ and $\bm s(t)=\sin\bm\theta(t)$, i.e., the angle difference identity. It is therefore a low-rank positive-semidefinite observation whose two nonzero eigenvalues are $\lambda_\pm=\frac{p}{2}\{1\pm\bar{R}_2(t)\}$, where $\bar{R}_2(t)=|p^{-1}\sum_j\exp\{2i\theta_j(t)\}|$ is the mean resultant length of the doubled phases~\cite{mardiaDirectionalStatistics1999}. The spectrum thus depends on the phases only through this single dispersion measure: $\bm A^{(c)}(t)$ has rank one when all phases agree modulo $\pi$ ($\bar{R}_2=1$), and the eigengap $p\bar{R}_2$ vanishes as $\bar{R}_2\to0$, leaving the leading eigenvector ill-defined. Further reductions emphasize different aspects of this structure. Retaining only the eigenbasis gives a point on the Grassmann manifold $\Gr(2,p)$ (for rank-two $\bm A^{(c)}$) and removes the eigenvalue weighting within that subspace. Alternatively, retaining only the leading eigenvector summarizes the dominant spatial coherence pattern, as in leading eigenvector dynamics analysis (LEiDA)~\cite{cabralCognitivePerformanceHealthy2017}, which is a particularly popular model for dynamic functional brain connectivity. LEiDA discards the secondary mode and the eigenvalues. Because the sign of an eigenvector is arbitrary, the retained direction belongs to real projective space $\RP^{p-1}$ although this distinction is often disregarded in the literature as we noted in Fig.~\ref{fig:3dspheres}.
%where $\bm c(t)=\cos\bm\theta(t)$ and $\bm s(t)=\sin\bm\theta(t)$, i.e., the angle difference identity. It is therefore a low-rank positive-semidefinite observation, fully specified by its nonzero eigenvalues and associated eigenspaces. Further reductions emphasize different aspects of this structure. Retaining only the eigenbasis gives a point on the Grassmann manifold $\Gr(2,p)$ (for rank-two $\bm A^{(c)}$) and removes the eigenvalue weighting within that subspace. Alternatively, retaining only the leading eigenvector summarizes the dominant spatial coherence pattern, as in leading eigenvector dynamics analysis (LEiDA)~\cite{cabralCognitivePerformanceHealthy2017}, which is a particularly popular model for dynamic functional brain connectivity. LEiDA discards the secondary mode and the eigenvalues. Because the sign of an eigenvector is arbitrary, the retained direction belongs to real projective space $\RP^{p-1}$ although this distinction is often disregarded in the literature as we noted in Fig.~\ref{fig:3dspheres}.

These representations thus remove different sources of variation: regional amplitudes, common phase, the sine, and aspects of the spectral structure of cosine coherence, see Fig.~\ref{fig:representation_ladder}. The resulting observations require different statistical models, and their suitability depends on which features distinguish the phase-coherence states of interest. We first examine these choices in a controlled synthetic benchmark, isolating the effects of global-phase treatment, retained information, and model complexity of the fitted model. We then translate these observations to an empirical study of brain phase-coherence states in multichannel fMRI.

\subsection{Synthetic benchmark}

%\begin{figure*}[t]
%    \centering
%    \includegraphics[width=1\linewidth]{figures_tables/synthetic_phase_mixture_oscillatory_and_fixed.png}
%    \caption{Recovery of two latent states from synthetic six-channel phase data. The left column shows the oscillatory time-series condition, in which a common rotating phase is added to all channels, making the marginal phases approximately uniform while preserving relative phases. The right column shows the fixed-phase, or stimulus-locked, condition, in which observations remain concentrated around the two state templates. The top row illustrates representative observations at noise arc width $\nu=\pi/2$: state 1 has synchronized relative phases, whereas state 2 has evenly spaced relative phases. Colored squares indicate the state-specific rank-3 perturbation structure. The middle and bottom rows report normalized mutual information between the true and estimated state assignments as a function of noise arc width for isotropic and state-specific rank-3 anisotropic noise, respectively. Each method was fitted with two components. Legends are ordered by area under the displayed NMI curve, reported in parentheses. Ordinary-torus methods retain absolute phase, whereas quotient-torus and projective representations are invariant to a common phase rotation. Higher NMI indicates better recovery; chance-level or failed recovery approaches zero.}
%    \label{fig:synthetic}
%\end{figure*}

\begin{figure}[t]
    \centering
    \includegraphics[width=1\linewidth]{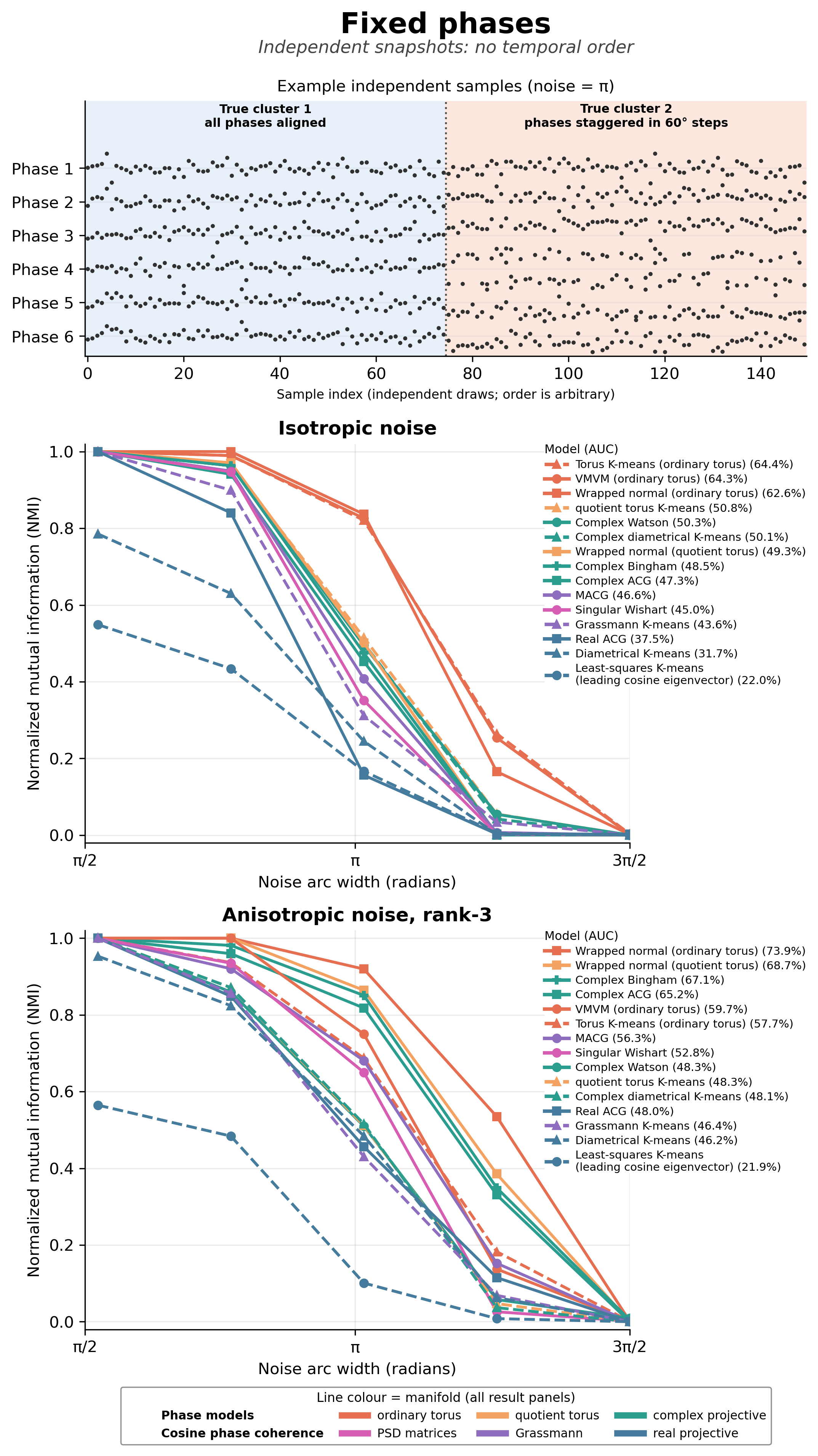}
    %\caption{Recovery of two latent states in synthetic six-channel phase data for fixed phases (as opposed to oscillatory) emulating a stimulus-locked condition. The top row illustrates the synthetic observations at noise arc width $\nu=\pi$. The middle and bottom rows report normalized mutual information between the true and estimated state assignments as a function of noise for isotropic and state-specific rank-3 anisotropic noise, respectively. Each mixture or $K$-means method was fitted with two components. Legends are ordered by area under the displayed NMI curve, reported in parentheses. Higher NMI indicates better recovery; chance-level or failed recovery approaches zero.}
    \caption{Recovery of two latent states in synthetic six-channel phase data for fixed phases (as opposed to oscillatory) emulating a stimulus-locked condition. The top row illustrates the synthetic observations at noise arc width $\nu=\pi$. The middle and bottom rows report normalized mutual information between the true and estimated state assignments as a function of noise for isotropic and state-specific rank-3 anisotropic noise, respectively. Line colour denotes the sample space and line style the method (solid: mixture model; dashed: $K$-means). Each method was fitted with two components. Legends are ordered by area under the displayed NMI curve, reported in parentheses. Higher NMI indicates better recovery; chance-level or failed recovery approaches zero.}
    \label{fig:synthetic_fixed}
\end{figure}

\begin{figure}[t]
    \centering
    \includegraphics[width=1\linewidth]{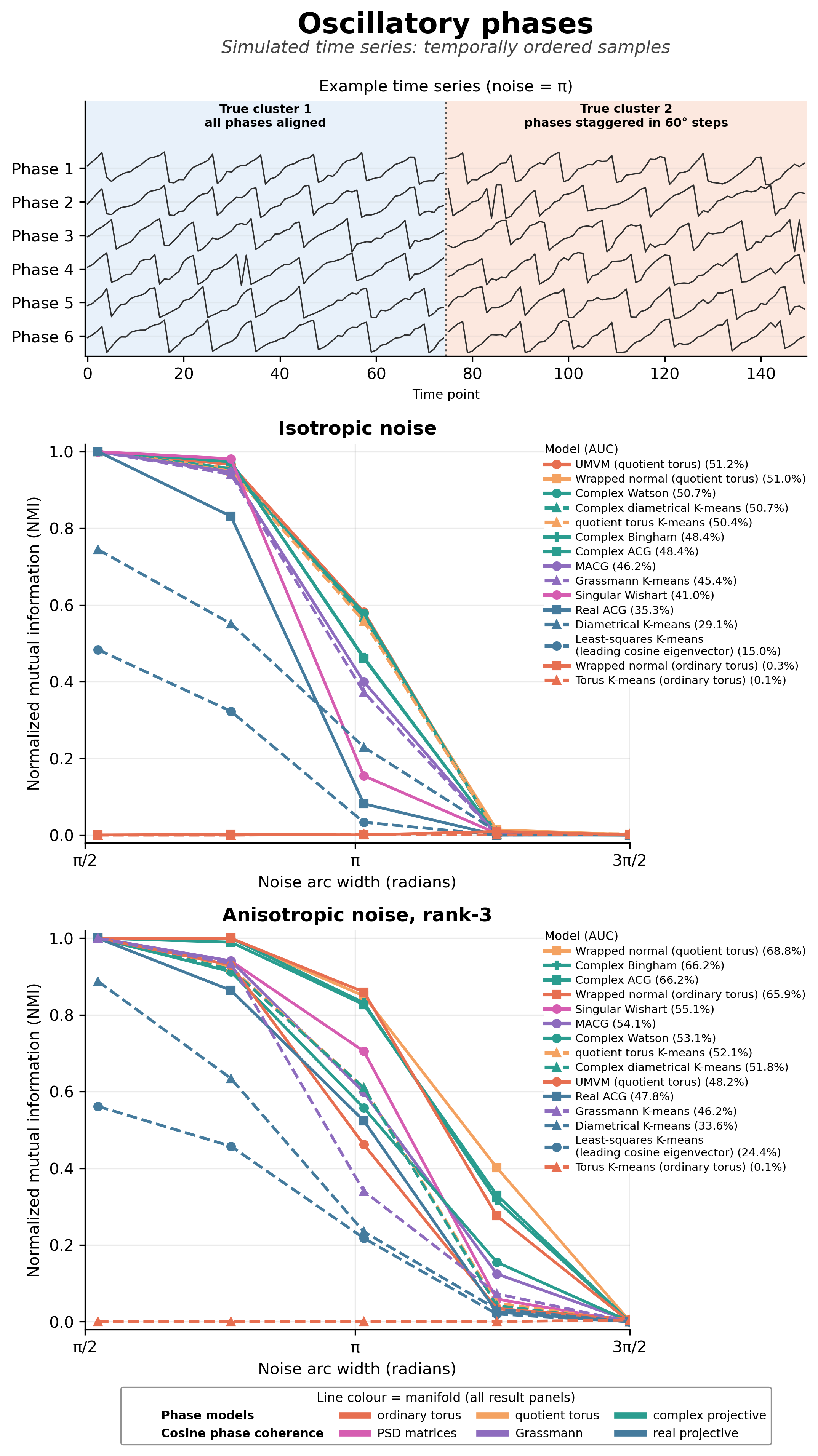}
    %\caption{Recovery of two latent states in synthetic six-channel phase data for oscillatory phases emulating a continuous recording condition. The top row illustrates the synthetic observations at noise arc width $\nu=\pi$. The middle and bottom rows report normalized mutual information between the true and estimated state assignments as a function of noise for isotropic and state-specific rank-3 anisotropic noise, respectively. Each mixture or $K$-means method was fitted with two components. Legends are ordered by area under the displayed NMI curve, reported in parentheses. Higher NMI indicates better recovery; chance-level or failed recovery approaches zero.}
    \caption{As Fig.~\ref{fig:synthetic_fixed}, but for oscillatory phases emulating a continuous recording condition, where a common carrier phase is added to all regions.}
    \label{fig:synthetic_oscillatory}
\end{figure}

%To reflect two common functional neuroimaging settings, we consider a time-locked condition and an oscillatory, resting-state-like condition, where in the latter, global phase is generally irrelevant. In both settings, we synthesized six-dimensional phase observations from an equally weighted mixture of two latent states, representing six arbitrary regions of the brain. The states were assigned strongly contrasting relative-phase patterns, with one state centered at zero phase difference and the other following a regularly spaced phase pattern. Variability around each state was introduced by adding Gaussian angular noise and wrapping the resulting phases to $(-\pi,\pi]$. We considered 5 noise levels and two noise covariance structures. In the isotropic condition, noise was independent and equally distributed across regions. In the anisotropic condition, a rank-3 structured component introduced state-specific dependencies between regions. The two states therefore differed in both their characteristic relative-phase patterns and the orientation of their covariance structure. In the oscillatory condition, the perturbations were temporally correlated, and a common rotating phase ``carrier" was added to all regions. 

To reflect two common functional neuroimaging settings, we consider a time-locked condition and an oscillatory, resting-state-like time series condition in which global phase is irrelevant; the oscillatory conditions mirror the fMRI data analyzed below and the audio data in Sec.~\ref{sec:audio}, respectively. In both settings, we drew $n_r=75$ phase vectors from each of two latent states $r\in\{1,2\}$, representing six arbitrary brain regions, by wrapping $\x\sim\mathcal{N}_6(\bm\mu^{(r)},\bm\Sigma^{(r)})$ to $(-\pi,\pi]^6$. The states had strongly contrasting relative-phase patterns, synchronized with $\bm\mu^{(1)}=\bm 0$ and evenly spaced with $\bm\mu^{(2)}=(0,\pi/3,\ldots,5\pi/3)^\top$. Isotropic noise used $\bm\Sigma^{(r)}=\sigma^2\bm I_6$, whereas anisotropic noise added a state-specific rank-3 component, so that the two states differed in both their relative-phase pattern and the orientation of their covariance structure. The noise magnitude was varied over five levels. In the oscillatory condition, the perturbations were temporally correlated and a common rotating carrier phase $\phi_t$ was added to all regions, $\bm\theta_t=\x_t+\phi_t\bm 1$ (wrapped), so that only relative phases identify the state.

%For each of five noise levels, we fitted two-component mixture models and $K$-means methods and quantified agreement between the estimated and true state assignments using normalized mutual information (NMI), which is a measure of weighted assignment overlap invariant to permutations of groups (see Sec.~\ref{app:nmi} for the definition). We tested mixtures of many of the non-matrix distributions mentioned in this paper, including the rank-1 von Mises copula-distribution on the torus with either von Mises marginals (VMVM) or uniform marginals (UMVM)~\cite{carriere2026circulabasedmultivariatedistributionsflat}. We also tested a least squares $K$-means on leading eigenvectors of cosine phase coherence matrices, corresponding to the LEiDA-model often used in functional neuroimaging, to exemplify the consequences of disregarding geometry. 

For each noise level, we fitted two-component mixture models and $K$-means methods and measured agreement with the true states using normalized mutual information (NMI; Appendix~\ref{app:nmi}). We tested mixtures of many of the non-matrix distributions mentioned in this paper, including the rank-1 von Mises copula-distribution on the torus with either von Mises marginals (VMVM) or uniform marginals (UMVM)~\cite{carriere2026circulabasedmultivariatedistributionsflat}. We also tested a least squares $K$-means on leading eigenvectors of cosine phase coherence matrices, corresponding to the LEiDA-model often used in functional neuroimaging, to exemplify the consequences of disregarding geometry.

Fig.~\ref{fig:synthetic_fixed} shows NMI as a function of noise in the fixed-phase condition and Fig~\ref{fig:synthetic_oscillatory} the oscillatory condition; within each panel, methods are ordered in the legend by their area under the displayed NMI curve. Across all four benchmarks (isotropic/anisotropic, fixed phase/oscillatory phase), toroidal models performed the best, closely followed by models on the complex projective space. These modeled sample spaces most closely align with the data generator's sample space, so this is not surprising. Models for cosine phase coherence performed slightly worse, indicating that these models discard useful information in the synthetic data. Models that disrespect the geometry of the data representation performed poorer than their counterparts, e.g., least-squares $K$-means for leading cosine eigenvectors. For data with anisotropic dispersion, models capable of modeling such rank performed better than their rank-one and $K$-means counterparts. 

When the synthetic data were oscillatory, quotient torus $K$-means, quotient Wrapped Normal, UMVM, and complex projective models performed the best, while ordinary toroidal $K$-means and Wrapped Normal models failed completely. Thus, a method can be geometrically appropriate while remaining scientifically inappropriate when a common phase rotation should be treated as a nuisance transformation. For fixed phases, the ordinary toroidal VMVM, torus $K$-means, and Wrapped Normal models performed better than their quotient counterparts. The projective and coherence-based representations are also insensitive to the common carrier because they are constructed from relative phases. Similarly, the SPD, Grassmannian, and real-projective methods are also constructed from phase coherence matrices and therefore less affected by whether phases are oscillatory or dispersed around a fixed centroid.

These results support three main conclusions. First, the appropriate invariance should be determined by the scientific question: global phase can be informative in a phase-locked setting but detrimental when a common oscillatory phase is a nuisance. Second, using models on sample spaces closer to the data-generating space is favorable in these synthetic data. Third, model complexity should match the data's underlying rank structure. 

\subsection{Experimental results: fMRI}
We now focus on the unsupervised identification of recurring phase-coherence states induced by task activity performed by the subjects being scanned. We use fMRI data from the Human Connectome Project~\cite{vanessenHumanConnectomeProject2012}, wherein subjects were scanned seven times, each during a different task probing different cognitive actions including emotion, gambling, language, motor, relational, social, and working memory. Each scan was between 200 and 400 observations long and consisted of between 2-10 task/rest blocks. We downloaded and used a minimally preprocessed version of the data and summarized voxel-wise data in 100 cortical and 16 subcortical regions~\cite{schaefer_local-global_2018,tianTopographicOrganizationHuman2020}. We fitted multivariate real Normal distributions on the original time-series $\bm x(t)$, complex Normal distributions on the analytic signal $\bm y(t)$, and many of the same distributions as in the synthetic benchmark test. This analysis extends an analysis in our recent work~\cite{olsen_uncovering_2026}.

We note that this setting corresponds to the \emph{oscillatory phases} (Fig.~\ref{fig:synthetic_oscillatory}), specifically the anisotropic setting. In contrast to the synthetic benchmark, no ground-truth brain states are available. The number of components was fixed a priori to K = 7, matching the number of tasks, and task labels were used only to evaluate the fitted models, never during estimation. Each scan contains interleaved task and rest blocks, and a given task may engage several recurring coherence states, so NMI with task labels measures how well unsupervised states align with experimental conditions rather than recovery of true latent states. This differs from the car experiment (Sec. VII), where K was varied and recovery peaked at the known number of sources.

\begin{figure*}
    \centering
    \includegraphics[width=1\linewidth]{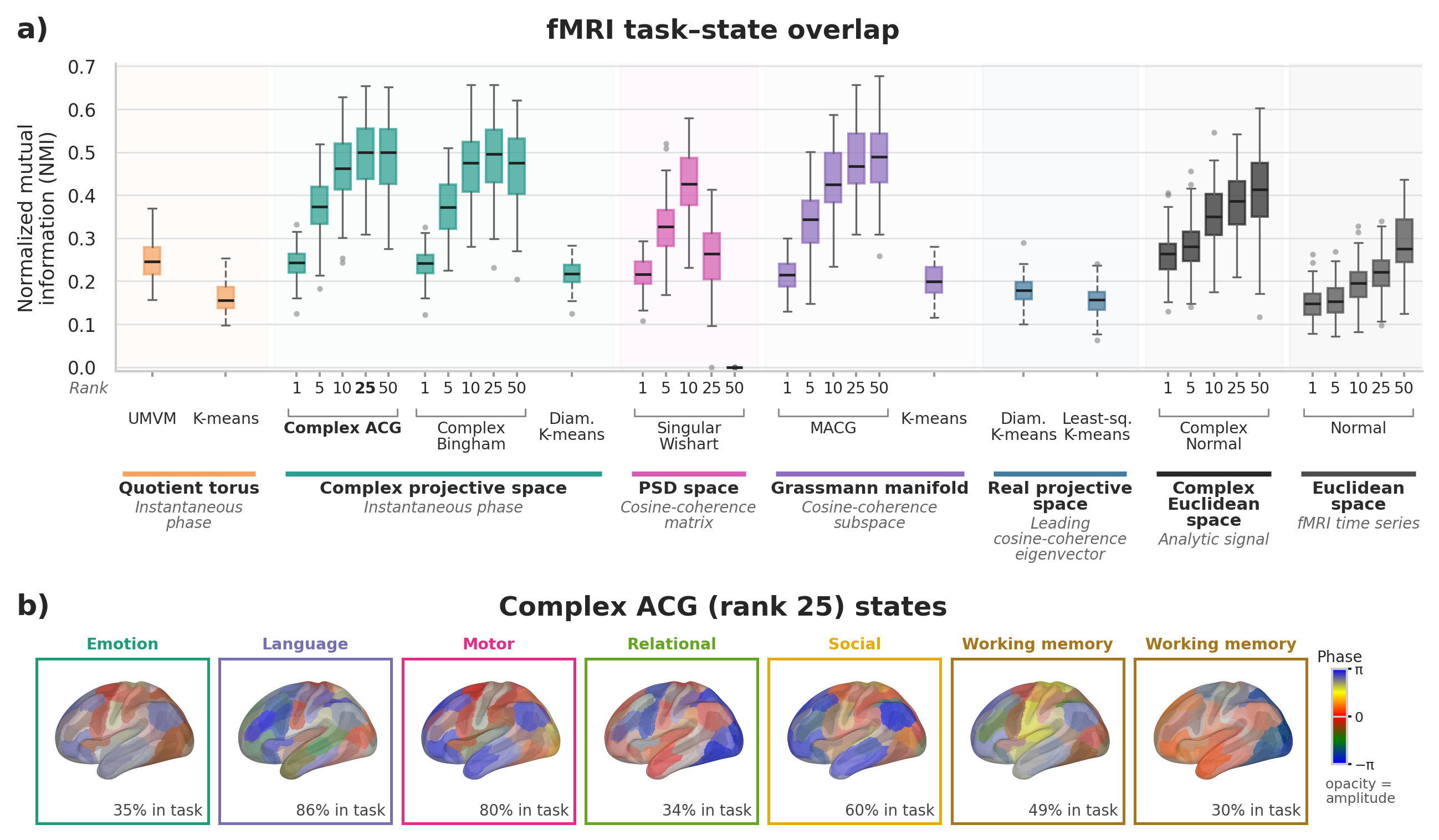}
    \caption{fMRI task–state overlap. (a) Subject-level normalized mutual information (NMI) between the fitted state partition and the task partition for the 99 held-out HCP subjects, with each model fitted with K = 7 states to 155 training subjects. Data were parcellated into 116 regions (Schaefer 100 cortical and Tian S1 subcortical). NMI was computed within task blocks, after pre-stimulus volumes were removed, between the model's state posterior and the seven task labels. The posterior is soft for mixture models and hard for K-means. K-means baselines are marked with dashed whiskers. (b) States of the best-performing model, Complex ACG (rank 25), each shown as the leading eigenvector of its scale matrix $\mathbf{I} + \mathbf{M}_k\mathbf{M}_k^{H}$ on the inflated left cortex (lateral view). Color shows phase, and opacity shows amplitude relative to the largest left-cortical amplitude across states. Because each eigenvector is defined only up to a global phase, it is rotated so that most regions lie near phase 0 (in-phase, red). Each state is assigned to, and framed in the color of, the task in which it is most occupied on held-out subjects. The percentage gives the share of the state's within-task occupancy that falls in that task.}
    \label{fig:fmri}
\end{figure*}

The results can be seen in Fig.~\ref{fig:fmri} and show that the rank-25 complex ACG and complex Bingham models for complex projective data identify task states better than other models. The Normal and complex Normal distribution were slightly worse, indicating that the amplitude contains less useful information than the phases alone. As in the oscillatory anisotropic benchmark, models that remove the common phase and represent anisotropic dispersion outperform rank-one, K-means, and geometry-ignoring alternatives, and least-squares K-means on leading eigenvectors performs among the worst. %Unlike the synthetic data, however, the generating states are unknown, and task labels serve only as an external reference.

%The matrix angular central Gaussian distribution was very close in performance to the complex projective models, indicating that the cosine contains information almost sufficient to explain task structure without the information in the sine. The singular Wishart distribution performed worse than the MACG, indicating that the added support and necessity of modeling the eigenvalues does not benefit this model. Distributions capable of modeling anisotropy performed much better than rank-one and $K$-means counterparts. However, of the low-rank directional methods, the UMVM model performed slightly better, indicating an improvement in using toroidal models. 

The matrix angular central Gaussian distribution was very close in performance to the complex projective models, indicating that the cosine contains information almost sufficient to explain task structure without the information in the sine. The singular Wishart distribution performed worse than the MACG, indicating that the added support and the need to model the eigenvalues, which carry only the scalar $\bar{R}_2(t)$, did not benefit this model. Distributions capable of modeling anisotropy performed much better than rank-one and $K$-means counterparts. However, of the low-rank directional methods, the UMVM model performed slightly better, indicating an improvement in using toroidal models.

%\begin{figure}
%    \centering
%    \includegraphics[width=1\linewidth]{figures_tables/fmri_task_state_nmi_boxplot.png}
%    \caption{Recovery of seven tasks in fMRI data using seven-component mixture and $K$-means models. 155 subjects were used for training and results were evaluated on 100 held-out subjects. Top: overlap between test-set posteriors or $K$-means labels and the true task labels, measured using normalized mutual information (NMI). Bottom: the seven states of the best-performing model (complex ACG, rank 25), each visualized as the leading eigenvector of its component's scale matrix, $I + M_kM_k^H$, shown on the inflated left cortex. Color indicates phase, after rotating each eigenvector so its entries sum to a positive real number. Opacity indicates amplitude, scaled by the largest amplitude across all states. States are framed and ordered by the task in which they were most occupied on held-out subjects, and the number in each panel gives the share of that state's posterior mass falling within that task. Two states were assigned to working memory and none to gambling. }
%    \label{fig:fmri}
%\end{figure}

\section{Speaker identification in a car environment}\label{sec:audio}
\begin{figure*}
    \centering
    \includegraphics[width=1\linewidth]{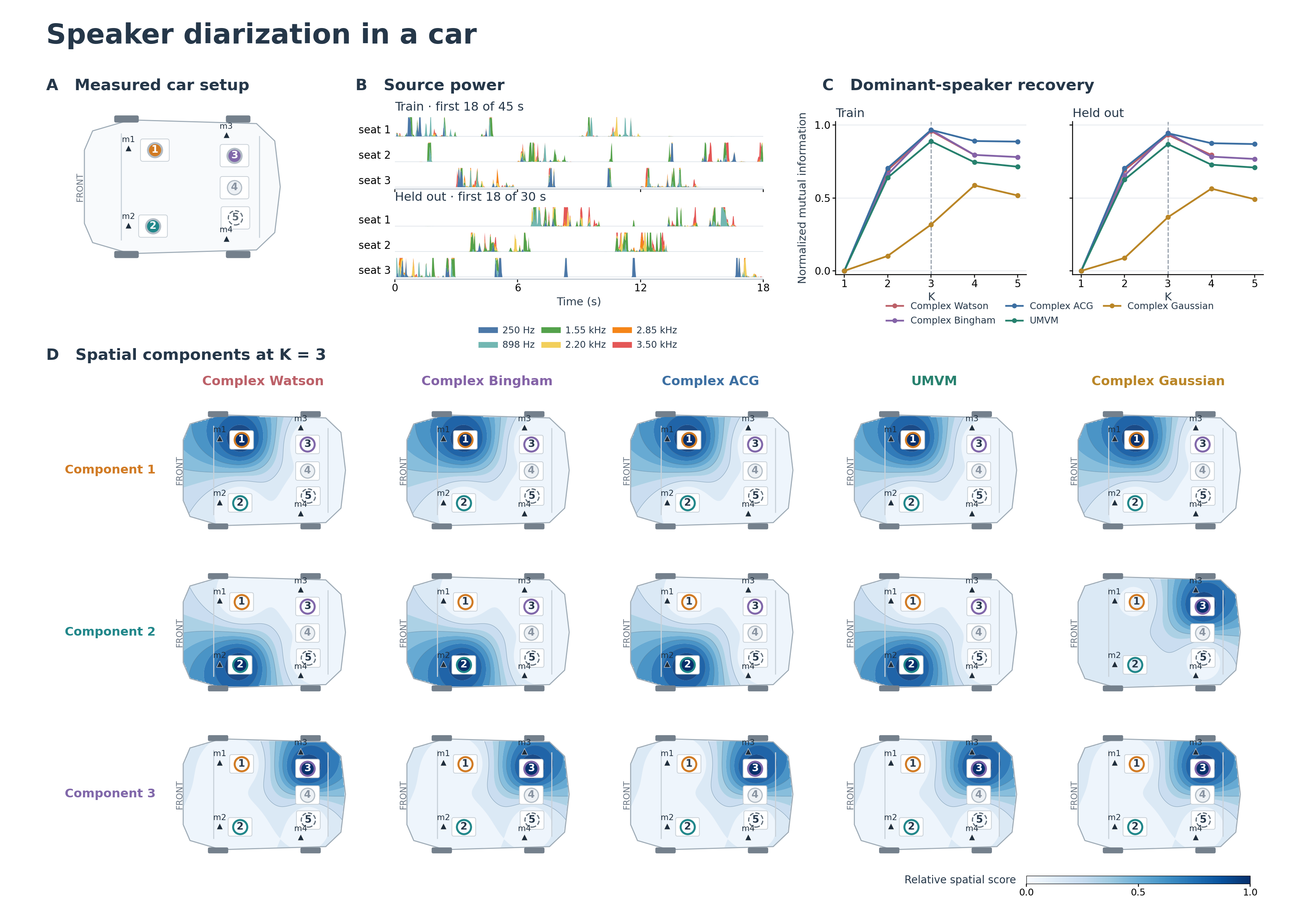}
    %\caption{Speaker diarization in measured car acoustics. A): A five-seat saloon layout cropped from the In-Car McVAMPIRE geometry with four microphones m1-m4. B): Isolated per-source power in the first 18 s of the training and held-out recordings. C): NMI between the dominant source and the shared component label in training and held-out frames as a function of the number of mixture components. D): Spatial scores for the $K=3$ components. Color denotes the component's relative score at the four measured candidate positions after pooling the six frequencies; contours are inverse-distance interpolations included only for visualization purposes, i.e., they are not actual location probabilities or fitted spatial covariance contours. UMVM denotes the uniform-marginal von Mises copula model.}
    \caption{Speaker diarization in simulated car acoustics. A): A five-seat saloon layout cropped from the In-Car McVAMPIRE geometry with four microphones m1-m4. B): Isolated per-source power in the first 18 s of the training and held-out recordings. C): NMI between the dominant source and the shared component label in training and held-out frames as a function of the number of mixture components. D): Spatial scores for the $K=3$ components. Color denotes the component's relative score at the four measured candidate positions after pooling the six frequencies; contours are inverse-distance interpolations included only for visualization purposes, i.e., they are not actual location probabilities or fitted spatial covariance contours. UMVM denotes the uniform-marginal von Mises copula model.}
    \label{fig:audio}
\end{figure*}

To provide a second use case for directional statistics we focus on speaker identification, a topic often occurring in signal processing and where directional statistics has previously shown to be advantageous~\cite{complex_ito_2016}. Specifically, we focus on a 5-seat car, with three seats occupied by a speaking person: two in the front and one in the back. Speakers were sampled from MiniLibriMix~\cite{joris_minilibrimix_2020} and projected to a 5-seat car using In-Car McVampire~\cite{muller_-car_2024} and sampled with $p=4$ microphones. The car recordings are therefore simulated rather than measured.
%Thus, while the speakers were actual people, these were not recorded in a car environment; instead, their audio was \emph{projected} into a car environment using the deep learning framework. 
The task is then to fit a mixture model with the correct number of components, which correctly identifies the spatial location of seats as well as the temporal activation sequence of speakers in a test set, see Fig.~\ref{fig:audio}. Typically, audio source recovery methods focus on a particular frequency band and computes the short time Fourier transform, such that the audio is represented by a time-series of $p$-dimensional complex observations. However, the energy in each observation can vary drastically. This has motivated the per-observation normalization to unit energy. Considering that global phase is irrelevant, these normalized complex-valued observations are naturally complex projective. We note that this case corresponds to the \emph{oscillatory phases} (Fig.~\ref{fig:synthetic_oscillatory}), the isotropic noise part, i.e., we don't expect methods modeling a covariance to perform better than methods merely modeling a mean. 

We simulated training and held-out recordings 45 s and 30 s long, respectively, and used disjoint source utterances. The conversation schedule was continuously active: every analysis frame contained at least one speaker. Approximately 90\% of the frames contained one active speaker, 8\% contained two, and 2\% contained all three. We retained six STFT bins, centered at 250, 898, 1547, 2203, 2852, and 3500 Hz. 

Here we compare the complex Gaussian distribution without normalizing STFT vectors, the complex Watson, Bingham, and ACG distributions after normalizing, and the UMVM model after extracting only the phase. For every model family, we fitted mixture orders $K=1$ to $5$ independently at each of the six frequencies. The frequency-specific component permutations were first aligned using their training responsibilities. We assessed component recovery using NMI. At each frequency, isolated source power was averaged over the four microphones and converted to a fraction of the total isolated-source power; the reference label was the speaker with the largest fraction after averaging over the six frequencies. We show training and held-out NMI rather than log likelihood because the model families are defined on different sample spaces and their absolute likelihoods are not easily comparable.

%The directional families showed a clear maximum in source recovery at the true number of occupied positions, $K=3$ (Fig.~\ref{fig:audio}C) for both the training and test sets. The same conclusion is visible in the spatial readout (Fig.~\ref{fig:audio}D): for each of these four models, the three displayed components had distinct maxima at occupied positions 1, 2, and 3. The complex Gaussian behaved differently. At $K=3$, the spatial readout contained a clear component for position 1 and two components whose angular preferences were concentrated near position 3. 

The directional families showed a clear maximum in source recovery at the true number of occupied positions, $K=3$ (Fig.~\ref{fig:audio}C) for both the training and test sets. The same conclusion is visible in the spatial readout (Fig.~\ref{fig:audio}D): for each of these four models, the three displayed components had distinct maxima at occupied positions 1, 2, and 3. The complex Gaussian behaved differently: its NMI was lower for all $K\geq2$ and peaked at $K=4$, and at $K=3$ its spatial readout contained one component for position 1 and two components near position 3, leaving position 2 unresolved.

Taken together, the results show that the measured multichannel phase and relative-amplitude structure is sufficient to identify the three reverberant source positions in held-out speech. Recovery is strongest when the component family treats global STFT magnitude as a nuisance and when the number of components matches the number of occupied positions. Because this experiment contains one fixed geometry and one constructed training--held-out scene, the reported NMI values are descriptive comparisons for this scene rather than population-level estimates; the STFT frames are overlapping and should not be treated as independent replicates.

\section{Discussion}
\label{sec:discussion}

Directional statistics provides a principled language for signal objects whose values are defined up to periodicity, normalization, sign, global phase, basis rotation, or positive-semidefinite structure. Practical value arises from the interaction of three considerations: whether the selected geometry reflects scientifically meaningful invariances in the data, whether the associated probability model adequately represents the observed variability (e.g., model complexity, elliptical vs quadratic, wrapped Normal vs von Mises), and whether inference remains computationally feasible at the relevant dimension, concentration, and sample size. The main opportunity for directional statistics in signal processing is not simply to replace Euclidean distributions with manifold-valued counterparts, but to make representation, invariance, uncertainty, and computation part of a single modeling decision.

\subsection{Benefits for signal modeling}

Respecting geometry changes both the quantities being estimated and the variation attributed to them. Circular averaging avoids discontinuities at phase-wrap boundaries; projective clustering groups sign- or phase-equivalent observations without representative-fixing heuristics; Grassmann models compare subspaces independently of their bases; and constrained matrix models retain positive-semidefinite structure. The resulting prototypes can be interpreted as phases, directions, axes, subspaces, or covariance objects of the same type as the observations. These operations address recurring tasks in signal processing such as phase tracking, brain connectivity, microphone-array processing, orientation estimation, and covariance-valued sensing. Mixtures translate variability into probabilistic state assignments, while directional emissions in hidden Markov models combine these assignments with state persistence and transition structure.

%The phase-coherence case study demonstrates how these choices affect the recovery of recurring signal states. In the synthetic experiments, toroidal and complex-projective models recovered states most effectively, while reductions to cosine coherence or leading eigenvectors discarded discriminative information. The treatment of common phase was important: absolute phase was informative in the fixed-phase condition, whereas removing the common carrier was essential in the oscillatory condition. Models capable of representing anisotropic dispersion also benefited from the structured variability in the data. In fMRI, rank-25 complex ACG and Bingham mixtures achieved the strongest task-state recovery, with MACG models performing closely. These results identify distinct sources of modeling improvement: preserving informative phase relationships, removing irrelevant variation, and allocating sufficient model complexity to within-state dependence.

In the phase-coherence case study, toroidal and complex-projective models recovered the synthetic states best, the treatment of common phase determined success or failure in the oscillatory condition, and anisotropic models benefited from structured within-state variability. In fMRI, rank-25 complex ACG and Bingham mixtures gave the strongest task-state recovery, with MACG close behind.

The audio experiment illustrates a related benefit for multichannel measurements. Normalizing the STFT observations retains relative amplitudes and phases while removing overall magnitude, and complex-projective models additionally identify observations differing only by a common phase. In the constructed three-source scene, the directional mixtures recovered the occupied source positions in held-out speech, with recovery peaking at the correct number of components. Here, directional modeling directs component capacity toward spatial source structure rather than variation in signal energy; the latter being more important for Euclidean models. Together, the two case studies show how representation and distributional flexibility can improve state discovery in different signal-processing settings.

These representations also make the information used by an analysis explicit. Removing amplitude, common phase, the imaginary part of complex coherence, or secondary eigendirections produces different statistical objects. Their suitability follows from the features needed for the task. The fMRI comparison, where MACG performed better than singular Wishart despite retaining less spectral information, illustrates that retaining additional information also requires sufficient data and an appropriate model for that information.

\subsection{Estimation and practical limitations}

The reviewed learning strategies provide several practical routes to employing directional statistics in modeling. Geometry-aware $K$-means gives representative states without evaluating a probability density; mixtures describe within-state variability and ambiguous membership; and hidden Markov models add temporal organization. Automatic differentiation supports likelihood optimization across these formulations and simplifies the incorporation of constraints and structured parameters. Low-rank-plus-isotropic parameterizations reduce the number of parameters, making quadratic and elliptical models more practical in high dimensions. The rank becomes an explicit complexity parameter. The \texttt{PCMM} toolbox implements these alternatives within a common framework, facilitating comparisons of representations, distribution families, and state-discovery methods.

An important distinction is between distributions that share a sample space but describe different forms of variability. Restricting an ambient density to a manifold, projecting out its radius, and wrapping its coordinates lead to different statistical constructions. Thus, Bingham and ACG models respect the same projective invariances but have different density shapes; their choice concerns both fit and computation. Likewise, a complex-projective model for phase-only data preserves relative phase while allowing support beyond the equal-magnitude observations. Such choices make it possible to retain the relevant invariance while selecting a computationally manageable family, provided its additional support and dispersion assumptions are appropriate for the application.

The main computational difficulty is the cost of evaluating some directional likelihoods. Matrix-argument normalizers, high-dimensional toroidal partition functions, and multivariate wrapping sums can become expensive, especially at high concentration. Mixtures multiply this cost through repeated component fitting, and hidden Markov models additionally require sequence recursions. Our timing comparisons illustrate substantial differences among families. Score matching and noise-contrastive estimation provide alternatives when normalizers are unavailable, with their own requirements: manifold differential operators for score matching, and suitable noise distributions for noise-contrastive estimation. In unnormalized mixtures, component normalizers and mixture weights are not separately recovered from their combined amplitudes alone.

\subsection{Topics beyond the scope of this review}

In our work, we did not cover dimensionality reduction and probabilistic matrix decomposition. Principal nested spheres and torus PCA summarize directional variation while respecting spherical or periodic structure~\cite{jung_analysis_2012,torus_eltzner_2018}. Directional statistics has also been used to describe probabilistic variation in Bayesian PCA using a matrix von Mises--Fisher distribution to describe uncertainty in principal directions~\cite{smidl_bayesian_2007}, and probabilistic PARAFAC2 uses the same family for orthogonal factors in multiway data~\cite{jorgensen_probabilistic_2024}. Directional statistics can therefore contribute both to constructing low-dimensional representations of directional observations and to quantifying uncertainty in the factors of a decomposition. 

Higher-order signal objects introduce further geometric structure. Moments, cumulants, and polyspectra can be represented by tensors whose constraints and factorization ambiguities differ from those of covariance matrices. Fixed-multilinear-rank Tucker and hierarchical tensor representations have associated manifold structures~\cite{falco_geometric_2015}; orthonormal factors can be represented on Stiefel manifolds, while their spans belong to Grassmann manifolds. The geometry depends on the selected tensor format, rank constraints, and equivalences among factors. Consequently, there is no single tensor counterpart to the positive-semidefinite matrix spaces reviewed here. A full treatment of tensor-valued observations and their probabilistic decompositions would require these additional choices.

A separate topic is the topology of the domain supporting the signal. Hypergraphs, simplicial complexes, and cell complexes encode relations among more than two entities, and topological signal processing uses incidence operators and Hodge Laplacians to describe signals on these domains~\cite{barbarossa_topological_2020,schaub_signal_2022}. This domain topology is distinct from the geometry of a tensor representation or of the signal values themselves. For example, an edge or simplex may carry a phase, orientation, or covariance matrix. Higher-order domains specify how observations are related, whereas directional geometry specifies their admissible values and invariances. Their joint statistical treatment lies outside the present review.

\subsection{Outlook}

Three developments are particularly relevant to the modeling choices examined here. First, vine copula and related pair-copula constructions offer a potential toward more flexible multivariate phase dependence than the one-factor circula used in our experiments~\cite{carriere2026circulabasedmultivariatedistributionsflat,ogata_pair_2023, kato_trivariate_2026}. Extending their practical use in signal processing could broaden the range of tractable toroidal models.

Second, score matching estimators (SMEs) remain to be explored fully in signal-processing applications. For exponential-family directional models on compact manifolds, such as the Fisher--Bingham, Bingham, multivariate sine, and torus-graph families, the score-matching objective is quadratic in the parameters, so the SME is obtained by solving a linear system without evaluating the normalizing constant~\cite{score_mardia_2016}. Moreover, the SME can retain high efficiency relative to maximum likelihood in concentrated regimes. Extending SMEs to mixtures and hidden Markov models, where normalizers are evaluated repeatedly, could make otherwise prohibitive toroidal and quadratic models practical in high dimensions.

Third, directional likelihoods can be incorporated further into representation learning and deep learning in general. Existing spherical embedding, rotation-estimation, and hyperspherical latent-variable models already demonstrate this connection~\cite{scott_vmf_2021,prokudin_deep_2018,davidson_hyperspherical_2018}. A learned feature direction can be accompanied by an estimated concentration or anisotropic dispersion, and a learned subspace can be modeled probabilistically rather than represented only by a deterministic basis. This connects feature learning to uncertainty and interpretable state models. The framework developed here provides the corresponding modeling sequence: identify the relevant signal information and invariances, choose a compatible representation and distribution, and fit it using an inference strategy suited to its statistical and computational requirements.

\section{Acknowledgements}
The authors acknowledge Guillaume Carrière and Prof. Frédéric Cazals at INRIA for their help in understanding and implementing the multivariate torus copula distribution. 

AI was used in the preparation of this manuscript. We used Litmaps for literature search, including citation graph searches and topic searches. We used ChatGPT 5.5 and 5.6Sol for brainstorming and drafting the initial structure of the manuscript and to mark factual errors. However, the bulk of the text was written by humans and AI was not used to correct language outside of detecting factual errors. Codex 5.5 and 5.6Sol was used to draft the Python code leading to Figs 4, 6, 8-12. The PCMM toolbox was mostly written before generative AI for coding became widely available, but the additional distributions introduced to the toolboxes along this paper were also developed using Codex - these are gathered in a single script called \texttt{PCMMtorchAdditional.py}. 

\appendices
\section{Hidden Markov modeling}\label{sec:app_hidden_markov_models}

Many signal-processing data sets are sequential. In time series, contiguous observations may have a higher probability of belonging to the same latent state than temporally distant observations. A hidden Markov model introduces the latent state sequence $z_i\in{1,\ldots,K}$ and a transition probability matrix $\bm T$, where
\begin{equation}
T_{k',k}=p(z_i=k\mid z_{i-1}=k')
\end{equation}
denotes the probability of transitioning from state $k'$ to state $k$. The transition probabilities satisfy $T_{k',k}\geq0$ and $\sum_{k=1}^K T_{k',k}=1$ for every $k'$. Each state is associated with a normalized emission distribution $f(\x_i\mid\boldsymbol\theta_k)$.

For $S$ independent observation sequences, let $\x_i^{(s)}$ denote observation $i$ in sequence $s$, and let $n_s$ be the length of that sequence. The initial state probabilities are
\begin{equation}
\bm\delta=\{\delta_k\}_{k=1}^K,\quad\delta_k\geq0,\quad\sum_{k=1}^K\delta_k=1.
\end{equation}
Unlike the mixture weights $\bm w$, the vector $\bm\delta$ governs only the first observation in each sequence.

The unscaled forward recursion is initialized by
\begin{equation}
\alpha_{1,k}^{(s)}=\delta_k f\left(\x_1^{(s)}\mid\boldsymbol\theta_k\right),
\end{equation}
and continued according to
\begin{equation}
\alpha_{i,k}^{(s)}=f\left(\x_i^{(s)}\mid\boldsymbol\theta_k\right)\sum_{k'=1}^K\alpha_{i-1,k'}^{(s)}T_{k',k},\quad i=2,\ldots,n_s.
\end{equation}
The corresponding log-likelihood is
\begin{equation}
\mathcal L_{\operatorname{HMM}}(\boldsymbol\theta,\boldsymbol\delta,\bm T\mid\mathcal X)=\sum_{s=1}^S\log\left(\sum_{k=1}^K\alpha_{n_s,k}^{(s)}\right).
\end{equation}
Parameters may be estimated using, e.g., the Baum--Welch EM algorithm or by direct gradient-based maximization of the forward log-likelihood. In the latter case, $\bm\delta$ and each row of $\bm T$ can be parameterized using softmax transformations.

After parameter estimation, the Viterbi decoder can be used to estimate the most likely latent state sequence $\bm z$~\cite{forney_viterbi_1973}. Dropping sequence-specific notation, the Viterbi recursion is initialized by
\begin{equation}
\xi_{1,k}=\delta_k f(\x_1\mid\boldsymbol\theta_k),
\end{equation}
and continued according to
\begin{equation}
\xi_{i,k}=f(\x_i\mid\boldsymbol\theta_k)\max_{k'}\xi_{i-1,k'}T_{k',k},\quad i=2,\ldots,n.
\end{equation}
The corresponding backpointer is
\begin{equation}
\psi_{i,k}=\argmax_{k'}\xi_{i-1,k'}T_{k',k}.
\end{equation}
The final state is
\begin{equation}
z_n=\argmax_k\xi_{n,k},
\end{equation}
and the most likely path is recovered through
\begin{equation}
z_i=\psi_{i+1,z_{i+1}},\quad i=n-1,\ldots,1.
\end{equation}

\section{Normalized mutual information}\label{app:nmi}
Given two binary or weighted responsibility matrices $\boldsymbol \beta_1$ and $\boldsymbol \beta_2$ of size $K_1\times N$ and $K_2\times N$, respectively, which sum to one for each column, normalized mutual information (NMI) is defined as~\cite{hinrichArchetypalAnalysisModeling2016}
\begin{align}
    \mathrm{NMI}(\boldsymbol \beta_1,\boldsymbol \beta_2)&= \frac{2 \mathrm{MI}(\boldsymbol \beta_1,\boldsymbol \beta_2)}{\mathrm{MI}(\boldsymbol \beta_1,\boldsymbol \beta_1)+\mathrm{MI}(\boldsymbol \beta_2,\boldsymbol \beta_2)},\nonumber\\
    \mathrm{MI}(\boldsymbol \beta_1,\boldsymbol \beta_2) &= \sum_{k_1,k_2}p(k_1,k_2)\log \frac{p(k_1,k_2)}{p(k_1)p(k_2)},\nonumber\\
    p(k_1,k_2) &= \frac{1}{N} \sum_n \beta_{k_1,n} \beta_{k_2,n}.\nonumber
\end{align}

\bibliographystyle{IEEEtran}
\bibliography{references/references_cleaned}

@IEEEtranBSTCTL{IEEEexample:BSTcontrol,
  CTLuse_forced_etal       = "yes",
  CTLmax_names_forced_etal = "6",
  CTLnames_show_etal       = "1",
  CTLdash_repeated_names   = "no"
}

@article{mardia_principal_2022,
	series = {50th {Anniversary} {Jubilee} {Edition}},
	title = {Principal component analysis and clustering on manifolds},
	volume = {188},
	issn = {0047-259X},
	doi = {10.1016/j.jmva.2021.104862},
	urldate = {2026-10-30},
	journal = {Journal of Multivariate Analysis},
	author = {Mardia, Kanti V. and Wiechers, Henrik and Eltzner, Benjamin and Huckemann, Stephan F.},
	month = mar,
	year = {2022},
	pages = {104862},
}

@article{Heggli_2021,
    author = {Heggli, Ole Adrian and Konvalinka, Ivana and Cabral, Joana and Brattico, Elvira and Kringelbach, Morten L and Vuust, Peter},
    title = {Transient brain networks underlying interpersonal strategies during synchronized action},
    journal = {Social Cognitive and Affective Neuroscience},
    volume = {16},
    number = {1-2},
    pages = {19-30},
    year = {2021},
    month = {01},
    doi = {10.1093/scan/nsaa056},
}

@Article{preti1701,
  author    = {Preti, Maria Giulia and Bolton, Thomas and Van De Ville, Dimitri},
  title     = {The dynamic functional connectome: State-of-the-art and perspectives},
  journal   = {NeuroImage},
  year      = {2017},
  volume    = {160},
  pages     = {41-54},
  doi       = {10.1016/j.neuroimage.2016.12.061},
}

@article{vonMises1918Ganzzahligkeit,
  author  = {von Mises, Richard},
  title   = {{\"U}ber die `{Ganzzahligkeit}' der {Atomgewichte} und verwandte {Fragen}},
  journal = {Physikalische Zeitschrift},
  year    = {1918}
}

@article{maximum_chen_2021,
	title = {Maximum likelihood estimation of the {Fisher}–{Bingham} distribution via efficient calculation of its normalizing constant},
	doi = {10.1007/s11222-021-10015-9},
	author = {Chen, Yici and Tanaka, Ken’ichiro},
	journal = {Statistics and Computing},
	year = {2021},
	litmapsId = {262572420}
}

@misc{carriere2026circulabasedmultivariatedistributionsflat,
      title={Circula-based multivariate distributions on the flat torus, with applications in structural biology}, 
      author={Guillaume Carrière and Alix Lhéritier and Frédéric Cazals},
      year={2026},
      eprint={2605.12577},
      archivePrefix={arXiv},
      primaryClass={stat.AP},
}

@article{fletcher_riemannian_2007,
  author  = {Fletcher, P. Thomas and Joshi, Sarang},
  title   = {Riemannian geometry for the statistical analysis of diffusion tensor data},
  journal = {Signal Processing},
  volume  = {87},
  number  = {2},
  pages   = {250--262},
  year    = {2007},
  doi     = {10.1016/j.sigpro.2005.12.018}
}

@article{basser_diffusion_1994,
  author  = {Basser, Peter J. and Mattiello, James and LeBihan, Denis},
  title   = {{MR} diffusion tensor spectroscopy and imaging},
  journal = {Biophysical Journal},
  volume  = {66},
  number  = {1},
  pages   = {259--267},
  year    = {1994},
  doi     = {10.1016/S0006-3495(94)80775-1}
}

@article{complete_bagyan_2024,
	title = {Complete asymptotic expansions for the normalizing constants of high-dimensional matrix {Bingham} and matrix {Langevin} distributions},
	doi = {10.3842/sigma.2024.094},
	author = {Bagyan, Armine and Richards, Donald},
	journal = {Symmetry, Integrability and Geometry: Methods and Applications},
	year = {2024},
	litmapsId = {279800397}
}

@article{HornikGrun2014movMF,
  author  = {Hornik, Kurt and Gr{\"u}n, Bettina},
  title   = {{movMF}: An {R} package for fitting mixtures of von {Mises}--{Fisher} distributions},
  journal = {Journal of Statistical Software},
  year    = {2014},
  volume  = {58},
  number  = {10},
  pages   = {1--31},
  doi     = {10.18637/jss.v058.i10},
}

@article{ChakrabortyWong2021BAMBI,
  author  = {Chakraborty, Saptarshi and Wong, Samuel W. K.},
  title   = {{BAMBI}: An {R} package for fitting bivariate angular mixture models},
  journal = {Journal of Statistical Software},
  year    = {2021},
  volume  = {99},
  number  = {11},
  pages   = {1--69},
  doi     = {10.18637/jss.v099.i11},
}

@article{JonaLasinioEtAl2020CircSpaceTime,
  author  = {Jona Lasinio, Giovanna and Santoro, Mario and
             Mastrantonio, Gianluca},
  title   = {{CircSpaceTime}: An {R} package for spatial and spatio-temporal modelling of circular data},
  journal = {Journal of Statistical Computation and Simulation},
  year    = {2020},
  volume  = {90},
  number  = {7},
  pages   = {1315--1345},
  doi     = {10.1080/00949655.2020.1725008},
}

@article{KurzEtAl2019libDirectional,
  author  = {Kurz, Gerhard and Gilitschenski, Igor and Pfaff, Florian
             and Drude, Lukas and Hanebeck, Uwe D. and
             Haeb-Umbach, Reinhold and Siegwart, Roland Y.},
  title   = {Directional statistics and filtering using {libDirectional}},
  journal = {Journal of Statistical Software},
  year    = {2019},
  volume  = {89},
  number  = {4},
  pages   = {1--31},
  doi     = {10.18637/jss.v089.i04},
}

@article{Berens2009CircStat,
  author  = {Berens, Philipp},
  title   = {{CircStat}: A {MATLAB} toolbox for circular statistics},
  journal = {Journal of Statistical Software},
  year    = {2009},
  volume  = {31},
  number  = {10},
  pages   = {1--21},
  doi     = {10.18637/jss.v031.i10},
}

@article{SablicaEtAl2025Watson,
  author  = {Sablica, Lukas and Hornik, Kurt and Leydold, Josef},
  title   = {{watson}: An {R} package for fitting mixtures of {Watson} distributions},
  journal = {Journal of Statistical Software},
  year    = {2025},
  volume  = {115},
  number  = {4},
  pages   = {1--30},
  doi     = {10.18637/jss.v115.i04},
}

@article{PewseyGarciaPortugues2021,
  author  = {Pewsey, Arthur and Garc{\'i}a-Portugu{\'e}s, Eduardo},
  title   = {Recent advances in directional statistics},
  journal = {TEST},
  year    = {2021},
  volume  = {30},
  number  = {1},
  pages   = {1--58},
  doi     = {10.1007/s11749-021-00759-x},
}

@Manual{Riemann,
  title = {{Riemann}: Learning with Data on Riemannian Manifolds},
  author = {Kisung You},
  year = {2025},
  note = {R package version 0.1.6},
  url = {https://www.kisungyou.com/Riemann/},
}

@book{topics_jammalamadaka_2001,
	title = {Topics in Circular Statistics},
	doi = {10.1142/4031},
	author = {Jammalamadaka, S. R. and SenGupta, A.},
	year = {2001},
	litmapsId = {175339393},
    publisher = {World Scientific}
}

@article{dispersion_fisher_1953,
	title = {Dispersion on a sphere},
	doi = {10.1098/rspa.1953.0064},
	author = {Fisher, R.},
	journal = {Proceedings of the Royal Society of London. Series A. Mathematical and Physical Sciences},
	year = {1953},
	litmapsId = {124956953}
}

@article{diametrical_dhillon_2003,
	title = {Diametrical clustering for identifying anti-correlated gene clusters},
	doi = {10.1093/bioinformatics/btg209},
	author = {Dhillon, I. and Marcotte, E. and Roshan, Usman},
	journal = {Bioinformatics},
	year = {2003},
	litmapsId = {222525734}
}

@article{matrix_chikuse_1990,
	title = {The matrix angular central {Gaussian} distribution},
	doi = {10.1016/0047-259x(90)90050-r},
	author = {Chikuse, Yasuko},
	journal = {Journal of Multivariate Analysis},
	year = {1990},
	litmapsId = {229908155}
}

@article{fundamental_mardia_2014,
	title = {Some fundamental properties of a multivariate von {Mises} distribution},
	doi = {10.1080/03610926.2012.670353},
	author = {Mardia, Kanti V. and Voss, Jochen},
	journal = {Communications in Statistics---Theory and Methods},
	year = {2014},
	litmapsId = {193495365}
}

@article{complex_ito_2016,
	title = {Complex angular central {Gaussian} mixture model for directional statistics in mask-based microphone array signal processing},
	doi = {10.1109/eusipco.2016.7760429},
	author = {Ito, N. and Araki, S. and Nakatani, T.},
	journal = {European Signal Processing Conference},
	year = {2016},
	litmapsId = {72937503}
}

@article{statistics_mardia_1972,
	title = {Statistics of Directional Data},
	doi = {10.2307/2529189},
	author = {Mardia, K.},
	year = {1972},
	litmapsId = {276809417},
    journal = {Biometrics}
}

@article{families_mardia_2020,
	title = {Families of discrete circular distributions with some novel applications},
	author = {Mardia, Kanti V. and Sriram, Karthik},
	journal = {arXiv},
	year = {2020},
	litmapsId = {207727116}
}

@article{protein_mardia_2007,
	title = {Protein bioinformatics and mixtures of bivariate von {Mises} distributions for angular data},
	doi = {10.1111/j.1541-0420.2006.00682.x},
	author = {Mardia, K. and Taylor, C. and Subramaniam, Ganesh K.},
	journal = {Biometrics},
	year = {2007},
	litmapsId = {55510111}
}

@article{simulation_hoff_2009,
	title = {Simulation of the matrix {Bingham}–von {Mises}–{Fisher} distribution, with applications to multivariate and relational data},
	doi = {10.1198/jcgs.2009.07177},
	author = {Hoff, Peter D.},
	journal = {Journal of Computational and Graphical Statistics},
	year = {2009},
	litmapsId = {262629540}
}

@article{angular_olsen_2023,
	title = {Angular central {Gaussian} and {Watson} mixture models for assessing dynamic functional brain connectivity during a motor task},
	doi = {10.1109/icasspw59220.2023.10193021},
	author = {Olsen, A. S. and Ortvald, Emil and Madsen, K.H. and Schmidt, Mikkel N. and Mørup, Morten},
	journal = {2023 IEEE International Conference on Acoustics, Speech, and Signal Processing Workshops (ICASSPW)},
	year = {2023},
	litmapsId = {267271268}
}

@article{blind_vu_2010,
	title = {Blind speech separation employing directional statistics in an expectation maximization framework},
	doi = {10.1109/icassp.2010.5495994},
	author = {Vu, Dang Hai Tran and Häb-Umbach, Reinhold},
	journal = {2010 IEEE International Conference on Acoustics, Speech and Signal Processing},
	year = {2010},
	litmapsId = {170378863}
}

@article{modeling_tabelow_2012,
	title = {Modeling the orientation distribution function by mixtures of angular central {Gaussian} distributions},
	doi = {10.1016/j.jneumeth.2011.09.001},
	author = {Tabelow, Karsten and Voss, Henning U. and Polzehl, Jörg},
	journal = {Journal of Neuroscience Methods},
	year = {2012},
	litmapsId = {98209051}
}

@article{modeling_ito_2016,
	title = {Modeling audio directional statistics using a complex {Bingham} mixture model for blind source extraction from diffuse noise},
	doi = {10.1109/icassp.2016.7471718},
	author = {Ito, N. and Araki, S. and Nakatani, T.},
	journal = {IEEE International Conference on Acoustics, Speech, and Signal Processing},
	year = {2016},
	litmapsId = {33927706}
}

@article{unsupervised_hasnat_2014,
	title = {Unsupervised clustering of depth images using {Watson} mixture model},
	doi = {10.1109/icpr.2014.46},
	author = {Hasnat, Md. Abul and Alata, Olivier and Trémeau, Alain},
	year = {2014},
	litmapsId = {82321383},
    journal = {2014 22nd International Conference on Pattern Recognition}
}

@article{singular_uhlig_1994,
	title = {On singular {Wishart} and singular multivariate beta distributions},
	doi = {10.1214/aos/1176325375},
	author = {Uhlig, H.},
	year = {1994},
	litmapsId = {96422250},
    journal = {The Annals of Statistics}
}

@article{directional_bouzid_2022,
	title = {Directional statistics-based deep metric learning for pedestrian tracking and re-identification},
	doi = {10.3390/drones6110328},
	author = {Bouzid, Abdelmalek and Sierra-Sosa, Daniel and Elmaghraby, Adel Said},
	journal = {Drones},
	year = {2022},
	litmapsId = {251344373}
}

@article{score_mardia_2016,
	title = {Score matching estimators for directional distributions},
	author = {Mardia, Kanti V. and Kent, John T. and Laha, Arnab Kumar},
	journal = {arXiv},
	year = {2016},
	litmapsId = {228356704}
}

@article{singular_srivastava_2003,
	title = {Singular {Wishart} and multivariate beta distributions},
	doi = {10.1214/aos/1065705118},
	author = {Srivastava, M.},
	year = {2003},
	litmapsId = {125671048},
    journal = {The Annals of Statistics}
}

@article{multivariate_sra_2013,
	title = {The multivariate {Watson} distribution: Maximum-likelihood estimation and other aspects},
	doi = {10.1016/j.jmva.2012.08.010},
	author = {Sra, Suvrit and Karp, Dmitrii},
	journal = {Journal of Multivariate Analysis},
	year = {2013},
	litmapsId = {178816128}
}

@article{gaussian_garcafernndez_2019,
	title = {Gaussian target tracking with direction-of-arrival von {Mises}–{Fisher} measurements},
	doi = {10.1109/tsp.2019.2911258},
	author = {Garcia-Fernandez, A. F. and Tronarp, Filip and Sarkka, Simo},
	journal = {IEEE Transactions on Signal Processing},
	year = {2019},
	litmapsId = {140362019}
}

@article{phase_cadieu_2010,
	title = {Phase coupling estimation from multivariate phase statistics},
	doi = {10.1162/neco_a_00048},
	author = {Cadieu, Charles F. and Koepsell, Kilian},
	journal = {Neural Computation},
	year = {2010},
	litmapsId = {221927232}
}

@article{torus_klein_2019,
	title = {Torus graphs for multivariate phase coupling analysis},
	doi = {10.1214/19-aoas1300},
	author = {Klein, Natalie and Orellana, J. and Brincat, S. and Miller, E. and Kass, R.},
	journal = {Annals of Applied Statistics},
	year = {2019},
	litmapsId = {267545310}
}

@article{data_kent_1997,
	title = {Data analysis for shapes and images},
	doi = {10.1016/s0378-3758(96)00043-2},
	author = {Kent, John T.},
	journal = {Journal of Statistical Planning and Inference},
	year = {1997},
	litmapsId = {33531889}
}

@article{applied_ley_2019,
	title = {Applied Directional Statistics: Modern Methods and Case Studies},
	doi = {10.1201/9781315228570},
	author = {Ley, Christophe and Verdebout, Thomas},
	journal = {Chapman and Hall/CRC eBooks},
	year = {2019},
	litmapsId = {70130323}
}

@article{statistical_dryden_2016,
	title = {Statistical Shape Analysis, with Applications in {R}},
	doi = {10.1002/9781119072492},
	author = {Dryden, Ian L. and Mardia, Kanti V.},
	journal = {Wiley Series in Probability and Statistics},
	year = {2016},
	litmapsId = {190957659}
}

@article{exact_kume_2018,
	title = {On the exact maximum likelihood inference of {Fisher}–{Bingham} distributions using an adjusted holonomic gradient method},
	doi = {10.1007/s11222-017-9765-3},
	author = {Kume, Alfred and Sei, Tomonari},
	journal = {Statistics and Computing},
	year = {2018},
	litmapsId = {198463883}
}

@article{complex_kent_1994,
	title = {The complex {Bingham} distribution and shape analysis},
	doi = {10.1111/j.2517-6161.1994.tb01978.x},
	author = {Kent, J.},
	year = {1994},
	litmapsId = {1348763},
    journal = {Journal of the Royal Statistical Society. Series B}
}

@article{antipodally_bingham_1974,
	title = {An antipodally symmetric distribution on the sphere},
	doi = {10.1214/aos/1176342874},
	author = {Bingham, Christopher},
	journal = {Annals of Statistics},
	year = {1974},
	litmapsId = {229538198}
}

@article{grassmann_gruber_2006,
	title = {Grassmann clustering},
	author = {Gruber, Peter and Theis, Fabian J.},
	journal = {European Signal Processing Conference},
	year = {2006},
	litmapsId = {72452985}
}

@article{ber_kummer_1836,
	title = {{\"U}ber die hypergeometrische {Reihe} $1+\frac{\alpha\cdot\beta}{1\cdot\gamma}x+\frac{\alpha(\alpha+1)\beta(\beta+1)}{1\cdot2\cdot\gamma(\gamma+1)}x^2+\dots$},
	doi = {10.1515/crll.1836.15.39},
	author = {Kummer, E.E.},
	journal = {Journal f{\"u}r die reine und angewandte Mathematik},
	year = {1836},
	litmapsId = {47265411}
}

@article{mixture_bijral_2007,
	title = {Mixture of {Watson} distributions: A generative model for hyperspherical embeddings},
	author = {Bijral, Avleen S. and Breitenbach, Markus and Grudic, Greg},
	journal = {Journal of Machine Learning Research},
	year = {2007},
	litmapsId = {1516126}
}

@article{von_khatri_1977,
	title = {The von {Mises}-{Fisher} matrix distribution in orientation statistics},
	author = {Khatri, C. G. and Mardia, K. V.},
	journal = {Journal of the Royal Statistical Society. Series B (Methodological)},
	year = {1977},
	litmapsId = {1516123}
}

@article{complex_mardia_1999,
	title = {The complex {Watson} distribution and shape analysis},
	author = {Mardia, K. V. and Dryden, I. L.},
	journal = {Journal of the Royal Statistical Society. Series B (Statistical Methodology)},
	year = {1999},
	litmapsId = {1516134}
}

@book{statistics_chikuse_2003,
	title = {Statistics on Special Manifolds},
	author = {Chikuse, Yasuko},
	year = {2003},
	litmapsId = {1516127},
    publisher = {Springer Nature}
}

@article{equatorial_watson_1965,
	title = {Equatorial distributions on a sphere},
	doi = {10.2307/2333824},
	author = {Watson, G. S.},
	journal = {Biometrika},
	year = {1965},
	litmapsId = {1516149}
}

@article{clustering_banerjee_2005,
	title = {Clustering on the unit hypersphere using von {Mises}-{Fisher} distributions},
	doi = {10.5555/1046920.1088718},
	author = {Banerjee, Arindam and Dhillon, Inderjit S. and Ghosh, Joydeep and Sra, Suvrit},
	journal = {Journal of Machine Learning Research},
	year = {2005},
	litmapsId = {156436195}
}

@article{statistical_tyler_1987,
	title = {Statistical analysis for the angular central {Gaussian} distribution on the sphere},
	doi = {10.1093/biomet/74.3.579},
	author = {Tyler, David E.},
	year = {1987},
	litmapsId = {9134234},
    journal = {Biometrika}
}

@article{estimation_nodehi_2020,
	title = {Estimation of parameters in multivariate wrapped models for data on a p-torus},
	doi = {10.1007/s00180-020-01006-x},
	author = {Nodehi, A. and Golalizadeh, M. and Maadooliat, Mehdi and Agostinelli, C.},
	journal = {Computational Statistics},
	year = {2020},
	litmapsId = {279352443}
}

@article{multivariate_mardia_2008,
	title = {A multivariate von {Mises} distribution with applications to bioinformatics},
	doi = {10.1002/cjs.5550360110},
	author = {Mardia, K. and Hughes, Gareth and Taylor, C. and Singh, Harshinder},
	year = {2008},
	litmapsId = {222670155},
    journal = {The Canadian Journal of Statistics}
}

@article{saddlepoint_kume_2013,
	title = {Saddlepoint approximations for the normalizing constant of {Fisher}--{Bingham} distributions on products of spheres and {Stiefel} manifolds},
	doi = {10.1093/biomet/ast021},
	author = {Kume, A. and Preston, S. and Wood, A.},
	year = {2013},
	litmapsId = {33785368},
    journal = {Biometrika}
}

@article{fisherbingham_kent_1982,
	title = {The {Fisher}-{Bingham} distribution on the sphere},
	doi = {10.1111/j.2517-6161.1982.tb01189.x},
	author = {Kent, J.},
	year = {1982},
	litmapsId = {261141553},
    journal = {Journal of the Royal Statistical Society. Series B}
}

@article{interpretable_kato_2025,
	title = {An interpretable family of projected normal distributions and a related copula model for {Bayesian} analysis of hypertoroidal data},
	author = {Kato, Shogo and Mastrantonio, Gianluca and Ishikawa, Masayuki},
	year = {2025},
	litmapsId = {290731222},
    journal = {arXiv}
}

@article{parametric_ali_2019,
  title   = {Parametric classification of {Bingham} distributions based on {Grassmann} manifolds},
  author  = {Ali, Muhammad and Gao, Junbin and Antolovich, Michael},
  journal = {IEEE Transactions on Image Processing},
  volume  = {28},
  number  = {12},
  pages   = {5771--5784},
  year    = {2019},
  doi     = {10.1109/TIP.2019.2922100}
}

@article{saddlepoint_kume_2005,
	title = {Saddlepoint approximations for the {Bingham} and {Fisher}–{Bingham} normalising constants},
	doi = {10.1093/biomet/92.2.465},
	author = {Kume, A. and Wood, A.},
	year = {2005},
	litmapsId = {91439868},
    journal = {Biometrika}
}

@article{directions_mardia_2005,
	title = {Directions and projective shapes},
	doi = {10.1214/009053605000000273},
	author = {Mardia, Kanti V. and Patrangenaru, Vic},
	journal = {arXiv},
	year = {2005},
	litmapsId = {41565643}
}

@article{elliptically_paine_2018,
	title = {An elliptically symmetric angular {Gaussian} distribution},
	doi = {10.1007/s11222-017-9756-4},
	author = {Paine, Phillip J. and Preston, Simon and Tsagris, Michail and Wood, Andrew T. A.},
	journal = {Statistics and Computing},
	year = {2018},
	litmapsId = {76795841}
}

@article{unified_kent_2018,
	title = {A new unified approach for the simulation of a wide class of directional distributions},
	doi = {10.1080/10618600.2017.1390468},
	author = {Kent, John T. and Ganeiber, Asaad M. and Mardia, Kanti V.},
	journal = {Journal of Computational and Graphical Statistics},
	year = {2018},
	litmapsId = {144673350}
}

@article{torus_eltzner_2018,
	title = {Torus principal component analysis with applications to {RNA} structure},
	doi = {10.1214/17-aoas1115},
	author = {Eltzner, B. and Huckemann, S. and Mardia, K.},
	journal = {Annals of Applied Statistics},
	year = {2018},
	litmapsId = {126670897}
}

@article{psilocybin_olsen_2022,
	title = {Psilocybin modulation of time-varying functional connectivity is associated with plasma psilocin and subjective effects},
	doi = {10.1016/j.neuroimage.2022.119716},
	author = {Olsen, Anders S. and Lykkebo-Velløe, Anders and Ozenne, Brice and Madsen, Martin K. and Stenbæk, Dea S. and Armand, Sophia and Mørup, Morten and Ganz, Melanie and Knudsen, Gitte Moos and Fisher, Patrick M.},
	journal = {NeuroImage},
	year = {2022},
	litmapsId = {250925116}
}

@article{statistical_turaga_2011,
	title = {Statistical computations on {Grassmann} and {Stiefel} manifolds for image and video-based recognition},
	doi = {10.1109/tpami.2011.52},
	author = {Turaga, P. and Veeraraghavan, A. and Srivastava, Anuj and Chellappa, R.},
	journal = {IEEE Transactions on Pattern Analysis and Machine Intelligence},
	year = {2011},
	litmapsId = {5600408}
}

@article{building_huang_2018,
	title = {Building deep networks on {Grassmann} manifolds},
	author = {Huang, Zhiwu and Wu, Jiqing and Gool, Luc Van},
	journal = {arXiv},
	year = {2018},
	litmapsId = {261355455}
}

@article{simulation_wood_1994,
	title = {Simulation of the von {Mises}--{Fisher} distribution},
	doi = {10.1080/03610919408813161},
	author = {Wood, A.},
	year = {1994},
	litmapsId = {79058479},
    journal = {Communications in Statistics---Simulation and Computation}
}

@article{source_drude_2014,
	title = {Source counting in speech mixtures using a variational {EM} approach for complex {Watson} mixture models},
	doi = {10.1109/icassp.2014.6854924},
	author = {Drude, Lukas and Chinaev, Aleksej and Vu, Dang Hai Tran and Häb-Umbach, Reinhold},
	journal = {2014 IEEE International Conference on Acoustics, Speech and Signal Processing (ICASSP)},
	year = {2014},
	litmapsId = {229935394}
}

@article{path_dodik_2021,
	title = {Path guiding using spatio-directional mixture models},
	doi = {10.1111/cgf.14428},
	author = {Dodik, Ana and Papas, Marios and Öztireli, Cengiz and Müller, Thomas},
	journal = {Computer Graphics Forum},
	year = {2021},
	litmapsId = {254168811}
}

@article{nonisotropic_kirchhof_2022,
	title = {A non-isotropic probabilistic take on proxy-based deep metric learning},
	doi = {10.48550/arxiv.2207.03784},
	author = {Kirchhof, M. and Roth, Karsten and Akata, Zeynep and Kasneci, Enkelejda},
	journal = {European Conference on Computer Vision},
	year = {2022},
	litmapsId = {234556809}
}

@article{multichannel_traa_2014,
	title = {Multichannel source separation and tracking with {RANSAC} and directional statistics},
	doi = {10.1109/taslp.2014.2365701},
	author = {Traa, Johannes and Smaragdis, Paris},
	journal = {IEEE/ACM Transactions on Audio, Speech, and Language Processing},
	year = {2014},
	litmapsId = {180270971}
}

@article{multiple_traa_2014,
	title = {Multiple speaker tracking with the factorial von {Mises}-{Fisher} filter},
	doi = {10.1109/mlsp.2014.6958891},
	author = {Traa, Johannes and Smaragdis, P.},
	journal = {International Workshop on Machine Learning for Signal Processing},
	year = {2014},
	litmapsId = {137681419}
}

@article{mardia_statistics_1975,
    title = {Statistics of directional data},
    volume = {37},
    issn = {0035-9246},
    number = {3},
    urldate = {2026-01-19},
    journal = {Journal of the Royal Statistical Society. Series B (Methodological)},
    publisher = {[Royal Statistical Society, Oxford University Press]},
    author = {Mardia, K. V.},
    year = {1975},
    pages = {349--393},
}

@book{mardiaDirectionalStatistics1999,
    title = {Directional Statistics},
    isbn = {978-0-470-31697-9},
    doi = {10.1002/9780470316979},
    urldate = {2021-07-27},
    publisher = {John Wiley and Sons Ltd.},
    author = {Mardia, Kanti V. and Jupp, Peter E.},
    month = may,
    year = {1999},
}

@article{cabralCognitivePerformanceHealthy2017,
    title = {Cognitive performance in healthy older adults relates to spontaneous switching between states of functional connectivity during rest},
    volume = {7},
    issn = {20452322},
    doi = {10.1038/s41598-017-05425-7},
    number = {1},
    journal = {Scientific Reports},
    author = {Cabral, Joana and Vidaurre, Diego and Marques, Paulo and Magalhães, Ricardo and Silva Moreira, Pedro and Miguel Soares, José and Deco, Gustavo and Sousa, Nuno and Kringelbach, Morten L.},
    year = {2017},
    pages = {1--13},
}

@article{wishart_generalised_1928,
    title = {The generalised product moment distribution in samples from a normal multivariate population},
    volume = {20A},
    issn = {0006-3444},
    doi = {10.1093/biomet/20A.1-2.32},
    number = {1-2},
    urldate = {2026-07-28},
    journal = {Biometrika},
    author = {Wishart, John},
    month = dec,
    year = {1928},
    pages = {32--52},
}

@article{glerean_functional_2012,
    title = {Functional magnetic resonance imaging phase synchronization as a measure of dynamic functional connectivity},
    volume = {2},
    issn = {2158-0014},
    doi = {10.1089/brain.2011.0068},
    number = {2},
    urldate = {2024-05-29},
    journal = {Brain Connectivity},
    author = {Glerean, Enrico and Salmi, Juha and Lahnakoski, Juha M. and Jääskeläinen, Iiro P. and Sams, Mikko},
    month = apr,
    year = {2012},
    pages = {91--101},
}

@book{abramowitz_handbook_1965,
    address = {New York, NY},
    edition = {9. Dover print.; [Nachdr. der Ausg. von 1972]},
    series = {Dover Books on Mathematics},
    title = {Handbook of Mathematical Functions: With Formulas, Graphs, and Mathematical Tables},
    isbn = {978-0-486-61272-0},
    shorttitle = {Handbook of mathematical functions},
    language = {eng},
    publisher = {Dover Publ},
    editor = {Abramowitz, Milton and Stegun, Irene A.},
    year = {1965},
}

@inproceedings{gutmann_noise-contrastive_2010,
    title = {Noise-contrastive estimation: A new estimation principle for unnormalized statistical models},
    issn = {1938-7228},
    shorttitle = {Noise-contrastive estimation},
    language = {en},
    urldate = {2025-02-18},
    booktitle = {Proceedings of the {Thirteenth} {International} {Conference} on {Artificial} {Intelligence} and {Statistics}},
    publisher = {JMLR Workshop and Conference Proceedings},
    author = {Gutmann, Michael and Hyvärinen, Aapo},
    month = mar,
    year = {2010},
    pages = {297--304},
}

@article{matsuda_estimation_2020,
    title = {Estimation of non-normalized mixture models},
    volume = {89},
    journal = {AISTATS 2019 - 22nd International Conference on Artificial Intelligence and Statistics},
    author = {Matsuda, Takeru and Hyvärinen, Aapo},
    year = {2020},
}

@article{downs_orientation_1972,
    title = {Orientation statistics},
    volume = {59},
    issn = {0006-3444},
    doi = {10.1093/biomet/59.3.665},
    number = {3},
    urldate = {2026-07-28},
    journal = {Biometrika},
    author = {Downs, Thomas D.},
    month = dec,
    year = {1972},
    pages = {665--676},
}

@article{kingma_adam_2014,
    title = {Adam: A method for stochastic optimization},
    doi = {10.48550/arxiv.1412.6980},
    urldate = {2023-03-02},
    journal = {3rd International Conference on Learning Representations, ICLR 2015 - Conference Track Proceedings},
    publisher = {International Conference on Learning Representations, ICLR},
    author = {Kingma, Diederik P. and Ba, Jimmy Lei},
    month = dec,
    year = {2014},
    note = {arXiv: 1412.6980},
}

@article{hyvarinen_estimation_2005,
    title = {Estimation of non-normalized statistical models by score matching},
    volume = {6},
    issn = {1533-7928},
    number = {24},
    urldate = {2026-07-28},
    journal = {Journal of Machine Learning Research},
    author = {Hyvärinen, Aapo},
    year = {2005},
    pages = {695--709},
}

@article{schaefer_local-global_2018,
    title = {Local-global parcellation of the human cerebral cortex from intrinsic functional connectivity {MRI}},
    volume = {28},
    issn = {1047-3211},
    doi = {10.1093/cercor/bhx179},
    number = {9},
    urldate = {2021-09-08},
    journal = {Cerebral Cortex},
    publisher = {Cereb Cortex},
    author = {Schaefer, Alexander and Kong, Ru and Gordon, Evan M and Laumann, Timothy O and Zuo, Xi-Nian and Holmes, Avram J and Eickhoff, Simon B and Yeo, B T Thomas},
    month = sep,
    year = {2018},
    pages = {3095--3114},
}

@inproceedings{arthur_k-means_2007,
    title = {K-means++: The advantages of careful seeding},
    volume = {07-09-Janu},
    isbn = {978-0-89871-624-5},
    urldate = {2021-07-23},
    booktitle = {Proceedings of the {Annual} {ACM}-{SIAM} {Symposium} on {Discrete} {Algorithms}},
    author = {Arthur, David and Vassilvitskii, Sergei},
    year = {2007},
    pages = {1027--1035},
}

@inproceedings{cetingul_intrinsic_2009,
    title = {Intrinsic mean shift for clustering on {Stiefel} and {Grassmann} manifolds},
    issn = {1063-6919},
    doi = {10.1109/CVPR.2009.5206806},
    urldate = {2026-08-31},
    booktitle = {2009 {IEEE} {Conference} on {Computer} {Vision} and {Pattern} {Recognition}},
    author = {Cetingul, Hasan Ertan and Vidal, Rene},
    month = jun,
    year = {2009},
    pages = {1896--1902},
}

@article{nielsen_modeling_2017,
    title = {Modeling dynamic functional connectivity using a {Wishart} mixture model},
    doi = {10.1109/PRNI.2017.7981505},
    journal = {2017 International Workshop on Pattern Recognition in Neuroimaging, PRNI 2017},
    author = {Nielsen, Soren F.V. and Madsen, Kristoffer H. and Schmidt, Mikkel N. and Morup, Morten},
    year = {2017},
}

@article{olsen_uncovering_2026,
    title = {Uncovering dynamic human brain phase coherence networks},
    volume = {123},
    doi = {10.1073/pnas.2518287123},
    number = {35},
    urldate = {2026-08-27},
    journal = {Proceedings of the National Academy of Sciences},
    publisher = {Proceedings of the National Academy of Sciences},
    author = {Olsen, Anders S. and Brammer, Anders and Fisher, Patrick M. and Mørup, Morten},
    month = sep,
    year = {2026},
    pages = {e2518287123},
}

@article{tianTopographicOrganizationHuman2020,
    title = {Topographic organization of the human subcortex unveiled with functional connectivity gradients},
    volume = {23},
    copyright = {2020 The Author(s), under exclusive licence to Springer Nature America, Inc.},
    issn = {1546-1726},
    doi = {10.1038/s41593-020-00711-6},
    language = {en},
    number = {11},
    urldate = {2024-05-15},
    journal = {Nature Neuroscience},
    publisher = {Nature Publishing Group},
    author = {Tian, Ye and Margulies, Daniel S. and Breakspear, Michael and Zalesky, Andrew},
    month = nov,
    year = {2020},
    pages = {1421--1432},
}

@article{habeck_geodesic_2025,
    title = {Geodesic slice sampling on the sphere},
    volume = {26},
    issn = {1533-7928},
    number = {297},
    urldate = {2026-08-31},
    journal = {Journal of Machine Learning Research},
    author = {Habeck, Michael and Hasenpflug, Mareike and Kodgirwar, Shantanu and Rudolf, Daniel},
    year = {2025},
    pages = {1--38},
}

@article{barbarossa_topological_2020,
    title = {Topological signal processing over simplicial complexes},
    volume = {68},
    issn = {1053-587X, 1941-0476},
    doi = {10.1109/TSP.2020.2981920},
    urldate = {2026-09-03},
    journal = {IEEE Transactions on Signal Processing},
    author = {Barbarossa, Sergio and Sardellitti, Stefania},
    year = {2020},
    note = {arXiv:1907.11577 [eess.SP]},
    pages = {2992--3007},
}

@misc{schaub_signal_2022,
    title = {Signal processing on simplicial complexes},
    doi = {10.48550/arXiv.2106.07471},
    urldate = {2026-09-03},
    publisher = {arXiv},
    author = {Schaub, Michael T. and Seby, Jean-Baptiste and Frantzen, Florian and Roddenberry, T. Mitchell and Zhu, Yu and Segarra, Santiago},
    month = feb,
    year = {2022},
    note = {arXiv:2106.07471 [eess.SP]},
}

@misc{joris_minilibrimix_2020,
    title = {{MiniLibriMix} dataset},
    url = {https://zenodo.org/records/3871592},
    doi = {10.5281/zenodo.3871592},
    language = {eng},
    urldate = {2026-09-07},
    publisher = {Zenodo},
    author = {Joris, Cosentino and Manuel, Pariente},
    month = jun,
    year = {2020},
}

@misc{muller_-car_2024,
    title = {{In-Car McVAMPIRE} -- {In-Car Multichannel Varying Mouth Position Impulse Response} dataset},
    url = {https://zenodo.org/records/12806684},
    doi = {10.5281/zenodo.12806684},
    urldate = {2026-09-07},
    publisher = {Zenodo},
    author = {Müller, Kaspar and Wolff, Tobias},
    month = aug,
    year = {2024},
}

@inproceedings{zemel_directional-unit_1992,
    title = {Directional-unit {Boltzmann} machines},
    volume = {5},
    urldate = {2026-08-31},
    booktitle = {Advances in {Neural} {Information} {Processing} {Systems}},
    publisher = {Morgan-Kaufmann},
    author = {Zemel, Richard and Williams, Christopher and Mozer, Michael C},
    year = {1992},
}

@article{canolty_detecting_2012,
    title = {Detecting event-related changes of multivariate phase coupling in dynamic brain networks},
    volume = {107},
    issn = {1522-1598},
    doi = {10.1152/jn.00610.2011},
    language = {eng},
    number = {7},
    journal = {Journal of Neurophysiology},
    author = {Canolty, Ryan T. and Cadieu, Charles F. and Koepsell, Kilian and Ganguly, Karunesh and Knight, Robert T. and Carmena, Jose M.},
    month = apr,
    year = {2012},
    pages = {2020--2031},
}

@article{kato_trivariate_2026,
    title = {A trivariate wrapped {Cauchy} copula},
    volume = {216},
    issn = {0047-259X},
    doi = {10.1016/j.jmva.2026.105676},
    urldate = {2026-09-11},
    journal = {Journal of Multivariate Analysis},
    author = {Kato, Shogo and Ley, Christophe and Loizidou, Sophia and Mardia, Kanti V.},
    month = nov,
    year = {2026},
    pages = {105676},
}

@article{smidl_bayesian_2007,
    title = {On {Bayesian} principal component analysis},
    volume = {51},
    issn = {0167-9473},
    doi = {10.1016/j.csda.2007.01.011},
    number = {9},
    urldate = {2026-09-11},
    journal = {Computational Statistics \& Data Analysis},
    author = {Šmídl, Václav and Quinn, Anthony},
    month = may,
    year = {2007},
    pages = {4101--4123},
}

@article{jorgensen_probabilistic_2024,
    title = {Probabilistic {PARAFAC2}},
    volume = {26},
    copyright = {http://creativecommons.org/licenses/by/3.0/},
    issn = {1099-4300},
    doi = {10.3390/e26080697},
    language = {en},
    number = {8},
    urldate = {2026-09-11},
    journal = {Entropy},
    publisher = {Multidisciplinary Digital Publishing Institute},
    author = {Jørgensen, Philip J. H. and Nielsen, Søren F. and Hinrich, Jesper L. and Schmidt, Mikkel N. and Madsen, Kristoffer H. and Mørup, Morten},
    month = aug,
    year = {2024},
    pages = {697},
}

@misc{mardia_projected_2026,
    title = {The projected isotropic normal distribution with applications in neuroscience},
    doi = {10.48550/arXiv.2603.03816},
    urldate = {2026-09-11},
    publisher = {arXiv},
    author = {Mardia, Kanti V. and Sa', Antonio Mauricio F. L. Miranda de},
    month = mar,
    year = {2026},
    note = {arXiv:2603.03816 [stat.ME]},
}

@article{jung_clustering_2021,
    title = {Clustering on the torus by conformal prediction},
    volume = {15},
    issn = {1932-6157, 1941-7330},
    doi = {10.1214/21-AOAS1459},
    language = {en},
    number = {4},
    urldate = {2026-08-31},
    journal = {The Annals of Applied Statistics},
    publisher = {Institute of Mathematical Statistics},
    author = {Jung, Sungkyu and Park, Kiho and Kim, Byungwon},
    month = dec,
    year = {2021},
    pages = {1583--1603},
}

@article{hinrichArchetypalAnalysisModeling2016,
    title = {Archetypal analysis for modeling multisubject {fMRI} data},
    volume = {10},
    issn = {19324553},
    doi = {10.1109/JSTSP.2016.2595103},
    number = {7},
    urldate = {2022-01-24},
    journal = {IEEE Journal on Selected Topics in Signal Processing},
    publisher = {Institute of Electrical and Electronics Engineers Inc.},
    author = {Hinrich, Jesper Love and Bardenfleth, Sophia Elizabeth and Roge, Rasmus Erbou and Churchill, Nathan William and Madsen, Kristoffer Hougaard and Morup, Morten},
    month = oct,
    year = {2016},
    pages = {1160--1171},
}

@article{sei_properties_2013,
    title = {Properties and applications of {Fisher} distribution on the rotation group},
    volume = {116},
    issn = {0047-259X},
    doi = {10.1016/j.jmva.2013.01.010},
    urldate = {2026-10-02},
    journal = {Journal of Multivariate Analysis},
    author = {Sei, Tomonari and Shibata, Hiroki and Takemura, Akimichi and Ohara, Katsuyoshi and Takayama, Nobuki},
    month = apr,
    year = {2013},
    pages = {440--455},
}

@article{jung_analysis_2012,
    title = {Analysis of principal nested spheres},
    volume = {99},
    issn = {0006-3444},
    doi = {10.1093/biomet/ass022},
    number = {3},
    urldate = {2026-10-02},
    journal = {Biometrika},
    author = {Jung, Sungkyu and Dryden, Ian L. and Marron, J. S.},
    month = sep,
    year = {2012},
    pages = {551--568},
}

@misc{ogata_pair_2023,
    title = {Pair circulas modelling for multivariate circular time series},
    doi = {10.48550/arXiv.2311.13131},
    urldate = {2026-10-02},
    publisher = {arXiv},
    author = {Ogata, Hiroaki},
    month = nov,
    year = {2023},
    note = {arXiv:2311.13131 [stat.ME]},
}

@article{tipping_probabilistic_1999,
    title = {Probabilistic principal component analysis},
    volume = {61},
    issn = {1369-7412},
    doi = {10.1111/1467-9868.00196},
    number = {3},
    urldate = {2026-10-03},
    journal = {Journal of the Royal Statistical Society Series B: Statistical Methodology},
    author = {Tipping, Michael E. and Bishop, Christopher M.},
    month = sep,
    year = {1999},
    pages = {611--622},
}

@article{wang_directional_2013,
    title = {Directional data analysis under the general projected normal distribution},
    volume = {10},
    issn = {1572-3127},
    doi = {10.1016/j.stamet.2012.07.005},
    number = {1},
    urldate = {2026-10-03},
    journal = {Statistical Methodology},
    author = {Wang, Fangpo and Gelfand, Alan E.},
    month = jul,
    year = {2013},
    pages = {113--127},
}

@inproceedings{dubey_speaker_2018,
  author    = {Dubey, Harishchandra and Sangwan, Abhijeet and Hansen, John H. L.},
  title     = {Robust speaker clustering using mixtures of von {Mises}--{Fisher} distributions for naturalistic audio streams},
  booktitle = {Proceedings of Interspeech},
  pages     = {3603--3607},
  year      = {2018},
  doi       = {10.21437/Interspeech.2018-50}
}

@inproceedings{nurminen_aoa_2018,
  author    = {Nurminen, Henri and Suomalainen, Laura and Ali-L{\"o}ytty, Simo and Pich{\'e}, Robert},
  title     = {{3D} angle-of-arrival positioning using von {Mises}--{Fisher} distribution},
  booktitle = {2018 21st International Conference on Information Fusion (FUSION)},
  pages     = {2036--2041},
  year      = {2018},
  doi       = {10.23919/ICIF.2018.8455205}
}

@article{gilitschenski_unscented_2016,
  author  = {Gilitschenski, Igor and Kurz, Gerhard and Julier, Simon J. and Hanebeck, Uwe D.},
  title   = {Unscented orientation estimation based on the {Bingham} distribution},
  journal = {IEEE Transactions on Automatic Control},
  volume  = {61},
  number  = {1},
  pages   = {172--177},
  year    = {2016},
  doi     = {10.1109/TAC.2015.2423831}
}

@article{lee_bayesian_2018,
  author  = {Lee, Taeyoung},
  title   = {Bayesian attitude estimation with the matrix {Fisher} distribution on {SO}(3)},
  journal = {IEEE Transactions on Automatic Control},
  volume  = {63},
  number  = {10},
  pages   = {3377--3392},
  year    = {2018},
  doi     = {10.1109/TAC.2018.2797162}
}

@inproceedings{prokudin_deep_2018,
  author    = {Prokudin, Sergey and Gehler, Peter V. and Nowozin, Sebastian},
  title     = {Deep directional statistics: Pose estimation with uncertainty quantification},
  booktitle = {Proceedings of the European Conference on Computer Vision (ECCV)},
  pages     = {534--551},
  year      = {2018}
}

@article{deng_deepbingham_2022,
  author  = {Deng, Haowen and Bui, Mai and Navab, Nassir and Guibas, Leonidas and Ilic, Slobodan and Birdal, Tolga},
  title   = {Deep {Bingham} networks: Dealing with uncertainty and ambiguity in pose estimation},
  journal = {International Journal of Computer Vision},
  volume  = {130},
  number  = {7},
  pages   = {1627--1654},
  year    = {2022},
  doi     = {10.1007/s11263-022-01612-w}
}

@inproceedings{yin_fishermatch_2022,
  author    = {Yin, Yingda and Cai, Yingcheng and Wang, He and Chen, Baoquan},
  title     = {{FisherMatch}: Semi-supervised rotation regression via entropy-based filtering},
  booktitle = {Proceedings of the IEEE/CVF Conference on Computer Vision and Pattern Recognition (CVPR)},
  pages     = {11164--11173},
  year      = {2022}
}

@inproceedings{straub_hypersphere_2015,
  author    = {Straub, Julian and Campbell, Trevor and How, Jonathan P. and Fisher, John W. III},
  title     = {Small-variance nonparametric clustering on the hypersphere},
  booktitle = {Proceedings of the IEEE Conference on Computer Vision and Pattern Recognition (CVPR)},
  year      = {2015},
  doi       = {10.1109/CVPR.2015.7298630}
}

@inproceedings{bae_surface_2021,
  author    = {Bae, Gwangbin and Budvytis, Ignas and Cipolla, Roberto},
  title     = {Estimating and exploiting the aleatoric uncertainty in surface normal estimation},
  booktitle = {Proceedings of the IEEE/CVF International Conference on Computer Vision (ICCV)},
  pages     = {13137--13146},
  year      = {2021},
  doi       = {10.1109/ICCV48922.2021.01289}
}

@inproceedings{shaban_directional_2022,
  author    = {Shaban, Amirreza and Rahimi, Amir and Ajanthan, Thalaiyasingam and Boots, Byron and Hartley, Richard},
  title     = {Few-shot weakly-supervised object detection via directional statistics},
  booktitle = {Proceedings of the IEEE/CVF Winter Conference on Applications of Computer Vision (WACV)},
  pages     = {3920--3929},
  year      = {2022}
}

@article{rathi_directional_2009,
  author  = {Rathi, Yogesh and Michailovich, Oleg and Shenton, Martha E. and Bouix, Sylvain},
  title   = {Directional functions for orientation distribution estimation},
  journal = {Medical Image Analysis},
  volume  = {13},
  number  = {3},
  pages   = {432--444},
  year    = {2009},
  doi     = {10.1016/j.media.2009.01.004}
}

@article{lee_wishart_1994,
  author  = {Lee, Jong-Sen and Grunes, Michael R. and Kwok, Ron},
  title   = {Classification of multi-look polarimetric {SAR} imagery based on complex {Wishart} distribution},
  journal = {International Journal of Remote Sensing},
  volume  = {15},
  number  = {11},
  pages   = {2299--2311},
  year    = {1994},
  doi     = {10.1080/01431169408954244}
}

@article{silva_wishart_2013,
  author  = {Silva, Wagner Barreto da and Freitas, Corina da Costa and Sant'Anna, Sidnei Jo{\~a}o Siqueira and Frery, Alejandro C.},
  title   = {Classification of segments in {PolSAR} imagery by minimum stochastic distances between {Wishart} distributions},
  journal = {IEEE Journal of Selected Topics in Applied Earth Observations and Remote Sensing},
  volume  = {6},
  number  = {3},
  pages   = {1263--1273},
  year    = {2013},
  doi     = {10.1109/JSTARS.2013.2248132}
}

@inproceedings{davidson_hyperspherical_2018,
  author    = {Davidson, Tim R. and Falorsi, Luca and De Cao, Nicola and Kipf, Thomas and Tomczak, Jakub M.},
  title     = {Hyperspherical variational auto-encoders},
  booktitle = {Proceedings of the 34th Conference on Uncertainty in Artificial Intelligence (UAI)},
  pages     = {856--865},
  year      = {2018}
}

@inproceedings{kobayashi_tvmf_2021,
  author    = {Kobayashi, Takumi},
  title     = {T-{vMF} similarity for regularizing intra-class feature distribution},
  booktitle = {Proceedings of the IEEE/CVF Conference on Computer Vision and Pattern Recognition (CVPR)},
  pages     = {6616--6625},
  year      = {2021}
}

@inproceedings{li_spherical_2021,
  author    = {Li, Shen and Xu, Jianqing and Xu, Xiaqing and Shen, Pengcheng and Li, Shaoxin and Hooi, Bryan},
  title     = {Spherical confidence learning for face recognition},
  booktitle = {Proceedings of the IEEE/CVF Conference on Computer Vision and Pattern Recognition (CVPR)},
  pages     = {15629--15637},
  year      = {2021}
}

@inproceedings{scott_vmf_2021,
  author    = {Scott, Tyler R. and Gallagher, Andrew C. and Mozer, Michael C.},
  title     = {von {Mises}--{Fisher} loss: An exploration of embedding geometries for supervised learning},
  booktitle = {Proceedings of the IEEE/CVF International Conference on Computer Vision (ICCV)},
  pages     = {10612--10622},
  year      = {2021}
}

@article{schmidt_music_1986,
  author  = {Schmidt, Ralph O.},
  title   = {Multiple emitter location and signal parameter estimation},
  journal = {IEEE Transactions on Antennas and Propagation},
  volume  = {34},
  number  = {3},
  pages   = {276--280},
  year    = {1986},
  doi     = {10.1109/TAP.1986.1143830}
}

@article{roy_esprit_1989,
  author  = {Roy, Richard and Kailath, Thomas},
  title   = {{ESPRIT}---Estimation of signal parameters via rotational invariance techniques},
  journal = {IEEE Transactions on Acoustics, Speech, and Signal Processing},
  volume  = {37},
  number  = {7},
  pages   = {984--995},
  year    = {1989},
  doi     = {10.1109/29.32276}
}

@inproceedings{balzano_grouse_2010,
  author    = {Balzano, Laura and Nowak, Robert and Recht, Benjamin},
  title     = {Online identification and tracking of subspaces from highly incomplete information},
  booktitle = {2010 48th Annual Allerton Conference on Communication, Control, and Computing},
  pages     = {704--711},
  year      = {2010},
  doi       = {10.1109/ALLERTON.2010.5706976}
}

@article{love_grassmannian_2003,
  author  = {Love, David J. and Heath, Robert W. Jr. and Strohmer, Thomas},
  title   = {Grassmannian beamforming for multiple-input multiple-output wireless systems},
  journal = {IEEE Transactions on Information Theory},
  volume  = {49},
  number  = {10},
  pages   = {2735--2747},
  year    = {2003},
  doi     = {10.1109/TIT.2003.817466}
}

@article{zheng_grassmann_2002,
  author  = {Zheng, Lizhong and Tse, David N. C.},
  title   = {Communication on the {Grassmann} manifold: A geometric approach to the noncoherent multiple-antenna channel},
  journal = {IEEE Transactions on Information Theory},
  volume  = {48},
  number  = {2},
  pages   = {359--383},
  year    = {2002},
  doi     = {10.1109/18.978730}
}

@article{shayovitz_tikhonov_2016,
  author  = {Shayovitz, Shachar and Raphaeli, Dan},
  title   = {Message passing algorithms for phase noise tracking using {Tikhonov} mixtures},
  journal = {IEEE Transactions on Communications},
  volume  = {64},
  number  = {1},
  pages   = {387--401},
  year    = {2016},
  doi     = {10.1109/TCOMM.2015.2506553}
}

@inproceedings{huber_mork_vonmises_2001,
  author    = {Huber-M{\"o}rk, Reinhold and Dutra, Luciano Vieira and Freitas, Corina da Costa},
  title     = {{SAR} interferogram phase filtering based on the von {Mises} distribution},
  booktitle = {Proceedings of the 2001 IEEE International Geoscience and Remote Sensing Symposium (IGARSS)},
  volume    = {6},
  year      = {2001},
  doi       = {10.1109/IGARSS.2001.978173}
}

@article{vu_cofi_phase_2025,
  author  = {Vu, Phan Viet Hoa and Breloy, Arnaud and Brigui, Fr{\'e}d{\'e}ric and Yan, Yajing and Ginolhac, Guillaume},
  title   = {Covariance fitting interferometric phase linking: Modular framework and optimization algorithms},
  journal = {IEEE Transactions on Geoscience and Remote Sensing},
  volume  = {63},
  pages   = {1--18},
  year    = {2025},
  doi     = {10.1109/TGRS.2025.3550978}
}

@misc{sato_probabilistic_2023,
    title = {A probabilistic rotation representation for symmetric shapes with an efficiently computable {Bingham} loss function},
    doi = {10.48550/arXiv.2305.18947},
    urldate = {2026-07-29},
    publisher = {arXiv},
    author = {Sato, Hiroya and Ikeda, Takuya and Nishiwaki, Koichi},
    month = may,
    year = {2023},
    note = {arXiv:2305.18947 [cs.CV]},
}

@article{jupp_maximum_1979,
    title = {Maximum likelihood estimators for the matrix von {Mises}-{Fisher} and {Bingham} distributions},
    volume = {7},
    issn = {0090-5364, 2168-8966},
    doi = {10.1214/aos/1176344681},
    language = {en},
    number = {3},
    urldate = {2026-07-29},
    journal = {The Annals of Statistics},
    publisher = {Institute of Mathematical Statistics},
    author = {Jupp, P. E. and Mardia, K. V.},
    month = may,
    year = {1979},
    pages = {599--606},
}

@misc{falco_geometric_2015,
    title = {Geometric structures in tensor representations (final release)},
    doi = {10.48550/arXiv.1505.03027},
    urldate = {2026-07-13},
    publisher = {arXiv},
    author = {Falco, Antonio and Hackbusch, Wolfgang and Nouy, Anthony},
    month = jun,
    year = {2015},
    note = {arXiv:1505.03027 [math.NA]},
}

@article{forney_viterbi_1973,
    title = {The {Viterbi} algorithm},
    volume = {61},
    issn = {1558-2256},
    doi = {10.1109/PROC.1973.9030},
    number = {3},
    urldate = {2024-11-13},
    journal = {Proceedings of the IEEE},
    author = {Forney, G.D.},
    month = mar,
    year = {1973},
    pages = {268--278},
}

@article{vanessenHumanConnectomeProject2012,
    title = {The {Human Connectome Project}: A data acquisition perspective},
    volume = {62},
    issn = {1095-9572},
    shorttitle = {The {Human} {Connectome} {Project}},
    doi = {10.1016/j.neuroimage.2012.02.018},
    language = {eng},
    number = {4},
    journal = {NeuroImage},
    author = {Van Essen, D. C. and Ugurbil, K. and Auerbach, E. and Barch, D. and Behrens, T. E. J. and Bucholz, R. and Chang, A. and Chen, L. and Corbetta, M. and Curtiss, S. W. and Della Penna, S. and Feinberg, D. and Glasser, M. F. and Harel, N. and Heath, A. C. and Larson-Prior, L. and Marcus, D. and Michalareas, G. and Moeller, S. and Oostenveld, R. and Petersen, S. E. and Prior, F. and Schlaggar, B. L. and Smith, S. M. and Snyder, A. Z. and Xu, J. and Yacoub, E. and {WU-Minn HCP Consortium}},
    month = oct,
    year = {2012},
    pages = {2222--2231},
}

@misc{koev_efficient_2005,
  title = {The efficient evaluation of the hypergeometric function of a matrix argument},
  doi = {10.48550/arXiv.math/0505344},
  urldate = {2025-02-17},
  publisher = {arXiv},
  author = {Koev, Plamen and Edelman, Alan},
  month = may,
  year = {2005},
  note = {arXiv:math/0505344},
}
\end{document}